\documentclass[acmsmall,screen,natbib=false]{acmart}

\AtBeginDocument{%
  }
\usepackage[
backend=biber,
style=numeric,
sorting=ynt
]{biblatex}
\setcopyright{acmcopyright}
\copyrightyear{2023}
\acmYear{2023}
\acmDOI{XXXXXXX.XXXXXXX}

\acmJournal{TOPS}

\usepackage{xfrac}
\usepackage{footnote}
\usepackage{subcaption}
\usepackage{enumitem}
\makesavenoteenv{tabular}
\usepackage{tabularx} % better tables
\usepackage{booktabs}
\usepackage[toc,page]{appendix}

\usepackage{xargs}
\usepackage{url}

\usepackage{amssymb}
\usepackage{wasysym} 
\usepackage{pifont}
\usepackage{algorithm2e}
\usepackage[font=small]{caption}
\usepackage{graphicx}
\usepackage{listings}
\usepackage{amsmath,bm}

\def\Snospace~{\S{}}

\newtheorem{mydef}{Definition}
\newtheorem{theorem}{Theorem}
\newtheorem{corollary}{Corollary}[theorem]

\newcommand{\head}[1]{\textbf{#1}} % same as \tableheadelement
\newcolumntype{L}[1]{>{\raggedright\arraybackslash}p{#1}}  % Column style
\newcolumntype{C}[1]{>{\centering\arraybackslash}p{#1}}  % Column style

\newcommand{\seqT}{\text{\bf T}}
\newcommand{\artifactremoval}{\text{\bf A}}
\newcommand{\eg}{e.g.,\xspace} % ',' if American english else ''
\usepackage{adjustbox}

\begin{document}

\title{Intriguing Properties of Adversarial ML Attacks in the Problem Space [Extended Version]}

\author{Jacopo Cortellazzi}
\email{jacopo.cortellazzi@kcl.ac.uk}
\affiliation{
  \institution{King's College London}
  \country{UK} 
}

\author{Feargus Pendlebury}
\email{f.pendlebury@ucl.ac.uk}
\affiliation{
  \institution{University College London}
  \country{UK} 
}
\author{Daniel Arp}
\email{d.arp@tu-berlin.de}
\affiliation{
  \institution{TU Berlin}
  \country{Germany} 
}

\author{Erwin Quiring}
\email{erwin.quiring@rub.de}
\affiliation{
  \institution{Ruhr University Bochum and ICSI}
  \country{Germany} 
}

\author{Fabio Pierazzi}
\email{fabio.pierazzi@kcl.ac.uk}
\affiliation{
  \institution{King's College London}
  \country{UK} 
}

\author{Lorenzo Cavallaro}
\email{l.cavallaro@ucl.ac.uk}
\affiliation{
  \institution{University College London}
  \country{UK} 
}

\renewcommand{\shortauthors}{Cortellazzi et al.}

\begin{abstract}
Recent research efforts on adversarial machine learning (ML) have investigated problem-space attacks, focusing on the generation of
real evasive objects in domains where, unlike images, there is no
clear inverse mapping to the feature space (e.g., software).  However,
the design, comparison, and real-world implications of problem-space attacks remain underexplored.

This article makes three major contributions.  Firstly, we propose a general
formalization for adversarial ML evasion attacks in the
problem-space, which includes the definition of a comprehensive set of
constraints on available transformations, preserved semantics, absent
artifacts, and plausibility.  We shed light on the relationship
between feature space and problem space, and we introduce the concept
of \emph{side-effect features} as the by-product of the inverse
feature-mapping problem.  This enables us to define and prove
necessary and sufficient conditions for the existence of problem-space
attacks.  

Secondly, building on our general formalization, we propose a novel
problem-space attack on Android malware that overcomes past
limitations in terms of semantics and artifacts.  We have tested our approach 
 on a dataset with 150K Android apps from 2016 and 2018 which show the practical
feasibility of evading a state-of-the-art malware classifier along with its hardened version.
Thirdly, we explore the effectiveness of adversarial training as a possible approach to enforce robustness against adversarial samples, evaluating its effectiveness on the considered machine learning models under different scenarios.

Our results demonstrate that ``adversarial-malware as a service’' is a
realistic threat, as we automatically generate thousands of realistic
and inconspicuous adversarial applications at scale, where on
average it takes only a few minutes to generate an adversarial instance.

\end{abstract}

\begin{CCSXML}
<ccs2012>
<concept>
<concept_id>10002978.10002997.10002998</concept_id>
<concept_desc>Security and privacy~Malware and its mitigation</concept_desc>
<concept_significance>500</concept_significance>
</concept>
<concept>
<concept_id>10002978.10003022.10003023</concept_id>
<concept_desc>Security and privacy~Software security engineering</concept_desc>
<concept_significance>500</concept_significance>
</concept>
<concept>
<concept_id>10010147.10010178</concept_id>
<concept_desc>Computing methodologies~Artificial intelligence</concept_desc>
<concept_significance>500</concept_significance>
</concept>
</ccs2012>
\end{CCSXML}



\maketitle

\section{Introduction}

Adversarial machine learning (ML) attacks have been extensively studied across various domains~\cite{Biggio:Wild}, and they present a significant threat to the widespread adoption of machine learning solutions in security-critical scenarios. This paper focuses on test-time evasion attacks in the so-called \emph{problem space}, wherein the challenge lies in modifying real input-space objects to correspond to an adversarial feature vector. 

The main challenge lies in the \emph{inverse feature-mapping problem}~\cite{maiorca2018towards, Maiorca:Bag, Konrad:Attribution, Tygar:Adversarial, Biggio:Evasion, biggio2013security}, as in many settings, converting a feature vector into a problem-space object is not feasible due to the non-invertible and non-differentiable nature of the feature-mapping function. Additionally, the modified problem-space object must be a valid and inconspicuous member of the considered domain while remaining robust against non-ML detection attempts, such as program analysis. Prior research has explored problem-space attacks in diverse areas, including text~\cite{alzantot2018generating, TextBugger}, malicious PDFs~\cite{Maiorca:Bag, Maiorca:PDF, Biggio:Evasion, Evans:EvadeML, Laskov:PDF, Laskov:PDF2}, Android malware~\cite{Battista:SecSVM, Rosenberg:Generic, Yang:Malware}, Windows malware~\cite{Battista:EXE}, source code attribution~\cite{Konrad:Attribution}, malicious Javascript~\cite{Fass:Javascript}, and eyeglass frames~\cite{sharif2016accessorize}. However, while there is a good understanding of how to perform feature-space attacks~\cite{Carlini:Robustness}, the requirements for an attack in the problem space and how to compare the strengths and weaknesses of existing solutions in a principled manner are less clear.

We introduce the concept of \emph{side-effect features} as the by-product of attempting to generate a problem-space transformation that perturbs the feature space in a specific direction. Defining side-effect features allows us to gain deeper insights into the relationships between the feature space and problem space. We establish necessary and sufficient conditions for the existence of problem-space attacks and identify two main types of search strategies (feature-driven and problem-driven) for generating adversarial objects in the problem space. Additionally, we analyze how these samples can enhance the robustness of a target model through adversarial training, considering scenarios where the attacker and defender capabilities depend on gadget characteristics. We explore how representing the same code through multiple representations may impact the potential robustness built during adversarial training. Furthermore, we employ our formalization to describe various interesting attacks proposed in both problem space and feature space. This analysis reveals that some promising prior problem-space attacks in the Android malware domain~\cite{Rosenberg:Generic, Yang:Malware, Papernot:ESORICS} have limitations, particularly concerning semantics and artifacts. The limitations of the mentioned works inspire us to propose a novel problem-space attack in the Android malware domain, addressing these issues and overcoming existing solutions' shortcomings.

At last, we evaluate adversarial training as a potential defense mechanism against problem space attacks and consider the impact and the differences on using multiple generative strategies for hardening. This approach is widely accepted as the most effective method in practice to improve the adversarial robustness of traditional machine learning and deep learning models. The fundamental premise of adversarial training, introduced by \cite{szegedy2013intriguing} \cite{goodfellow2014explaining}, is to train a model to withstand adversarial attacks by exposing it to carefully crafted adversarial examples, which are designed to perturb the model's predictions. Investigating adversarial training methods is crucial in the current landscape of AI and machine learning, where models are increasingly deployed in safety-critical applications such as autonomous driving \cite{zhang2019theoretically} and medical diagnostics \cite{finlayson2019adversarial}. By understanding and improving upon adversarial training techniques, researchers can enhance the security and reliability of AI systems, ultimately contributing to the responsible deployment of artificial intelligence.

In summary, this paper presents three major contributions:
\begin{itemize}
\item We propose a novel general formalization of problem-space attacks (\autoref{sec:formalization}), which defines key requirements and commonalities across different domains, establishes necessary and sufficient conditions for problem-space attacks, and facilitates the comparison of strengths and weaknesses of prior approaches. Notably, existing strategies for adversarial malware generation are found to be among the weakest in terms of attack robustness. We introduce the concept of \emph{side-effect features}, revealing connections between feature space and problem space, enabling principled reasoning about search strategies for problem-space attacks.
\item Building on our formalization, we propose a novel problem-space attack in the Android malware domain, utilizing automated software transplantations~\cite{barr2015automated}, which addresses limitations of prior work regarding semantics and artifacts. Through experiments on a dataset of ~150k apps from 2016-2018 (\autoref{sec:experiments}), we demonstrate the feasibility of evading a state-of-the-art malware classifier, DREBIN~\cite{Arp:Drebin}, and its hardened version, Sec-SVM~\cite{Battista:SecSVM}. 
Compared to the conference version, we have enhanced the gadgets dataset, being able to extract the whole set of DREBIN features. The time required to generate an adversarial example is in the order of minutes, highlighting the realistic threat of the "adversarial-malware as a service" scenario and the insufficiency of existing defenses.
\item We further expand our reasoning from the conference version and propose a comprehensive study of whether different types of adversarial hardening techniques can serve as a valuable defense against this type of adversarial attacks. We consider both adversarial retraining~\cite{chen2020explore} and adversarial training~\cite{madry2017towards}. We show that adversarial training can improve model robustness but the impact of it heavily depends on the considered model. 
\end{itemize}

\section{Problem-Space Adversarial ML Attacks}
\label{sec:formalization}

Our research is centered on \emph{evasion attacks}~\cite{Biggio:Evasion, Carlini:Robustness, Tygar:Adversarial}, where adversaries manipulate objects during the testing phase to produce specific misclassifications.
We draw upon previous literature to give insights into \emph{feature-space} attacks (see \autoref{sec:fs-attack}), and subsequently present a fresh perspective on \emph{problem-space} attacks (\autoref{sec:ps-attack}).
Subsequently, we shed light on the key aspects of our framework, illustrating its application to both the well-established feature-space and the newer problem-space attacks, as informed by related studies across multiple domains (\autoref{sec:instances}).
Our approach to threat modeling, rooted in understanding the attacker's knowledge and abilities, aligns with existing studies\cite{Biggio:Wild, FAIL, Carlini:Evaluating}. A comprehensive overview can be found in Appendix~\ref{app:threatmodel}.
For reader convenience, we have cataloged symbols in Appendix~\ref{app:symbol}.

\subsection{Feature-Space Attacks}
\label{sec:fs-attack}

It's worth noting that the various definitions of feature-space attacks (refer to \autoref{sec:fs-attack}) have been previously summarized in existing literature~\cite{Biggio:Wild,Carlini:Robustness,Battista:SecSVM,Papernot:ESORICS,szegedy2013intriguing,huang2011adversarial,dalvi2004adversarial,lowd2005good}. We present them here for clarity and to set the stage for examining the connections between feature-space and problem-space attacks in subsequent sections.

We introduce a \emph{problem space} $\mathcal{Z}$, also known as the \emph{input space}, encompassing entities of a specific domain, such as images~\cite{Carlini:Robustness}, sound recordings~\cite{Carlini:Audio}, software~\cite{Konrad:Attribution}, or PDF documents~\cite{Maiorca:PDF}. Every entity $z \in \mathcal{Z}$ is paired with a corresponding authentic label~$y \in \mathcal{Y}$, with $\mathcal{Y}$ symbolizing the realm of potential labels. As most machine learning approaches predominantly operate on vectorized numerical data~\cite{Bishop:ML}, entities in $\mathcal{Z}$ must undergo a transformation to be suitable for machine learning procedures.

\begin{mydef}[Feature Mapping]
A \textit{feature mapping} denotes a function $\varphi: \mathcal{Z} \longrightarrow \mathcal{X} \subseteq \mathbb{R}^n$ that transforms a problem-space entity $z \in \mathcal{Z}$ into an $n$-dimensional feature vector $\bm{x} \in \mathcal{X}$, resulting in $\varphi(z)=\bm{x}$. This encapsulates \emph{implicit/latent} mappings as well, where features are not immediately discernible upon input but are internally computed by the model, for instance, in deep learning frameworks~\cite{goodfellow2016deep}.
\end{mydef}

\begin{mydef}[Discriminant Function]
For an $m$-class machine learning classifier represented as $g:\mathcal{X} \longrightarrow \mathcal{Y}$, a \emph{discriminant function} $h: \mathcal{X} \times \mathcal{Y} \longrightarrow \mathbb{R}$ yields a real value $h(\bm{x}, i)$, often expressed succinctly as $h_i(\bm{x})$. This value characterizes the suitability of object $\bm{x}$ for class $i \in \mathcal{Y}$. Elevated returns from the discriminant function $h_i$ indicate a stronger association with class $i$. Specifically, the anticipated label for an entity $\bm{x}$ is determined as $g(\bm{x}) = \hat{y} = \arg\max_{i \in \mathcal{Y}} h_i(\bm{x})$.
\end{mydef}

The goal of a \emph{targeted} feature-space assault is to alter an entity $\bm{x} \in \mathcal{X}$ assigned with the label~$y \in \mathcal{Y}$ into another entity~$\bm{x}'$ that correlates with a designated class~$t \in \mathcal{Y}$, where $t \neq y$ (meaning, to adapt $\bm{x}$ to be wrongly classified under target class $t$). Attackers can discern a disturbance $\bm{\delta}$ to tweak $\bm{x}$ such that $g(\bm{x}+\bm{\delta})=t$ by refining a meticulously designed \emph{attack objective function}. For a comprehensive definition of this function, readers are directed to~\cite{Carlini:Robustness} and \cite{Biggio:Wild}, where \emph{high-confidence} assaults and multi-class scenarios are taken into consideration.

\begin{mydef}[Attack Objective Function]
\label{eq:objfun}
Given an object $\bm{x} \in \mathcal{X}$ and a target label $t \in \mathcal{Y}$, an \emph{attack objective function} $f: \mathcal{X} \times \mathcal{Y} \longrightarrow \mathbb{R}$ is defined as follows:
\begin{align}
	 f(\bm{x}, t) =  \max_{i \neq t} \{ h_i(\bm{x}) \} - h_t(\bm{x})\,,
\end{align}
{for which we use the shorthand $f_{t}(\bm{x})$}. Generally, $\bm{x}$ is classified as a member of $t$ if and only if $f_{t}(\bm{x}) < 0$. An adversary can also enforce a \emph{desired attack confidence} $\kappa \in \mathbb{R}$ such that the attack is considered successful if and only if $f_{t}(\bm{x}) < - \kappa$.
\end{mydef}

The parameter $\kappa$ allows to specify the desired confidence of an adversarial attack. If $\kappa =0$, a \emph{low-confidence} attack is performed where the object is merely misclassified; if $\kappa > 0$ then the object is misclassified with higher confidence. It is worth observing that it is a common misconceptions of adversarial attacks that the feature-space perturbation must be minimal~\cite{Biggio:Wild}. This was derived from the popularity of adversarial attacks in the image domain, in which you want to minimize the perceptual perturbation. It has been already shown in works~\cite{Carlini:Robustness,Carlini:Audio} that \emph{high-confidence} attacks are much more effective in evading a classifier.
The intuition is to minimize $f_{t}$ by modifying $\bm{x}$ in directions that follow the negative gradient of $f_{t}$, i.e., to get $\bm{x}$ closer to the target class $t$.

In addition to the attack objective function, a considered problem-space domain may also come with constraints on the modification of the feature vectors.
For example, in the image domain the value of pixels must be bounded between 0 and 255~\cite{Carlini:Robustness}; in software, some features in $\bm{x}$ may only be added but not removed (e.g., API calls~\cite{Battista:SecSVM}).

\begin{mydef}[Feature-Space Constraints]
	We define $\Omega$ as the set of \emph{feature-space constraints}, i.e., a set of constraints on the possible feature-space modifications. The set $\Omega$ reflects the requirements of realistic problem-space objects.
	Given an object $\bm{x} \in \mathcal{X}$, any modification of its feature values can be represented as a \emph{perturbation vector} $\bm{\delta} \in \mathbb{R}^n$; if $\bm{\delta}$ satisfies $\Omega$, we {borrow notation from \textit{model theory}~\cite{weissmello2015modeltheory}} and write $\bm{\delta} \models \Omega$.
\end{mydef}

These are in part due to domain constraints (e.g., the features are bounded between 0 and 255 in images, and some features cannot be changed) and in part due to allowed modifications constraints (e.g., only addition of software permissions may be allowed, or a maximum perturbation $\delta_{max}$).

As examples of feature-space constraints, in the image domain~\cite[e.g.,][]{Carlini:Robustness, Biggio:Wild} the perturbation $\bm{\delta}$ is subject to an upper bound based on $l_p$ norms ($|| \bm{\delta}||_p \leq \delta_{max}$), to preserve similarity to the original object; in the software domain~\cite[e.g.,][]{Battista:SecSVM, Papernot:ESORICS}, only some features of $\bm{x}$ may be modified, such that $\bm{\delta}_{lb} \preceq \bm{\delta} \preceq \bm{\delta}_{ub}$ (where $\bm{\delta}_1 \preceq \bm{\delta}_2$ implies that each element of $\bm{\delta}_1$ is $\leq$ the corresponding i-th element in $\bm{\delta}_2$).

Common instances of $\Omega$ in the image domain are obtained by constraining the maximum number of perturbations with $l_p$ norms ($|| \bm{\delta}||_p \leq \delta_{max}$); we are more interested in constraints in the feature space, that determine which amount of perturbations the attacker do on each features ($\bm{\delta}_{lb} \preceq \bm{\delta} \preceq \bm{\delta_{ub}}$, where $\bm{v_1} \preceq \bm{v_2}$ implies that each element of $\bm{v_1}$ is $\leq$ of the corresponding i-th element in $\bm{v_2}$).

We can now formalize the traditional feature-space attack as in related work~\cite{Carlini:Robustness, Biggio:Evasion, Battista:SecSVM, Biggio:Wild, papernot2016limitations}.
\begin{mydef}[Feature-Space Attack]
Given a machine learning classifier $g$, an object $\bm{x} \in \mathcal{X}$ with label $y \in \mathcal{Y}$, and a target label $t \in \mathcal{Y}, \, t \neq y$, the adversary aims to identify a perturbation vector $\bm{\delta} \in \mathbb{R}^n$ such that $g(\bm{x}+\bm{\delta})=t$. The desired perturbation can be achieved by solving the following optimization problem:
\begin{align}
	\bm{\delta}^* = \arg\min_{\bm{\delta} \in \mathbb{R}^n} \quad & f_{t}(\bm{x}+\bm{\delta})\\
	\text{subject to:} \quad & \bm{\delta} \models \Omega\,.
\end{align}
A feature-space attack is successful if $f_{t}(\bm{x}+\bm{\delta}^*) < 0$ (or less than $-\kappa$, if the attacker enforces a desired attack confidence).
\end{mydef}

Without loss of generality, we observe that the feature-space attacks definition can be extended to ensure that the adversarial example is closer to the training data points ({e.g., through the tuning of a parameter $\lambda$ that penalizes adversarial examples generated in low density regions, as in the mimicry attacks of~\cite{Biggio:Evasion}}).

\subsection{Problem-Space Attacks}
\label{sec:ps-attack}

This section presents a novel formalization of problem-space attacks and introduces insights into the relationship between feature space and problem space.

{\bf Inverse Feature-Mapping Problem.}
The major challenge that complicates (and, in most cases, prevents) the direct applicability of gradient-driven feature-space attacks to find problem-space adversarial examples is the so-called \emph{inverse feature-mapping problem}~\cite{maiorca2018towards, Maiorca:Bag, Konrad:Attribution, Tygar:Adversarial, Biggio:Evasion, biggio2013security}. As an extension, \cite{Konrad:Attribution} discuss the \emph{feature-problem space dilemma}, which highlights the difficulty of moving in both directions: from feature space to problem space, and from problem space to feature space.
In most cases, the feature mapping function $\varphi$ is not bijective, i.e., \emph{not injective} and \emph{not surjective}.
This means that given $z \in \mathcal{Z}$ with features $\bm{x}$, and a feature-space perturbation $\bm{\delta}^*$, there is no one-to-one mapping that allows going from $\bm{x}+\bm{\delta}^*$ to an adversarial problem-space object $z'$. Nevertheless, there are two additional scenarios.
If $\varphi$ is not invertible but is \emph{differentiable}, then it is possible to backpropagate the gradient of $f_t(\bm{x})$ from $\mathcal{X}$ to $\mathcal{Z}$ to derive how the input can be changed in order to follow the negative gradient (e.g., to know which input pixels to perturbate to follow the gradient in the deep-learning latent feature space).
If $\varphi$ is not invertible and not differentiable, then the challenge is to find a way to map the adversarial feature vector $\bm{x}' \in \mathcal{X}$ to an adversarial object $z' \in \mathcal{Z}$, by applying a transformation to $z$ in order to produce $z'$ such that $\varphi(z')$ is ``as close as possible'' to $\bm{x}'$; i.e., to follow the gradient towards the transformation that most likely leads to a successful evasion~\cite{Battista:EXE}.
In problem-space settings such as software, the function $\varphi$ is typically not invertible and not differentiable, so the search for transforming $z$ to perform the attack cannot be purely gradient-based.

In this section, we consider the general case in which the feature mapping $\varphi$ is not differentiable and not invertible (i.e., the most challenging setting), and we refer to this context to  formalize problem-space evasion attacks.

First, we define a \emph{problem-space transformation} operator through which we can alter problem-space objects.
Due to their generality, we adapt the code transformation definitions from the \emph{compiler engineering} literature~\cite{aho1986compilers,Konrad:Attribution} to formalize general problem-space transformations.

\begin{mydef}[Problem-Space Transformation]
	A problem-space transformation $T:\mathcal{Z}\longrightarrow\mathcal{Z}$ takes a problem-space object $z \in \mathcal{Z}$ as input and modifies it to~$z' \in \mathcal{Z}$. We refer to the following notation: $T(z)=z'$.
\end{mydef}

The possible problem-space transformations are either \emph{addition}, \emph{removal}, or \emph{modification} (i.e., combination of addition and removal). In the case of computer programs,  \emph{obfuscation} is a special case of modification.

\begin{mydef}[Transformation Sequence]
	{A transformation sequence $\seqT = T_n \circ T_{n-1} \circ \dots \circ T_1$ is the subsequent application of problem-space transformations to an object $z \in \mathcal{Z}$}.
\end{mydef}

Intuitively, given a problem-space object $z \in \mathcal{Z}$ with label $y\in\mathcal{Y}$, the purpose of the adversary is to find a transformation sequence $\seqT$ such that the transformed object $\seqT(z)$ is classified into any target class $t$ chosen by the adversary ($t\in\mathcal{Y}$, $t \neq y$). One way to achieve such a transformation is to first compute a feature-space perturbation $\bm{\delta}^*$, and then modify the problem-space object~$z$ so that features corresponding to $\bm{\delta}^*$ are carefully altered. However, in the general case where the feature mapping $\varphi$ is neither invertible nor differentiable, the adversary must perform a search in the problem-space that approximately follows the negative gradient in the feature space. However, this search is not unconstrained, because the adversarial problem-space object $\seqT(z)$ must be realistic.

{\bf Problem-Space Constraints.} Given a problem-space object $z \in \mathcal{Z}$, a transformation sequence $\seqT$ must lead to an object~$z'=\seqT(z)$ that is valid and realistic. To express this formally, we identify four main types of constraints common to any problem-space attack:
\begin{enumerate}
	\item \emph{Available transformations}, which describe which modifications can be performed in the problem-space by the attacker (e.g., only addition and not removal).
	\item \emph{Preserved semantics}, the semantics to be preserved while mutating $z$ to $z'$, with respect to specific feature abstractions which the attacker aims to be resilient against (e.g., in programs, the transformed object may need to produce the same dynamic call traces). Semantics may also be preserved by construction~\cite[e.g.,][]{Konrad:Attribution}.
	\item \emph{Plausibility} (or \emph{Inconspicuousness}), which describes which (qualitative) properties must be preserved in mutating $z$ to $z'$, so that $z$ appears realistic upon manual inspection. For example, often an adversarial image must look like a valid image from the training distribution~\cite{Carlini:Robustness}; a program's source code must look manually written and not artificially or inconsistently altered~\cite{Konrad:Attribution}. In the general case, verification of plausibility may be hard to automate and may require human analysis.
	\item {\emph{Robustness to preprocessing}, which determines which non-ML techniques could disrupt the attack (e.g., filtering in images, dead code removal in programs)}.
\end{enumerate}
These constraints have been sparsely mentioned in prior literature~\cite{Biggio:Evasion, Konrad:Attribution, Evans:EvadeML, Biggio:Wild}, but have never been identified together as a set for problem-space attacks. When designing a novel problem-space attack, it is fundamental to explicitly define these four types of constraints, to clarify strengths and weaknesses. {While we believe that this framework captures all nuances of the current state-of-the-art for a thorough evaluation and comparison, we welcome future research that uses this as a foundation to identify new constraints.}

We now introduce formal definitions for the constraints. First, similarly to~\cite{Battista:SecSVM, Biggio:Wild}, we define the space of available transformations.

\begin{mydef}[Available Transformations]
	We define $\mathcal{T}$ as the space of \emph{available transformations}, which determines which types of automated problem-space transformations $T$ the attacker can perform. In general, it determines if and how the attacker can add, remove, or edit parts of the original object $z \in \mathcal{Z}$ to obtain a new object $z' \in \mathcal{Z}$. We write $\seqT \in \mathcal{T}$ if a transformation sequence consists of available transformations.

\end{mydef}

For example, the pixels of an image may be modified only if they remain within the range of integers 0 to 255~\cite[e.g.,][]{Carlini:Robustness}; in programs, an adversary may only add valid no-op API calls to ensure that modifications preserve functionality~\cite[e.g.,][]{Rosenberg:Generic}.

Moreover, the attacker needs to ensure that some semantics are preserved during the transformation of $z$, according to some feature abstractions. Semantic equivalence is known to be generally undecidable~\cite{barr2015automated, Konrad:Attribution}; hence, as in~\cite{barr2015automated}, we formalize semantic equivalence through \emph{testing}, by borrowing notation from \emph{denotational semantics}~\cite{pierce2002types}.

\begin{mydef}[Preserved Semantics]
Let us consider two problem-space objects $z$ and $z' = \seqT(z)$, and a suite of automated tests $\Upsilon$ to verify preserved semantics. We define $z$ and $z'$ to be \emph{semantically equivalent} with respect to $\Upsilon$ if they satisfy all its tests $\tau \in \Upsilon$, where {$\tau:\mathcal{Z} \times \mathcal{Z} \longrightarrow \mathbb{B}$}. In particular, we denote semantics equivalence with respect to a test suite $\Upsilon$ as follows:
\begin{align}
	\llbracket z \rrbracket^{\tau} = \llbracket z' \rrbracket^{\tau}, \, \forall \tau \in \Upsilon\,,
\end{align}
{where $\llbracket z \rrbracket^{\tau}$ denotes the semantics of $z$ induced during test $\tau$.}
\end{mydef}

Informally, $\Upsilon$ consists of tests that are aimed at evaluating whether $z$ and $z'$ (or parts of them) lead to the same abstract representations in a certain feature space. In other words, the tests in $\Upsilon$ model preserved semantics.
For example, in programs a typical test aims to verify that malicious functionality is preserved; this is done through tests where, given a certain test input, the program produces exactly the same output~\cite{barr2015automated}. Additionally, the attacker may want to ensure that an adversarial program ($z'$) leads to the same instruction trace as its benign version ($z$)---so as not to raise suspicion in feature abstractions derived from dynamic analysis.

Plausibility is more subjective than semantic equivalence, but in many scenarios it is critical that an adversarial object is inconspicuous when manually audited by a human. In order to be plausible, an analyst must believe that the adversarial object is a valid member of the problem-space distribution.

\begin{mydef}[Plausibility]
We define $\Pi$ as the set of {(typically)} manual tests to verify \emph{plausibility}. We say $z$ looks like a valid member of the data distribution to a human being if it satisfies all tests $\pi \in \Pi$, where {$\pi:\mathcal{Z} \longrightarrow \mathbb{B}$}.
\end{mydef}

Plausibility is often hard to verify automatically; previous work has often relied on user studies with domain experts to judge the plausibility of the generated objects (e.g., program plausibility in~\cite{Konrad:Attribution}, realistic eyeglass frames in~\cite{sharif2016accessorize}). Plausibility in software-related domains may also be enforced by construction during the transformation process, e.g., by relying on automated software transplantation~\cite{barr2015automated,Yang:Malware}.

{In addition to semantic equivalence and plausibility, the adversarial problem-space objects need to ensure they are robust to non-ML automated \textit{preprocessing} techniques that could alter properties on which the adversarial attack depends, thus compromising the attack.}

\begin{mydef}[{Robustness to Preprocessing}]
	{We define $\Lambda$ as the set of preprocessing operators an object $z'=\seqT(z)$ should be resilient to. We say $z'$ is robust to preprocessing if $\artifactremoval(\seqT(z))=\seqT(z)$ for all $\artifactremoval \in \Lambda$, where $\artifactremoval:\mathcal{Z}\longrightarrow \mathcal{Z}$ simulates an expected preprocessing.}

\end{mydef}

{Examples of preprocessing operators in $\Lambda$ include compression to remove pixel artifacts (in images), filters to remove noise (in audio), and program analysis to remove dead or redundant code (in programs).}

{Properties affected by preprocessing are often related to \textit{fragile and spurious features} learned by the target classifier. While taking advantage of such features may be necessary to demonstrate the weaknesses of the target model, an attacker should be aware that these brittle features are usually the first to change when a model is improved. Given this, a stronger attack is one that does not rely on them.}

{As a concrete example, in an attack on authorship attribution,~\cite{Konrad:Attribution} purposefully omit layout features (such as the use of spaces vs. tabs) which are trivial to change. Additionally,~\cite{Evans:EvadeML} discovered the presence of font objects was a critical (but erroneously discriminative) feature following their problem-space attack on PDF malware. These are features that are cheap for an attacker to abuse but can be easily removed by the application of some preprocessing. As a defender, investigation of this constraint will help identify features that are weak to adversarial attacks. Note that knowledge of  preprocessing can also be exploited by the attacker (e.g., in \emph{scaling attacks}~\cite{xiao2019scaling}).}

We can now define a {fundamental} set of problem-space constraint elements from the previous definitions.

\begin{mydef}[Problem-Space Constraints]
	We define the \emph{problem-space constraints} $\Gamma = \{ \mathcal{T}, \Upsilon, \Pi, \Lambda \}$ as the set of all constraints satisfying $\mathcal{T}, \Upsilon, \Pi, \Lambda$. We write $\seqT(z) \models \Gamma$ if a transformation sequence applied to object $z\in \mathcal{Z}$ satisfies all the problem-space constraints, and we refer to this as a \emph{valid} transformation sequence. The problem-space constraints $\Gamma$ determine the feature-space constraints $\Omega$, and we denote this relationship as $\Gamma \vdash \Omega$ (i.e., $\Gamma$ determines $\Omega$); with a slight abuse of notation, we can also write that $\Omega \subseteq  \Gamma$, because some constraints may be specific to the problem space (e.g., program size similar to that of benign applications) and may not be possible to enforce in the feature space $\mathcal{X}$.
\end{mydef}

{\bf Side-Effect Features.} Satisfying the problem-space constraints $\Gamma$ further complicates the inverse feature mapping, as $\Gamma$ is a superset of~$\Omega$. Moreover, enforcing $\Gamma$ may require substantially altering an object $z$ to ensure satisfaction of all constraints during mutations. Let us focus on an example in the software domain, so that $z$ is a program with features $\bm{x}$; if we want to transform $z$ to $z'$ such that $\varphi(z')=\bm{x}+\bm{\delta}$, we may want to add to $z$ a program $o$ where $\varphi(o)=\bm{\delta}$. However, the union of $z$ and $o$ may have features different from $\bm{x}+\bm{\delta}$, because other consolidation operations are required (e.g., name deduplication, class declarations, resource name normalization)---which cannot be feasibly computed in advance for each possible object in $\mathcal{Z}$. Hence, after modifying $z$ in an attempt to obtain a problem-space object $z'$ with certain features (e.g., close to $\bm{x}+\bm{\delta}$), the attacker-modified object may have some additional features that are not related to the intended transformation (e.g., adding an API which maps to a feature in $\bm{\delta}$), but are required to satisfy all the problem-space constraints in $\Gamma$ (e.g., inserting valid parameters for the API call, and importing dependencies for its invocation). We call \emph{side-effect features} $\bm{\eta}$ the features that are altered in $z' = \seqT(z)$ specifically for the satisfaction of problem-space constraints. We observe that these features do not follow any particular direction of the gradient, and hence they could have both a positive or negative impact on the classification score.

\begin{figure}[t]
	\centering
	\includegraphics[scale=0.25]{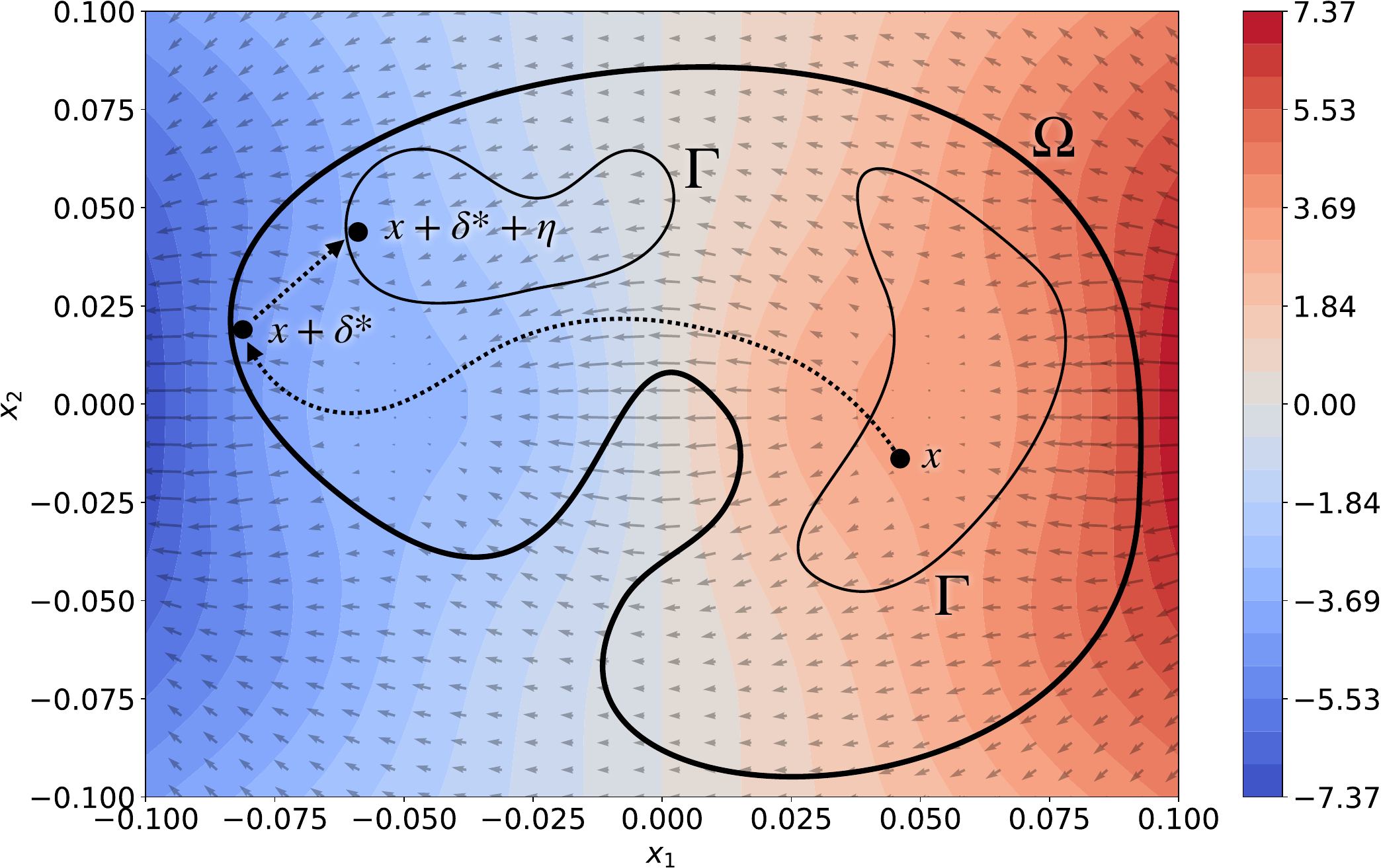}
	\caption{Example of projection of the feature-space attack vector $\bm{x}+\bm{\delta}^*$ in the \emph{feasible} problem space, resulting in side-effect features $\bm{\eta}$. {The background displays the value of the discriminant function $h(\bm{x})$, where negative values indicate the target class of the evasion attack.} Small arrows represent directions of the negative gradient. The thick solid line represents the \emph{feasible} feature space determined by $\Omega$, and the thin solid line that determined by $\Gamma$ (which is more restrictive). The dotted arrow represents the gradient-based attack $\bm{x}+\bm{\delta}^*$ derived from $\bm{x}$, which is then projected into $\bm{x}+\bm{\delta}^*+\bm{\eta}$ to fit into the feasible problem space.}
	\label{fig:projection}
\end{figure}

{\bf Analogy with Projection.}
\autoref{fig:projection} presents an analogy between side-effect features $\bm{\eta}$ and the notion of \emph{projection} in numerical optimization~\cite{Bishop:ML}, which helps explain the nature and impact of $\bm{\eta}$ in problem-space attacks.
The right half corresponds to higher values of a discriminant function $h(\bm{x})$ and the left half to lower values.
The vertical central curve (where the heatmap value is equal to zero) represents the decision boundary: objects on the left-half are classified as negative (e.g., benign), and objects on the right-half as positive (e.g., malicious).
The goal of the adversary is to conduct a \emph{maximum confidence attack} that has an object misclassified as the negative class.
The thick solid line represents the \emph{feasible feature space} determined by constraints $\Omega$, and the thin solid line the \emph{feasible problem space} determined by $\Gamma$ (which corresponds to two unconnected areas).
We assume that the initial object $\bm{x} \in \mathcal{X}$ is always within the feasible problem space.
In this example, the attacker first conducts a gradient-based attack in the feature space on object $\bm{x}$, which results in a feature vector $\bm{x}+\bm{\delta}^*$, which is classified as negative with high-confidence. However, this point is not in the feasibility space of constraints $\Gamma$, which is more restrictive than that of $\Omega$. Hence, the attacker needs to find a \emph{projection} that maps $\bm{x}+\bm{\delta}^*$ back to the feasible problem-space regions, which leads to the addition of a side-effect feature vector $\bm{\eta}$.

\begin{mydef}[Side-Effect Feature Vector]
	We define $\bm{\eta}$ as the \emph{side-effect feature vector} that results from enforcing $\Gamma$ while choosing a sequence of transformations $\seqT$ such that $\seqT(z) \models \Gamma$. In other words, $\bm{\eta}$ are the features derived from the \emph{projection} of a feature-space attack onto a feasibility region that satisfies problem-space constraints $\Gamma$.

\end{mydef}

{We observe that in settings where the feature mapping $\varphi$ is neither differentiable nor invertible, and where the problem-space representation is very different from the feature-space representation (e.g., unlike in images or audio), it is generally infeasible or impossible to compute the exact impact of side-effect features on the objective function in advance---because the set of problem-space constraints $\Gamma$ cannot be expressed analytically in closed-form.} Hence the attacker needs to find a transformation sequence $\seqT$ such that $\varphi(\seqT(z))=\varphi(z')$ is within the feasibility region of problem-space constraints~$\Gamma$.

It is relevant to observe that, in the general case, if an object $z_o$ is added to (or removed from) two different objects $z_1$ and $z_2$, it is possible that the resulting side-effect feature vectors $\bm{\eta}_1$ and $\bm{\eta}_2$ are different (e.g., in the software domain~\cite{Konrad:Attribution}).

{\bf Considerations on Attack Confidence.} There are some important characteristics of the impact of the side-effect features $\bm{\eta}$ on the attack objective function. If the attacker performs a \emph{maximum-confidence attack} in the feature space under constraints $\Omega$, then the confidence of the problem-space attack will always be \emph{lower or equal} than the one in the feature-space attack. This is intuitively represented in \autoref{fig:projection}, where the point is moved to the maximum-confidence attack area within $\Omega$, and the attack confidence is reduced after projection to the feasibility space of the problem space, induced by $\Gamma$. In general, the confidence of the feature- and problem-space attacks could be equal, depending on the constraints $\Omega$ and $\Gamma$, and on the shape of the discriminant function $h$, which is also not necessarily convex (e.g., in deep learning~\cite{goodfellow2016deep}).
In the case of \emph{low-confidence} feature-space attacks, projecting into the problem-space feasibility constraint may result in a positive or negative impact (not known a priori) on the value of the discriminant function. This can be seen from \autoref{fig:projection}, where the object $\bm{x}+\bm{\delta}^*$ would be found close to the center of the plot, where $h(\bm{x})=0$.

{\bf Problem-Space Attack.} We now have all the components that are required to formalize a problem-space attack.

\begin{mydef}[Problem-Space Attack]
	We define a \emph{problem-space attack} as the problem of finding the sequence of valid transformations $\seqT$ for which the object $z \in \mathcal{Z}$ with label $y \in \mathcal{Y}$ is misclassified to a target class $t \in \mathcal{Y}$ as follows:
\begin{align}
	\text{argmin}_{\seqT \in \mathcal{T}} \quad & f_{t}(\varphi(\seqT(z))) = f_{t}(\bm{x} + \bm{\delta}^* + \bm{\eta}) \\ 
	\text{subject to:} \quad
	& \llbracket z \rrbracket^{\tau} = \llbracket \seqT(z) \rrbracket^{\tau}, \quad \forall \tau \in \Upsilon \\
		& \pi(\seqT(z))=1, \quad \forall \pi \in \Pi \\
		& {\artifactremoval(\seqT(z)) = \seqT(z), \quad \forall \artifactremoval \in \Lambda}
\end{align}
where $\bm{\eta}$ is a side-effect feature vector that separates the feature vector generated by $\seqT(z)$ from the theoretical feature-space attack $\bm{x}+\bm{\delta}^*$ (under constraints $\Omega$). An equivalent, more compact, formulation is as follows:
\begin{align}
	\text{argmin}_{\seqT \in \mathcal{T}} \quad & f_{t}(\varphi(\seqT(z))) = f_{t}(\bm{x} + \bm{\delta}^* + \bm{\eta})\\
		\text{subject to:} \quad & \seqT(z) \models \Gamma\,.
\end{align}
\end{mydef}

{\bf Search Strategy.}
The typical search strategy for adversarial perturbations in feature-space attacks is based on following the negative gradient of the objective function through some numerical optimization algorithm, such as stochastic gradient descent~\cite{Biggio:Wild,Carlini:Robustness,Carlini:Audio}.
However, it is not possible to directly apply gradient descent in the general case of problem-space attacks, when the feature space is not invertible nor differentiable~\cite{Konrad:Attribution, Biggio:Wild}; and it is even more complicated if a transformation sequence $\seqT$ produces side-effect features $\bm{\eta} \neq \bm{0}$.

In the problem space, we identify two main types of search strategy: \emph{problem-driven} and \emph{feature-driven}.
In the problem-driven approach, the search of the optimal $\seqT$ proceeds heuristically by beginning with random mutations of the object~$z$, and then learning from experience how to appropriately mutate it further in order to misclassify it to the target class (e.g., using Genetic Programming~\cite{Evans:EvadeML} or variants of Monte Carlo tree search~\cite{Konrad:Attribution}).
This approach iteratively uses local approximations of the negative gradient to mutate the objects.
The feature-driven approach attempts to identify mutations that follow some signal from the feature space. This can be, for example, a target feature vector (e.g., a mimicry attack~\cite{Laskov:PDF2}) or the negative gradient by relying on an approximate inverse feature mapping (e.g., in PDF malware~\cite{Maiorca:Bag}, in Android malware~\cite{Yang:Malware}).
If a search strategy equally makes extensive use of both problem-driven and feature-driven methods, we call it a {\emph{hybrid} strategy}.
We note that search strategies may have different trade-offs in terms of \emph{effectiveness} and \emph{costs}, depending on the time and resources they require. {While there are some promising avenues in this challenging but important line of research~\cite{bogdan2018troncoso}, it warrants further investigation in future work.}

Feature-space attacks can still give us some useful information: before searching for a problem-space attack, we can verify whether a feature-space attack exists, which is a necessary condition for realizing the problem-space attack.

\begin{theorem}[Necessary Condition for Problem-Space Attacks]
	\label{eq:nc}
	Given a problem-space object $z \in \mathcal{Z}$ of class $y \in \mathcal{Y}$, with features $\varphi(z)=\bm{x}$, and a target class $t \in \mathcal{Y}$, $t \neq y$, {there exists} a transformation sequence $\seqT$ that causes $\seqT(z)$ to be misclassified as $t$ \emph{only if} there is a solution for the feature-space attack under constraints $\Omega$.
	More formally, only if:
	\begin{align}\label{eq:necessary-condition}
		\exists \bm{\delta}^* = \arg\min_{\bm{\delta} \in \mathbb{R}^n : \bm{\delta} \models \Omega} f_{t}(\bm{x} + \bm{\delta}) : f_{t}(\bm{x}+\bm{\delta}^*) < 0\,.
	\end{align}
\end{theorem}

The proof of \autoref{eq:nc} is in Appendix~\ref{app:proofs}. We observe that \autoref{eq:nc} is necessary but \emph{not sufficient} because, {although it is not required to be invertible or differentiable, some sort of ``mapping'' between problem- and feature-space perturbations needs to be known by the attacker.  A \textit{sufficient condition} for a problem-space attack, reflecting the attacker's ideal scenario, is knowledge of a set of problem-space transformations which can alter feature values arbitrarily. This describes the scenario for some domains, such as images~\cite{Carlini:Robustness,goodfellow2015adversarial}, in which the attacker can modify any pixel value of an image independently.}

\begin{theorem}[{Sufficient Condition for Problem-Space Attacks}]
\label{eq:nsc}

	Given a problem-space object $z \in \mathcal{Z}$ of class $y \in \mathcal{Y}$, with features $\varphi(z)=\bm{x}$, and a target class $t \in \mathcal{Y}$, $t \neq y$, {there exists} a transformation sequence $\seqT$ that causes $\bm{x}$ to be misclassified as $t$ \emph{if} \autoref{eq:necessary-condition} and \autoref{eq:sufficient-condition} are satisfied:
	\begin{align}
	\tag{\ref{eq:necessary-condition}}
		&\exists \bm{\delta}^* = \arg\min_{\bm{\delta} \in \mathbb{R}^n: \bm{\delta} \models \Omega} f_{t}(\bm{x} + \bm{\delta}) : f_{t}(\bm{x}+\bm{\delta}^*) < 0\\
		&\forall \bm{\delta} \in \mathbb{R}^n: \bm{\delta} \models \Omega, \quad \exists \seqT:\seqT(z) \models \Gamma, \varphi(\seqT(z))=\bm{x}+\bm{\delta} \label{eq:sufficient-condition}
	\end{align}
Informally, an attacker is always able to find a problem-space attack if a feature-space attack exists (necessary condition) and they know problem-space transformations that can modify any feature by any value (sufficient condition).

\end{theorem}

The proof of \autoref{eq:nsc} is in Appendix~\ref{app:proofs}. 
{In the general case, while there may exist an optimal feature-space perturbation $\bm{\delta}^*$, there may \textit{not} exist a problem-space transformation sequence $\seqT$ that alters the feature space of $\seqT(z)$ exactly so that $\varphi(\seqT(z)) = \bm{x} + \bm{\delta}^*$.}

{This is because, in practice, given a target feature-space perturbation $\bm{\delta}^*$, a problem-space transformation may generate a vector $\varphi(\seqT(z))=\bm{x}+\bm{\delta}^*+\bm{\eta^}*$, where $\bm{\eta}^* \neq \bm{0}$  (i.e., where there may exist at least one $i$ for which $\eta_i \neq 0$) due to the requirement that problem-space constraints $\Gamma$ must be satisfied. This prevents easily finding a problem-space transformation that follows the negative gradient. Given this, the attacker is forced to apply some search strategy based on the available transformations.}

\begin{corollary}
	If \autoref{eq:nsc} is satisfied only on a subset of feature dimensions $X_i$ in $\mathcal{X}$, which collectively create a subspace $\mathcal{X}_{eq} \subset \mathcal{X}$, then the attacker can restrict the search space to $\mathcal{X}_{eq}$, for which they know that an equivalent problem/feature-space manipulation exists.
\end{corollary}

\subsection{Describing problem-space attacks in different domains}
\label{sec:instances}

\autoref{tab:instantiations} briefly illustrates the main parameters of problem-space attacks by considering a representative set of adversarial attacks in different domains: 
Android (our attack proposed in \autoref{sec:apg-android} and one recent proposal from the literature~\cite{he2023efficient}), code attribution~\cite{Konrad:Attribution, Matyukhina2019adversarial}, Windows~\cite{Battista:EXE, lucas2021makeover}, text~\cite{TextBugger, ren19generating}, and PDF~\cite{Evans:EvadeML, Laskov:PDF2}.  The references contained in the conference version have been omitted from this document. For a thorough understanding, please consider also the table from the conference version.

This table shows how our formalization can be used to describe and compare problem-space attacks by highlighting their key concepts, strengths and weaknesses.
For example, although transformations obviously depend on the domain, we observe a general trend that modifications adding features are more common than modifications changing or removing features. This can be attributed to the mere complexity of preserving semantics if content is rewritten, and motivates our approach in \autoref{sec:apg-android} to use automated software transplantation with opaque predicates.  
Moreover, various search strategies are used. The strategy is also often linked with the threat scenario. A feature-driven or hybrid strategy is rather used with more knowledge while a problem-driven strategy is chosen with zero knowledge.  
Finally, the semantics are usually preserved while plausibility and robustness to preprocessing are more often neglected. 
For instance, attacks that add content at the end or in unused areas of a file~\cite[e.g.,][]{Battista:EXE,Laskov:PDF2} can be reverted by considering the actual file payload. Likewise, layout modifications and added comments in source code~\cite[e.g.,][]{Matyukhina2019adversarial} can be easily reverted by using code formatting tools or even a simple regular expression, respectively.

\begin{table}[t]
	\centering
	\tiny 
\begin{tabularx}{\textwidth}{L{0.7cm}L{0.7cm}|L{3.75cm}L{0.0cm}L{4.0cm}C{0.1cm}C{0.1cm}|L{2.3cm}}
	 \toprule
        \head{Domain} & \head{Work} & \multicolumn{4}{c}{\head{Problem-Space Constraints}} && 
        \head{Threat Model \&} \vspace{0.06cm} \\
  		& 
        & {Available\ Transformers $\mathcal{T}$}
        & \multicolumn{2}{l}{Plausibility $\Pi$}
		& {$\Upsilon$}
		& {$\Lambda$}
		& \head{Search Strategy} \\ 
  \midrule

Android & \emph{Our Attack} & Code addition through automated software transplantation
& \CIRCLE &(i) Code is realistic by construction through use of exhaustive automated software transplantation. (ii) Mutated apps install and start on an emulator.
&\CIRCLE & \CIRCLE & PK: Feature-driven \\
& \cite{he2023efficient} & Code addition of targeted Android components & \LEFTcircle &(i) Code is realistic by construction through use of limited set of perturbations. (ii) Mutated apps install and start on an emulator. (iii) Added intents could lead to unintended effects & \CIRCLE & \CIRCLE & ZK: Problem-driven \\
\midrule
Code \mbox{Attrib.} & \cite{Konrad:Attribution} & Rewrite source code with targeted code transformations (excl.\ layout on purpose) 
& \CIRCLE & The code does not look suspicious and seems written by a human (survey with developers)
& \CIRCLE & \CIRCLE & ZK: Problem-driven\\
            & \cite{Matyukhina2019adversarial} & Rewrite code by exploiting layout, comments, and control-flow flattening & 
\Circle & Modification visible by e.g.\ adding comments from target or using control-flow flattening & \CIRCLE & \Circle & LK: Feature-driven \\ 
   \midrule	
Windows & \cite{Battista:EXE} & Append bytes at the end & \LEFTcircle & Appended bytes not visible by user, but easily by security expert & \CIRCLE & \Circle & PK: Feature-driven \\
& \cite{lucas2021makeover} & Binary diversification: code is rewritten either through in-place randomization (IPR) or code displacement (DISP) & \CIRCLE & Executed code is modified which might be difficult to spot & \CIRCLE & \CIRCLE & PK: Hybrid, ZK: Problem-driven \\
\midrule
 Text & \cite{TextBugger} & Apply character-level (\eg spelling mistakes) and word-level perturbations (\eg synonyms) & \LEFTcircle & Modifications such as typos are visible, but might be overlooked & \CIRCLE & \LEFTcircle & PK: Hybrid, ZK: Problem-driven \\
 & \cite{ren19generating} & Apply word-level perturbations (\eg synonyms) & \LEFTcircle & Synonyms should preserve meaning, but no guarantee possible & \CIRCLE & \CIRCLE & ZK: Problem-driven \\
 \midrule
 PDF & \cite{Evans:EvadeML} & Addition/Removal of elements in the PDF tree structure & \LEFTcircle & PDF can be opened, but PDF elements are modified which might be visible & \CIRCLE & \LEFTcircle & ZK: Problem-driven \\
 & \cite{Laskov:PDF2} & Add content in unused area between cross-reference table and trailer & \LEFTcircle & Modifications not visible by user, but easily by security expert & \CIRCLE & \Circle & LK, PK:~Feature-driven\\  
\bottomrule
\end{tabularx}
\caption{
Problem-space attacks in different domains (with $\mathcal{Z} \neq \mathcal{F}$), summarized with our formalization.\\
{\footnotesize Recall notation: $\Upsilon$=preserved semantics; $\Lambda$=robustness to preprocessing; ZK=zero knowledge, PK=perfect knowledge, and LK=limited knowledge}
}
\label{tab:instantiations}
\end{table}

\section{Attack on Android}
\label{sec:apg-android}

Our formalization of problem-space attacks reveals weaknesses in prior approaches. 
Hence, we propose---through our formalization---a novel problem-space attack in this domain that overcomes these limitations, especially in terms of preserved semantics and {preprocessing robustness} (see \autoref{sec:instances} and \autoref{sec:related} for a detailed comparison).

\subsection{Threat Model}
\label{sec:apg-threatmodel}

We assume an attacker with \emph{perfect knowledge} $\theta_{PK} = (\mathcal{D}, \mathcal{X}, g, \bm{w})$ (see \autoref{app:threatmodel} for details on threat models). This follows Kerckhoffs' principle~\cite{KerckhoffsLaCM} and ensures a defense does not rely on ``security by obscurity" by unreasonably assuming some properties of the defense can be kept secret~\cite{Carlini:Evaluating}.  Although deep learning has been extensively studied in adversarial attacks, recent research~\cite[e.g.,][]{Tesseract} has shown that---if retrained frequently---the DREBIN classifier~\cite{Arp:Drebin} achieves state-of-the-art performance for Android malware detection, {which makes it a suitable target classifier for our attack}. DREBIN relies on a linear SVM, and embeds apps in a \emph{binary} feature-space $\mathcal{X}$ which captures the presence/absence of components in Android applications in $\mathcal{Z}$ (such as permissions, URLs, Activities, Services, strings).
 We assume to know classifier $g$ and feature-space $\mathcal{X}$, and train the parameters $\bm{w}$ with SVM hyperparameter $C=1$, as in the original DREBIN paper~\cite{Arp:Drebin}.
{Using DREBIN also enables us to evaluate the effectiveness of our problem-space attack against a recently proposed hardened variant, Sec-SVM~\cite{Battista:SecSVM}}. Sec-SVM enforces more evenly distributed feature weights, which require an attacker to modify more features to evade detection.
Although the authors of~\cite{Battista:SecSVM} evaluated the robustness of Sec-SVM in the feature-space $\mathcal{X}$ for increasing $l_0$ attack norms, in~\autoref{sec:experiments} we demonstrate that it is feasible to evade Sec-SVM in the problem-space.

We consider an attacker intending to evade detection based on \emph{static analysis}, without relying on code obfuscation as it may increase suspiciousness of the apps~\cite{Balzarotti:Packer, Balzarotti:DeepPacker} (see \autoref{sec:discussion}).

\subsection{Available Transformations}

We use \emph{automated software transplantation}~\cite{barr2015automated} to extract slices of bytecode (i.e., \emph{gadgets}) from benign \emph{donor} applications and inject them into a malicious \emph{host}, to {mimic the appearance of benign apps and} induce the learning algorithm to misclassify the malicious host as benign.\footnote{Our approach is generic and it would be immediate to do the opposite, i.e., transplant malicious code into a benign app. However, this would require a dataset with \emph{annotated} lines of malicious code. For this practical reason and for the sake of clarity of this section, we consider only the scenario of adding benign code parts to a malicious app.}
An advantage of this process is that we avoid relying on a hardcoded set of transformations~\cite[e.g.,][]{Konrad:Attribution}; this ensures adaptability across different application types and time periods.
In this work, we consider only \emph{addition} of bytecode to the malware---which ensures that we do not hinder the malicious functionality of the app.

\textbf{Organ Harvesting.}
In order to augment a malicious host with a given \emph{benign feature} $X_i$, we must first extract a bytecode gadget $\rho$ corresponding to $X_i$ from some donor app.
As we intend to produce realistic examples, we use \textit{program slicing}~\cite{weiser1981slicing} to extract a functional set of statements that includes a reference to $X_i$.
The final gadget consists of the target reference (\textit{entry point} $L_o$), a forward slice (\textit{organ} $o$), and a backward slice (\textit{vein} $v$).

We first search for $L_o$, corresponding to an appearance of code corresponding to the desired feature in the donor.
Then, to obtain $o$, we perform a context-insensitive forward traversal over the donor's System Dependency Graph (SDG), starting at the entry point, transitively including all of the functions called by any function whose definition is reached.
Finally, we extract $v$, containing all statements needed to construct the parameters at the entry point. To do this, we compute a backward slice by traversing the SDG in reverse.
Note that while there is only one organ, there are usually multiple veins to choose from, but only one is necessary for the transplantation.
When traversing the SDG, class definitions that will certainly be already present in the host are excluded (e.g., system packages such as \lstinline{android} and \lstinline{java}).

For example, for an Activity feature where the variable \lstinline{intent} references the target Activity of interest, we might extract the invocation \lstinline{startActivity(intent)} (entry point $L_o$), the class implementation of the Activity itself along with any referenced classes (organ $o$), and all statements necessary to construct \lstinline{intent} with its parameters (vein $v$).

There is a special case for Activities that have no corresponding vein in the bytecode (e.g., a \lstinline{MainActivity} or an Activity triggered by an intent filter declared in the Manifest); here, we provide an \textit{adapted vein}, a minimal Intent creation and \lstinline{startActivity()} call adapted from a previously mined benign app that will trigger the Activity. Note that organs with original veins are always prioritized above those without.

In this work we have extended the feature extraction capabilities modifying the existing implementation with the goal to amplify the total amount of available gadgets. 
Indeed, our framework currently supports the extraction of all types of features supported by the considered feature space~\cite{Arp:Drebin}. Specifically, while in the conference paper we were supporting the extracting of Activity and URL components, now we also support the extraction of Suspicious calls, API calls, Provider, Receivers and Permissions. We do not focus on intent-filters due to possible disruptions of the original semantic \autoref{sec:semantics}.

With each additional component, we were required to devise a unique strategy for its extraction, thereby augmenting our existing capabilities. Take, for example, the process of extracting services, which necessitated the identification of 'startService' methods as a starting point, followed by tracking the subsequent connections from these methods. This technique was equally applied to other components, including Receivers and Providers. In the case of other categories such as APIs or permissions, our approach centered on pinpointing the specific functions tied to these elements and employing program slicing for identify and extract the whole set of dependencies. This systematic and tailored approach enabled us to effectively dissect and analyze the complexities inherent in each component and category, enhancing the overall scope and precision of our extraction process.

In order to preserve the final logic behind every type of feature we have personalized the extraction mechanism. As explained before, we need to extract pieces of code that reference or use the feature we are interested to extract in order to avoid potential dead code, e.g. for Activity we focus on extracting also the intent which invokes the target feature. This could be more difficult for permissions because we need to identify which piece of code may actually use that and it is not directly referenced. In that specific case, we need indeed a mapping between the actual permission and the possible system calls which would imply its usage.

\textbf{Organ Implantation.}
In order to implant some gadget $\rho$ into a host, it is necessary to identify an injection point $L_H$ where $v$ should be inserted.
Implantation at $L_H$ should fulfill two criteria: firstly, it should maintain the syntactic validity of the host; {secondly, it should be as unnoticeable as possible so as not to contribute to any violation of plausibility}.

To maximize the probability of fulfilling the first criterion, we restrict $L_H$ to be between two statements of a class definition in a non-system package.

For the second criterion, we take a heuristic approach by using \emph{Cyclomatic Complexity} (CC)---a software metric that quantifies the code complexity of components within the host---and choosing $L_H$ such that we maintain existing homogeneity of CC across all components.
Finally, the host entry point $L_H$ is inserted into a \emph{randomly chosen} function among those of the selected class, to avoid creating a pattern that might be identified by an analyst.

\subsection{Preserved Semantics}
\label{sec:semantics}
Given an application $z$ and its modified (adversarial) version $z'$, we aim to ensure that $z$ and $z'$ lead to the same dynamic execution, i.e., the malicious behavior of the application is preserved.
We enforce this by construction by wrapping the newly injected execution paths in conditional statements that always return {\tt False}.
This guarantees the newly inserted code is never executed at runtime---so users will not notice anything odd while using the modified app.
In~\autoref{sec:opaque}, we describe how we generate such conditionals without leaving artifacts.

To further preserve semantics, we also decide to omit {\tt intent-filter} elements as transplantation candidates.
For example, an {\tt intent-filter} could declare the app as an eligible option for reading PDF files; consequently, whenever attempting to open a PDF file, the user would be able to choose the host app, which (if selected) would trigger an Activity defined in the transplanted benign bytecode---violating our constraint of preserving dynamic functionality.

\subsection{{Robustness to Preprocessing}}
\label{sec:opaque}

{Program analysis techniques that perform redundant code elimination would remove unreachable code.} Our evasion attack relies on features associated with the transplanted code, and to preserve semantics we need conditional statements that always resolve to {\tt False} at runtime; so, we must subvert static analysis techniques that may identify that this code is never executed.
We achieve this by relying on \emph{opaque predicates}~\cite{Moser:opaque}, i.e., carefully constructed obfuscated conditions where the outcome is always known at design time (in our case, \texttt{False}), but the actual truth value is difficult or impossible to determine during a static analysis.

We refer the reader to Appendix~\ref{app:opaque} for a detailed description of how we generate strong opaque predicates and make them look legitimate.

\subsection{Plausibility}

In our model, an example is satisfactorily plausible if it resembles a real, functioning Android application (i.e., is a valid member of the problem-space $\mathcal{Z}$).
Our methodology aims to maximize the plausibility of each generated object by injecting full slices of bytecode from \emph{real} benign applications.
There is only one case in which we inject artificial code: the opaque predicates that guard the entry point of each gadget (see Appendix \ref{app:opaque} for an example).

In general, we can conclude that plausibility is guaranteed \emph{by construction} thanks to the use of automated software transplantation~\cite{barr2015automated}.
This contrasts with other approaches that inject \emph{standalone} API calls and URLs or \emph{no-op} operations~\cite[e.g.,][]{Rosenberg:Generic} that are completely orphaned and unsupported by the rest of the bytecode (e.g., an API call result that is never used).

We also practically assess that each mutated app still functions properly after modification by installing and running it on an Android emulator.
Although we are unable to thoroughly explore every path of the app in this automated manner, it suffices as a smoke test to ensure that we have not fundamentally damaged the structure of the app.

\subsection{Search Strategy}

We propose a \emph{feature-driven} search strategy based on a \emph{greedy algorithm}, which aims to follow the gradient direction by transplanting a gadget with benign features into the malicious host.
There are two main phases: \emph{Initialization} (Ice-Box Creation) and \emph{Attack} (Adversarial Program Generation).
This section offers an overview of the proposed search strategy, and the detailed steps are reported in Appendix~\ref{app:algos}.

{\bf Initialization Phase (Ice-Box Creation).}
We first harvest gadgets from potential donors and collect them in an \emph{ice-box} $G$, which is used for transplantation at attack time. The main reason for this, instead of looking for gadgets on-the-fly, is to have an immediate estimate of the \emph{side-effect features} when each gadget is considered for transplantation.
Looking for gadgets on-the-fly is possible, but may lead to less optimal solutions and uncertain execution times.

For the initialization we aim to gather gadgets that move the score of an object towards the benign class (i.e., negative score), hence we consider the classifier's top $n_f$ benign features (i.e., with negative weight).
For each of the top-$n_f$ features, we extract $n_d$ candidate gadgets, excluding those that lead to an overall positive (i.e., malicious) score.
We recall that this may happen even for benign features since the context extracted through forward and backward slicing may contain many other features that are indicative of maliciousness.
We empirically verify that with $n_f=500$ and $n_d=5$ we are able to create a successfully evasive app for all the malware in our experiments.
To estimate the side-effect feature vectors for the gadgets, we inject each into a \emph{minimal app}, i.e., an Android app we developed with minimal functionality (see Appendix~\ref{app:algos}).
It is important to observe that the ice-box can be expanded over time, as long as the target classifier does not change its weights significantly.
Algorithm~\ref{alg:initialization} in Appendix~\ref{app:algos} reports the detailed steps of the initialization phase.

{\bf Attack Phase.}
We aim to automatically mutate $z$ into $z'$ so that it is misclassified as goodware, i.e., $h(\varphi(z'))<0$, by transplanting harvested gadgets from the ice-box $G$.
First we search for the list of ice-box gadgets that should be injected into $z$.
Each gadget $\rho_j$ in the ice-box $G$ has feature vector $\bm{r}_j$ which includes the desired feature and side-effect features.
We consider the actual feature-space contribution of gadget $i$ to the malicious host $z$ with features $\bm{x}$ by performing the set difference of the two binary vectors, $\bm{r}_j \wedge \neg \bm{x}$.
We then sort the gadgets in order of decreasing negative contribution, which ideally leads to a faster convergence of $z$'s score to a benign value.

Next we filter this candidate list to include gadgets \emph{only if} they satisfy some practical feasibility criteria.
We define a \emph{check\_feasibility} function which implements some heuristics to limit the excessive increase of certain statistics which would raise suspiciousness of the app.
Preliminary experiments revealed a tendency to add too many permissions to the Android Manifest, hence, we empirically enforce that candidate gadgets add no more than 1 new permission to the host app.
Moreover, we do not allow addition of permissions listed as \emph{dangerous} in the Android documentation~\cite{android:dangerous}.
The other app statistics remain reasonably within the distribution of benign apps (more discussion in~\autoref{sec:experiments}), and so we decide not to enforce a limit on them.

The remaining candidate gadgets are iterated over and for each candidate $\rho_j$, we combine the gadget feature vector $\bm{r}_j$ with the input malware feature vector $\bm{x}$, such that $\bm{x}' = \bm{x} \vee \bm{r}_j$.
We repeat this procedure until the updated $\bm{x}'$ is classified as goodware (for low-confidence attacks) or until an attacker-defined confidence level is achieved (for high-confidence attacks).

Finally, we inject all the candidate gadgets at once through automated software transplantation, and check that problem-space constraints are verified and that the app is still classified as goodware.

Algorithm~\ref{alg:attack} in Appendix~\ref{app:algos} reports the detailed steps of the attack phase.

\section{Experimental Evaluation}
\label{sec:experiments}

We evaluate the effectiveness of our novel problem-space Android attack, in terms of  success rate and required time---and also when in the presence of feature-space defenses.

\subsection{Experimental Settings}
\label{sec:exp-settings}

{\hspace{\parindent} \bf Prototype.}
We create a prototype of our novel problem-space attack (\autoref{sec:apg-android}) using a combination of Python for the ML functionality and Java for the program analysis operations; in particular, to perform transplantations in the problem-space we rely on FlowDroid~\cite{arzt2014flowdroid}, which is based on Soot~\cite{vallee2010soot}.
We release the code of our prototype to other academic researchers (see~\autoref{sec:availability}).
We ran all experiments on an Ubuntu VM with 36 vCPUs, 252GB of RAM, and NVIDIA Tesla V100 GPU.

{\bf Classifiers.}
As defined in the threat model (\autoref{sec:apg-threatmodel}), we consider the DREBIN classifier~\cite{Arp:Drebin}, based on a binary feature space and a linear SVM, and its recently proposed hardened variant, Sec-SVM~\cite{Battista:SecSVM}, which requires the attacker to modify more features to perform an evasion. This last model is implemented in PyTorch. In this work, we apply a feature selection selecting the 10,000 most important features out of the $\approx$1,5 million features of the original representation, maintaining the same clean performance.
We use hyperparameter C=1 for the linear SVM as in~\cite{Arp:Drebin} with the dual problem formulation, and identify the optimal Sec-SVM parameter $k=0.25$ (i.e., the maximum feature weight) in our setting by enforcing a maximum performance loss of 2\% AUC. See Appendix~\ref{app:secsvm} for implementation details.

{\bf Attack Confidence.}
\label{attack_confidence}
We consider two attack settings: \emph{low-confidence} ({\bf L}) and \emph{high-confidence} ({\bf H}). The (L) attack merely overcomes the decision boundary (so that $h(\bm{x})<0$). The (H) attack maximizes the distance from the hyperplane into the goodware region; while generally this distance is unconstrained, {here we set it to be $\leq$ the negative scores of 25\% of the benign apps (i.e., within their interquartile range). This avoids making superfluous modifications, which may only increase suspiciousness or the chance of transplantation errors, while being closer in nature to past mimicry attacks~\cite{Biggio:Evasion}.}

{\bf Dataset.}
We collect apps from AndroZoo~\cite{Allix:AndroZoo}, a large-scale dataset with timestamped Android apps crawled from different stores, and with VirusTotal summary reports.
We use the same labeling criteria as Tesseract~\cite{Tesseract} (which is derived from~\cite{Miller:Reviewer}): an app is considered \emph{goodware} if it has 0 VirusTotal detections, as \emph{malware} if it has 4+ VirusTotal detections, and is discarded as \emph{grayware} if it has between 1 and 3 VirusTotal detections.
For the dataset composition, we follow the example of Tesseract and use an average of 10\% malware~\cite{Tesseract}.
The final dataset contains $\approx$150K recent Android applications, dated between Jan 2016 and Dec 2018, specifically 135,708 goodware and 15,765 malware.

{\bf Dataset Split.}
Tesseract~\cite{Tesseract} demonstrated that, in non-stationary contexts such as Android malware, if time-aware splits are not considered, then the results may be inflated due to \emph{concept drift} (i.e., changes in the data distribution).
However, here we aim to specifically evaluate the effectiveness of an adversarial attack. Although it likely exists, the relationship between adversarial and concept drift is still unknown and is outside the scope of this work.
If we were to perform a time-aware split, it would be impossible to determine whether the success rate of our ML-driven adversarial attack was due to an intrinsic weakness of the classifier or due to natural evolution of malware (i.e., the introduction of new non-ML techniques malware developers rely on to evade detection).
Hence, we perform a \emph{random split} of the dataset to simulate \emph{absence of concept drift}~\cite{Tesseract}; this also represents the most challenging scenario for an attacker, as they aim to mutate a test object coming from the same distribution as the training dataset (on which the classifier likely has higher confidence).
In particular, we consider a 66\% training and 34\% testing random split.\footnote{We consider only one split due to the overall time required to run the experiments. Including some prototype overhead, it requires about one month to run all configurations.}

{\bf Testing.} The test set contains a total of 5,203 malware.
The statistics reported in the remainder of this section refer only to \emph{true positive} malware (3,897 for SVM and 3,762 for Sec-SVM), i.e., we create adversarial variants only if the app is detected as malware by the classifier under evaluation.
Intuitively, it is not necessary to make an adversarial example of a malware application that is already misclassified as goodware; hence, we avoid inflating results by removing false negative objects from the dataset.
For the evaluation we apply a restriction on the harvested gadgets, selecting the stealthiest ones: we are indeed considering only the gadgets with maximum 20 classes in total and maximum one permission.
During the transplantation phase of our problem-space attack some errors occur due to bugs and corner-case errors in the FlowDroid framework~\cite{arzt2014flowdroid}.
Since these errors are related on implementation limitations of the FlowDroid research prototype, and not conceptual errors, the success rates in the remainder of this section refer only to applications that did not throw FlowDroid exceptions during the transplantation phase.

\subsection{Evaluation}

We analyze the performance of our Android problem-space attack in terms of runtime cost and successful evasion rate.
An attack is successful if an app $z$, originally classified as malware, is mutated into an app $z'$ that is classified as goodware and satisfies the problem-space constraints.

\begin{figure}
\centering
\begin{subfigure}{.4\columnwidth}
  \centering
  \includegraphics[width=\columnwidth]{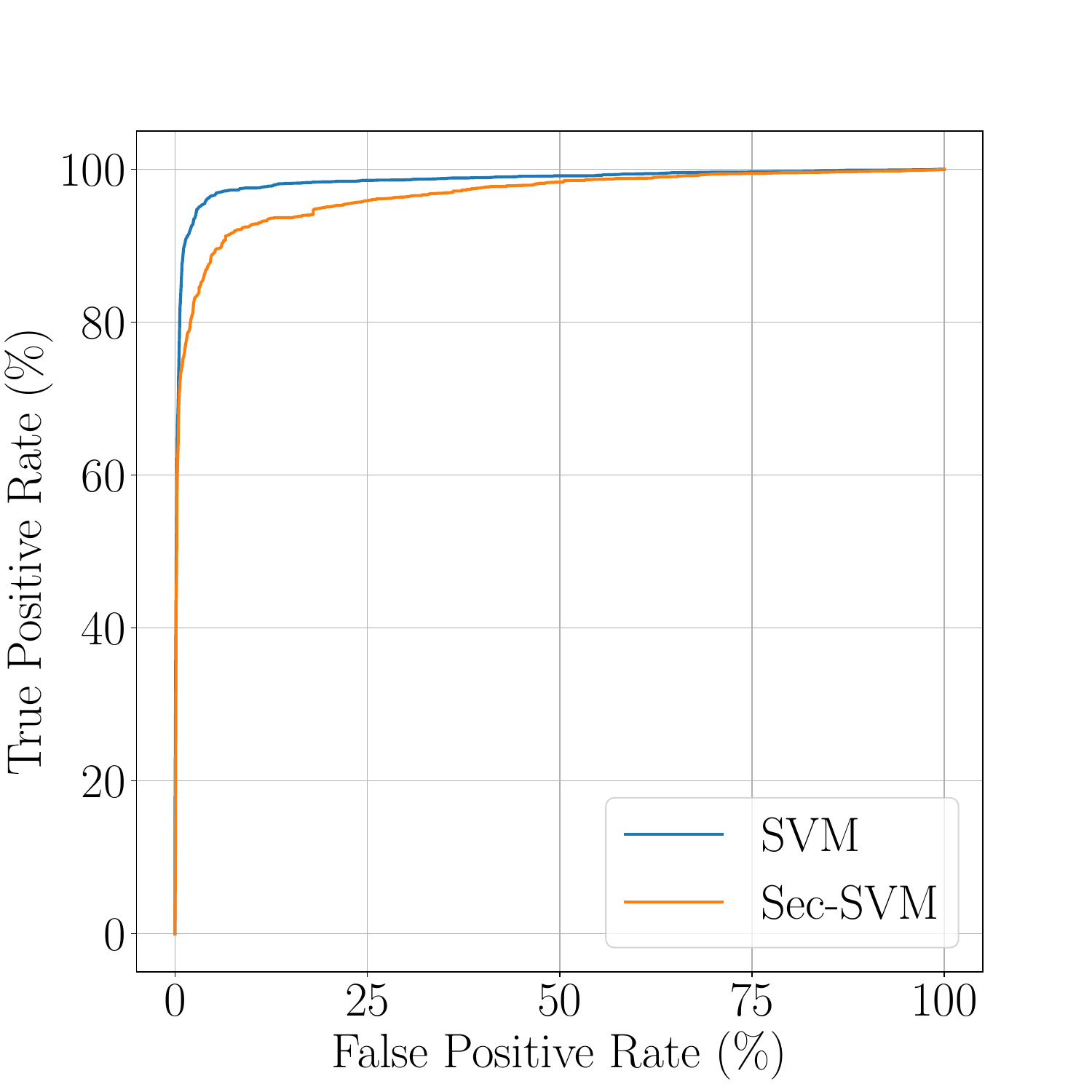}
\end{subfigure}
\begin{subfigure}{.4\columnwidth}
  \centering
  \includegraphics[width=\columnwidth]{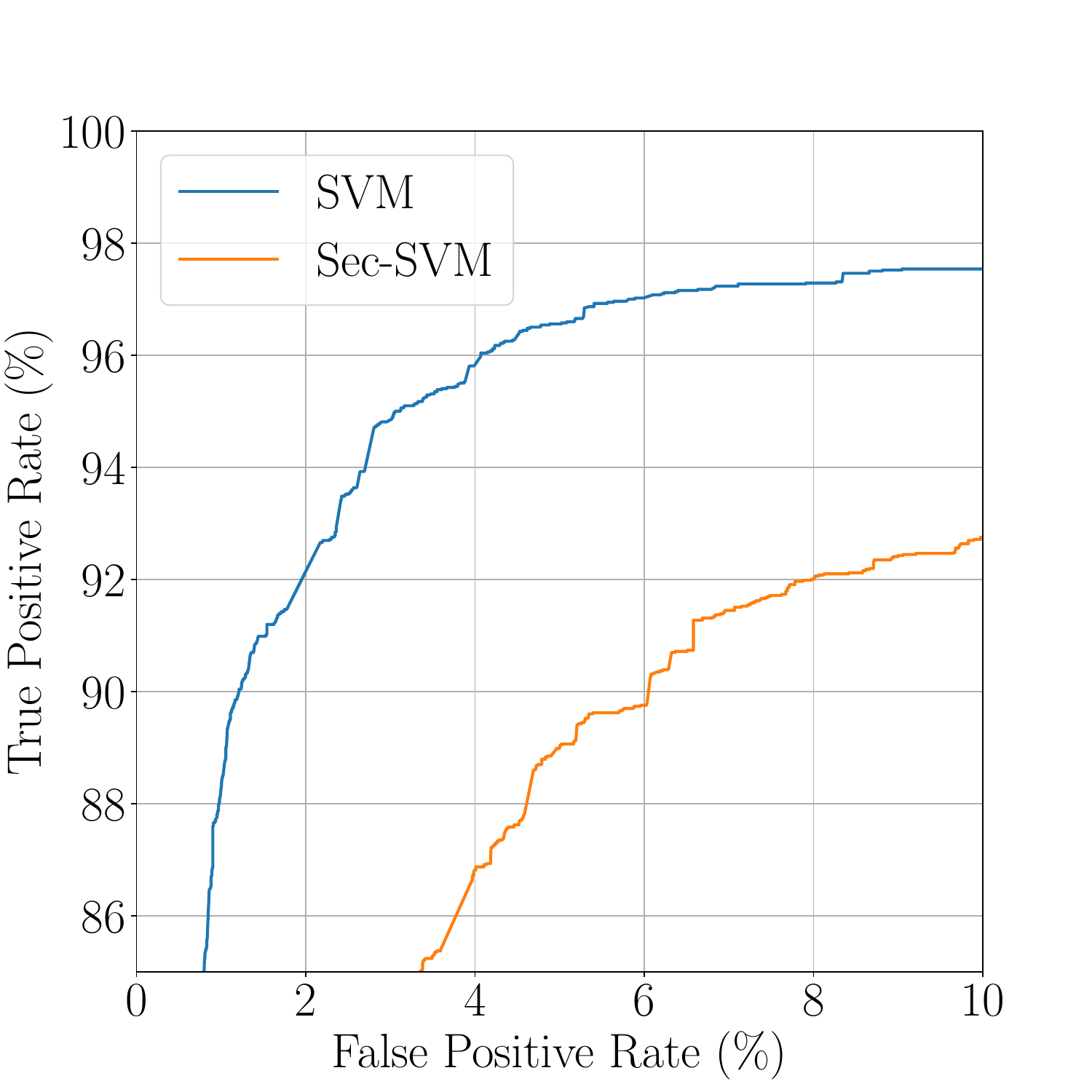}
\end{subfigure}
\caption{Performance of SVM and Sec-SVM in absence of adversarial attacks.}
\label{fig:roc}
\end{figure}

\autoref{fig:roc} reports the AUROC of SVM and Sec-SVM on the DREBIN feature space in absence of attacks.
As expected~\cite{Battista:SecSVM}, Sec-SVM sacrifices some detection performance in return for greater feature-space adversarial robustness.

\begin{figure}
\centering
\begin{subfigure}{.3\columnwidth}
  \centering
  \includegraphics[width=\columnwidth]{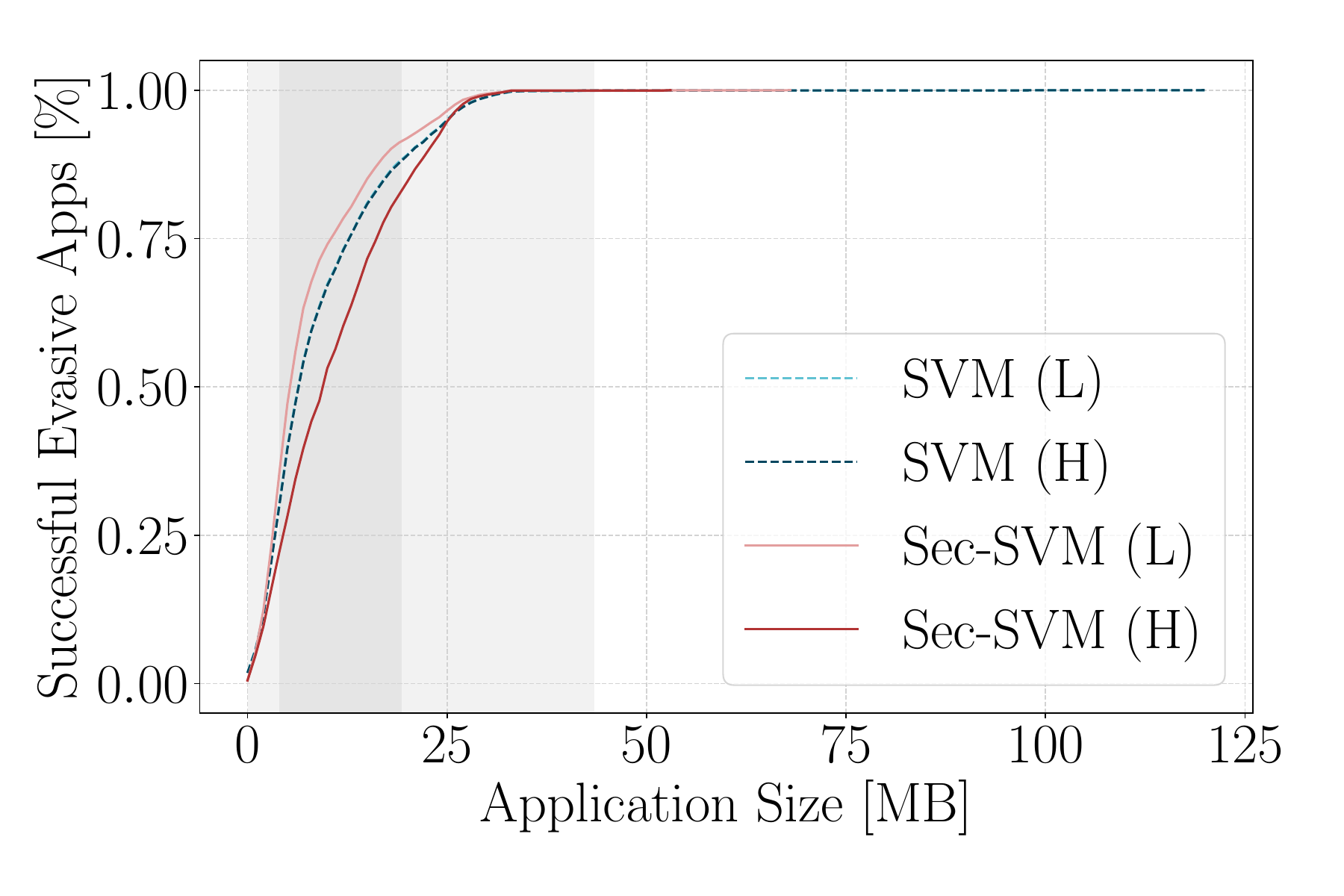}
\end{subfigure}
\begin{subfigure}{.3\columnwidth}
  \centering
  \includegraphics[width=\columnwidth]{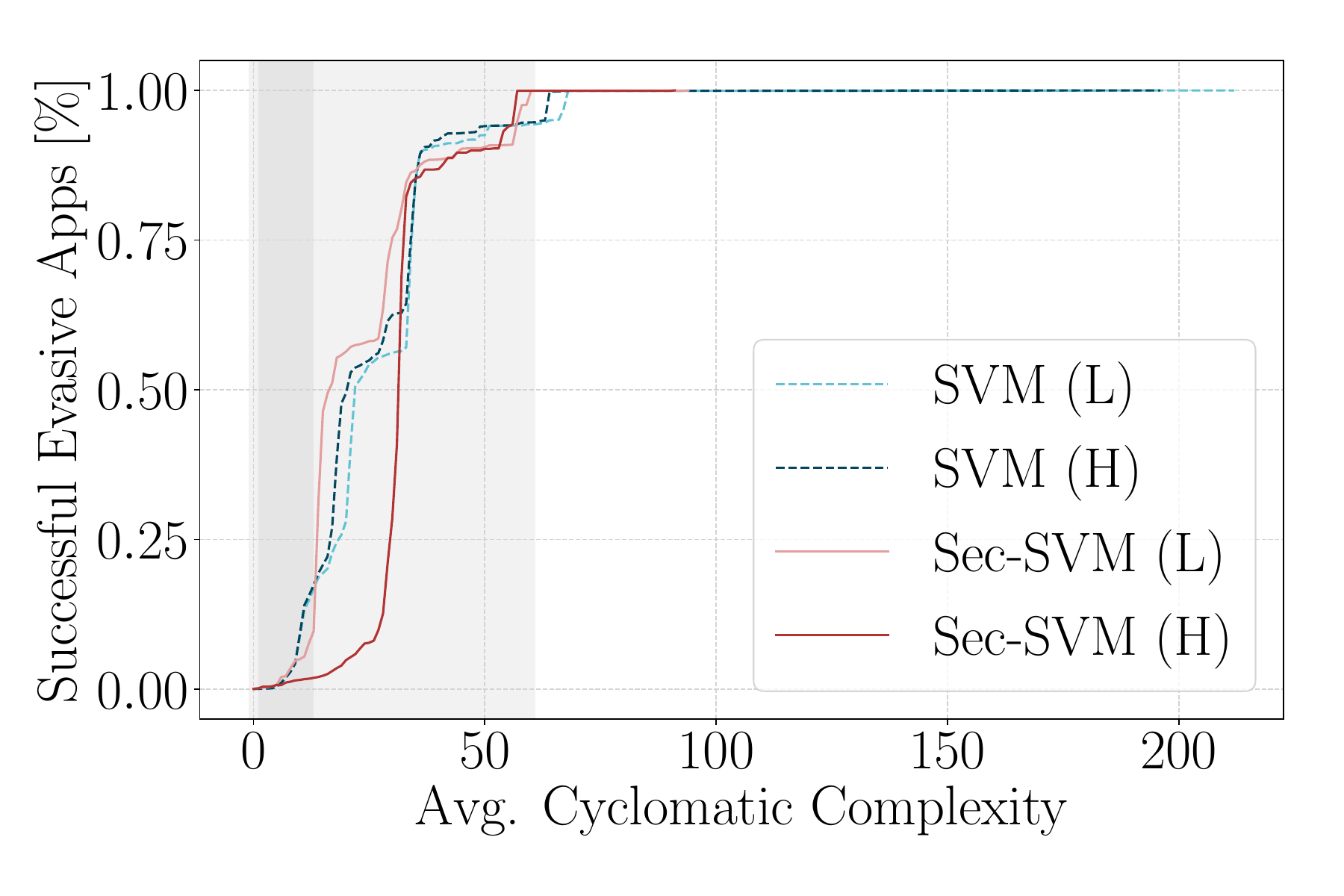}
\end{subfigure}
\begin{subfigure}{.3\columnwidth}
  \centering
  \includegraphics[width=\columnwidth]{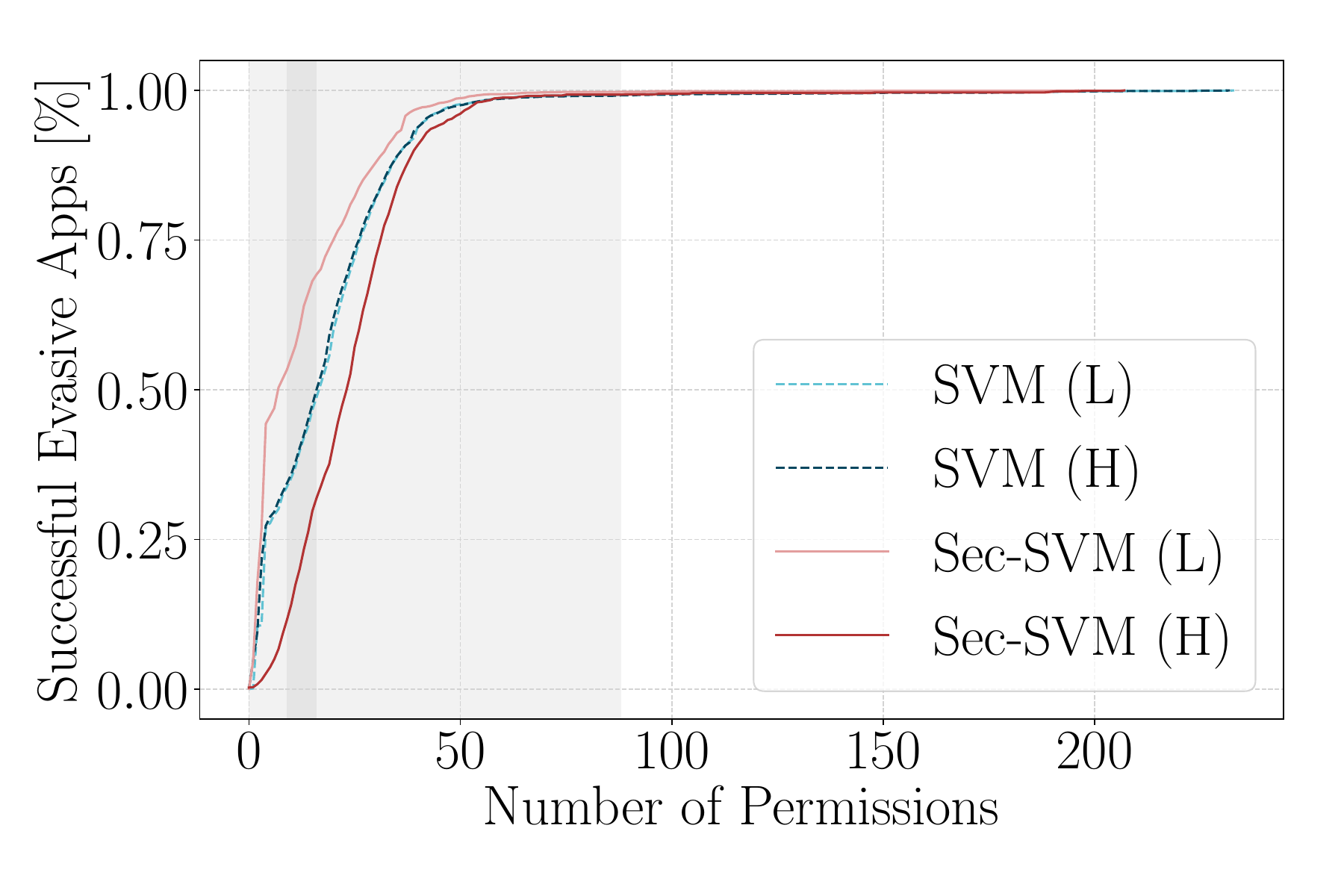}
\end{subfigure}
\begin{subfigure}{.3\columnwidth}
  \centering
  \includegraphics[width=\columnwidth]{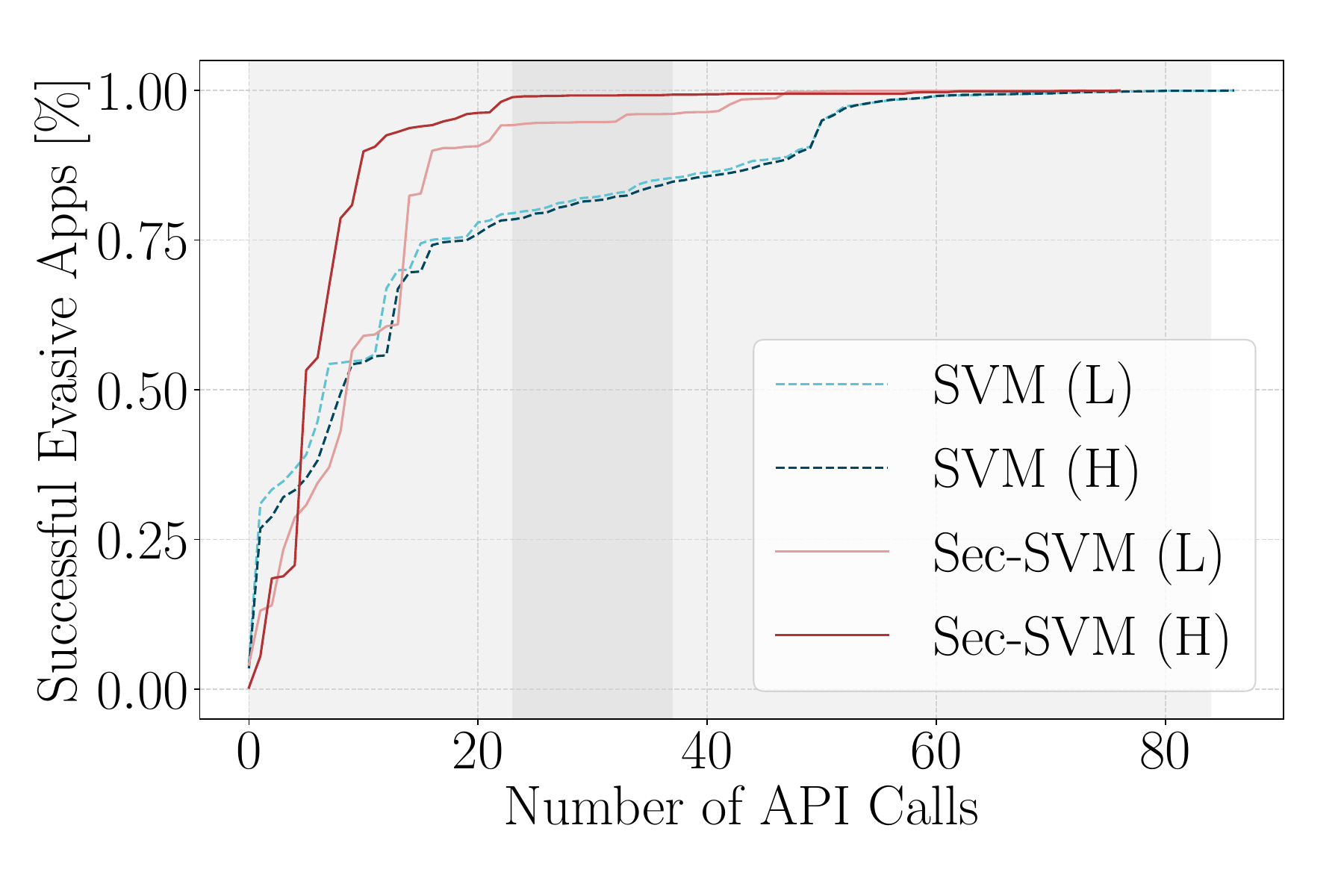}
\end{subfigure}
\begin{subfigure}{.3\columnwidth}
  \centering
  \includegraphics[width=\columnwidth]{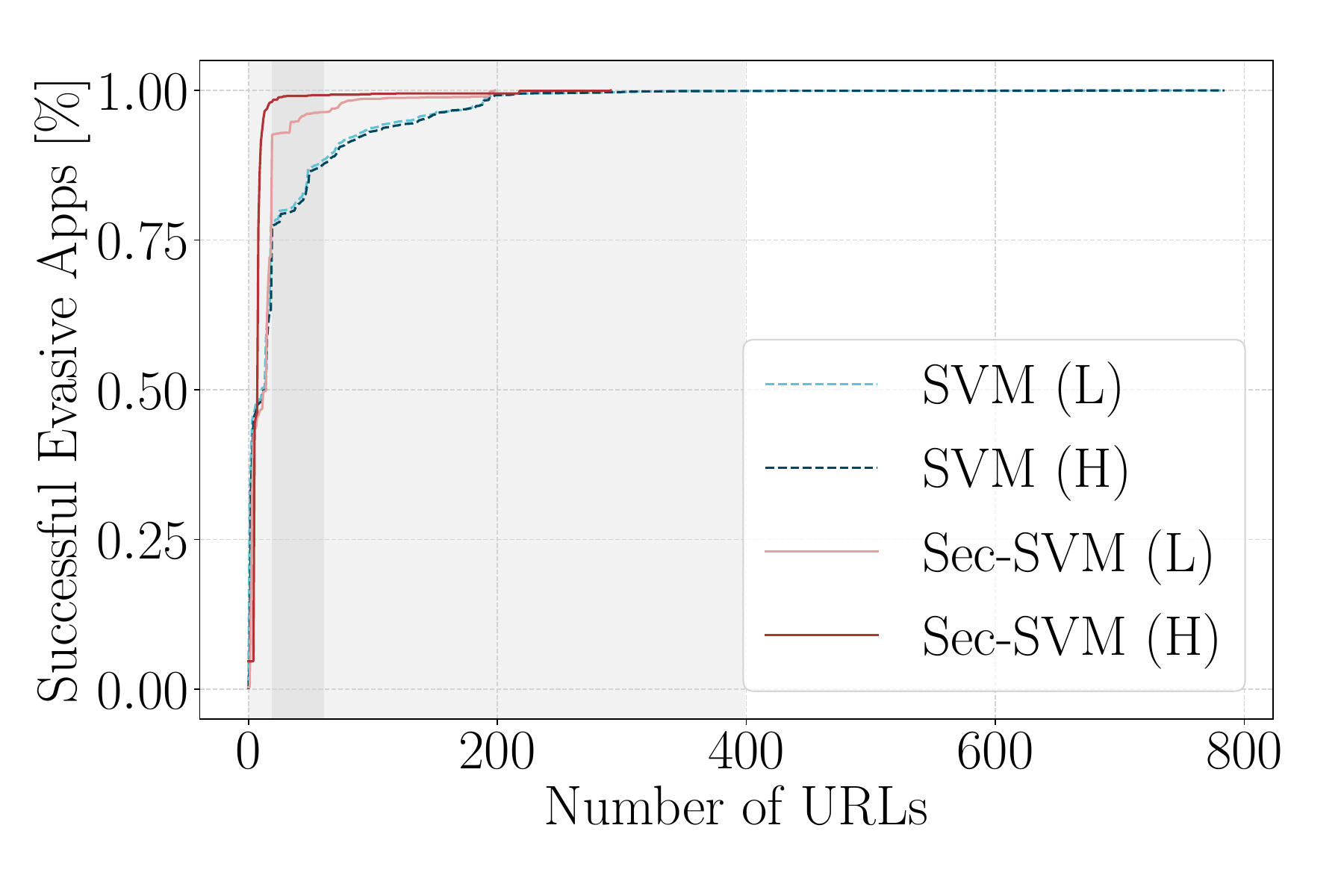}
\end{subfigure}
\begin{subfigure}{.3\columnwidth}
  \centering
  \includegraphics[width=\columnwidth]{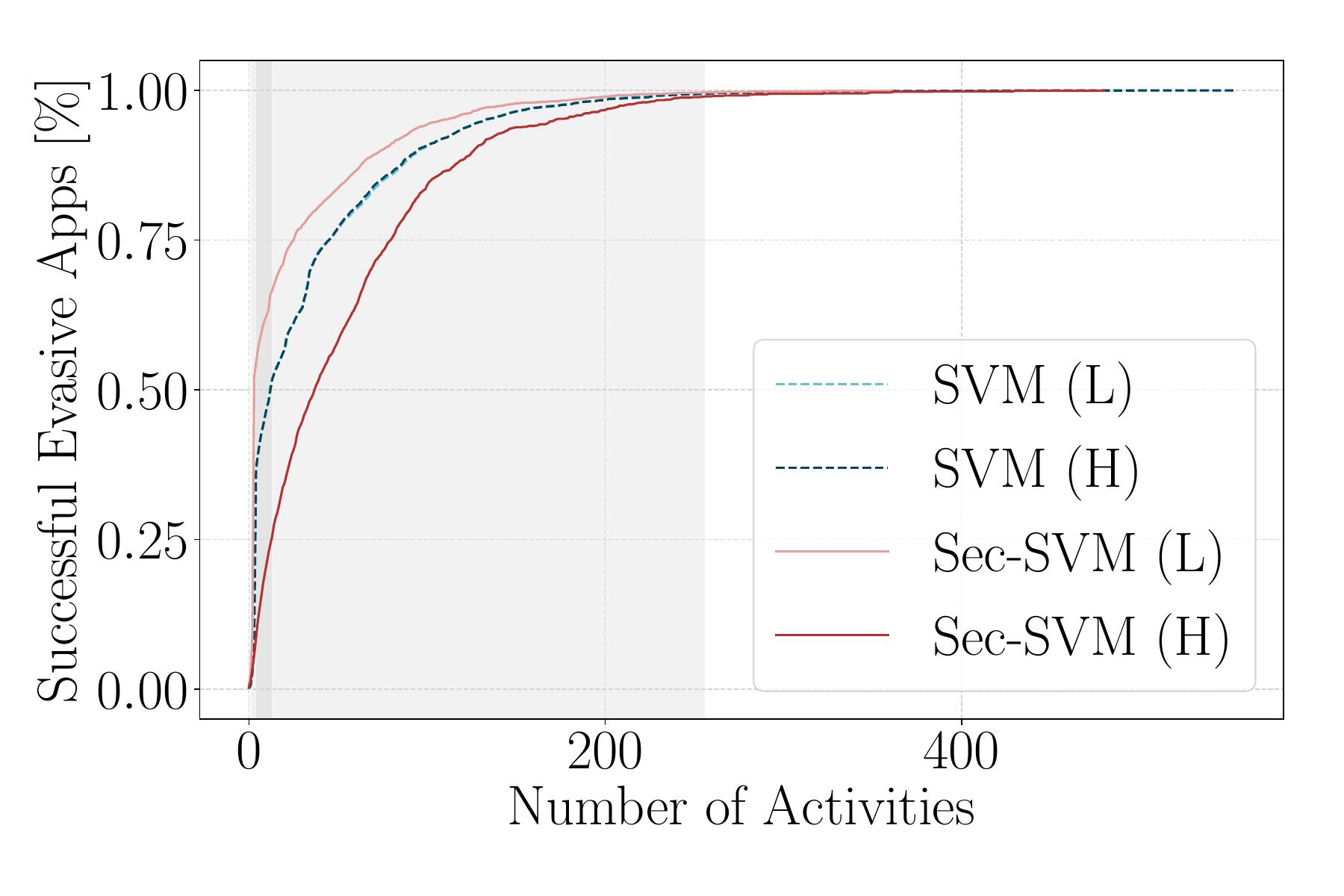}
\end{subfigure}
\begin{subfigure}{.3\columnwidth}
  \centering
  \includegraphics[width=\columnwidth]{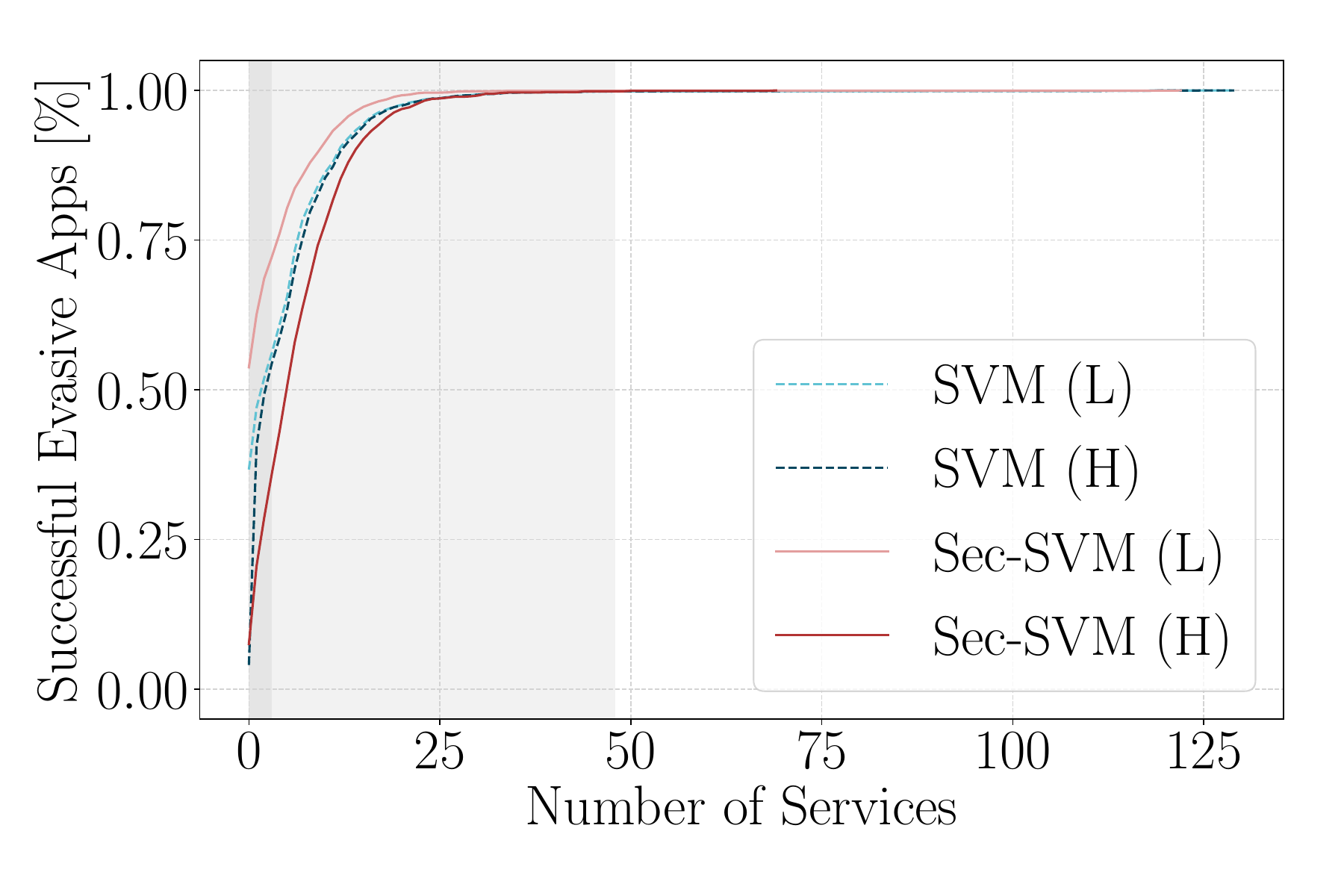}
\end{subfigure}
\begin{subfigure}{.3\columnwidth}
  \centering
  \includegraphics[width=\columnwidth]{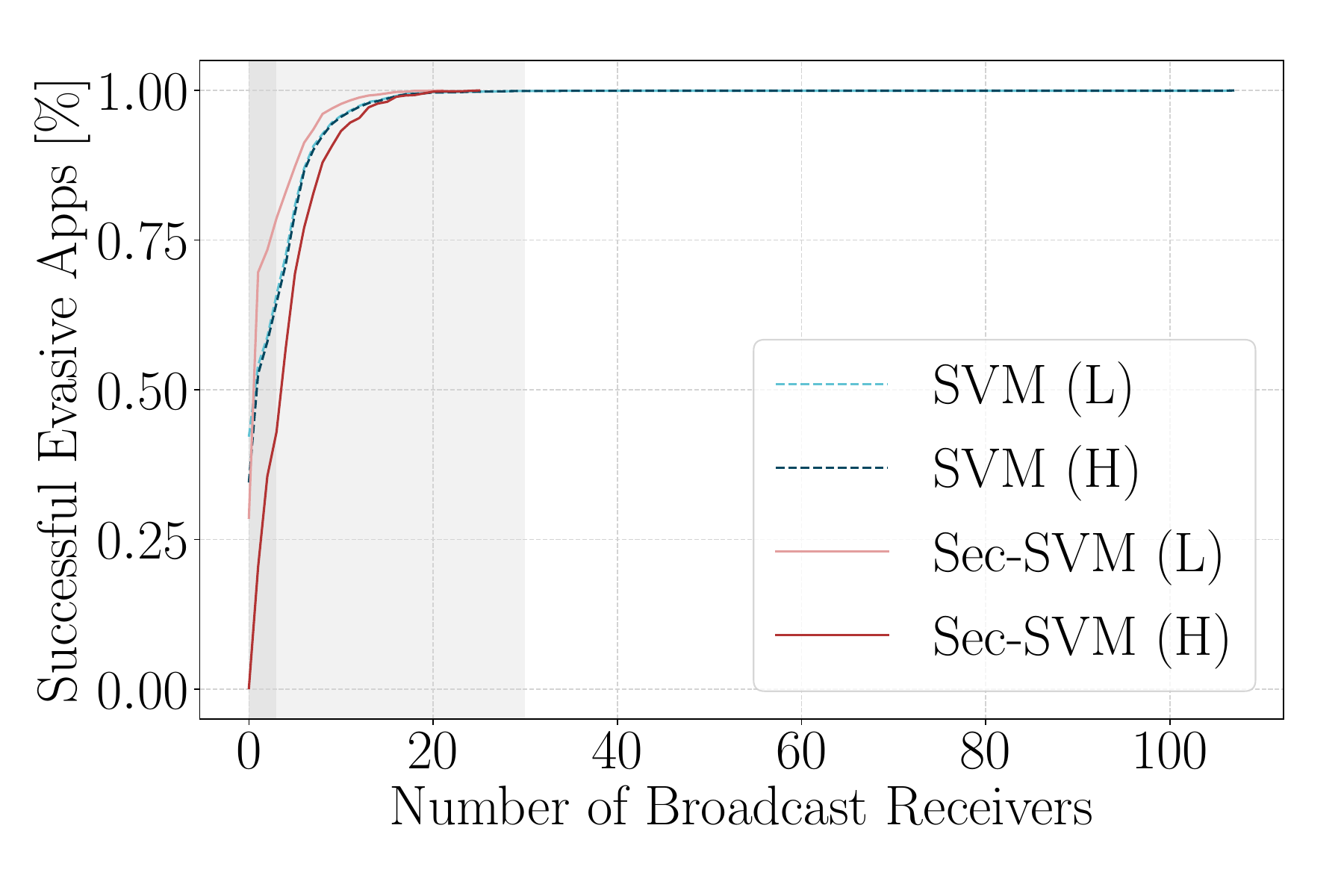}
\end{subfigure}
\begin{subfigure}{.3\columnwidth}
  \centering
  \includegraphics[width=\columnwidth]{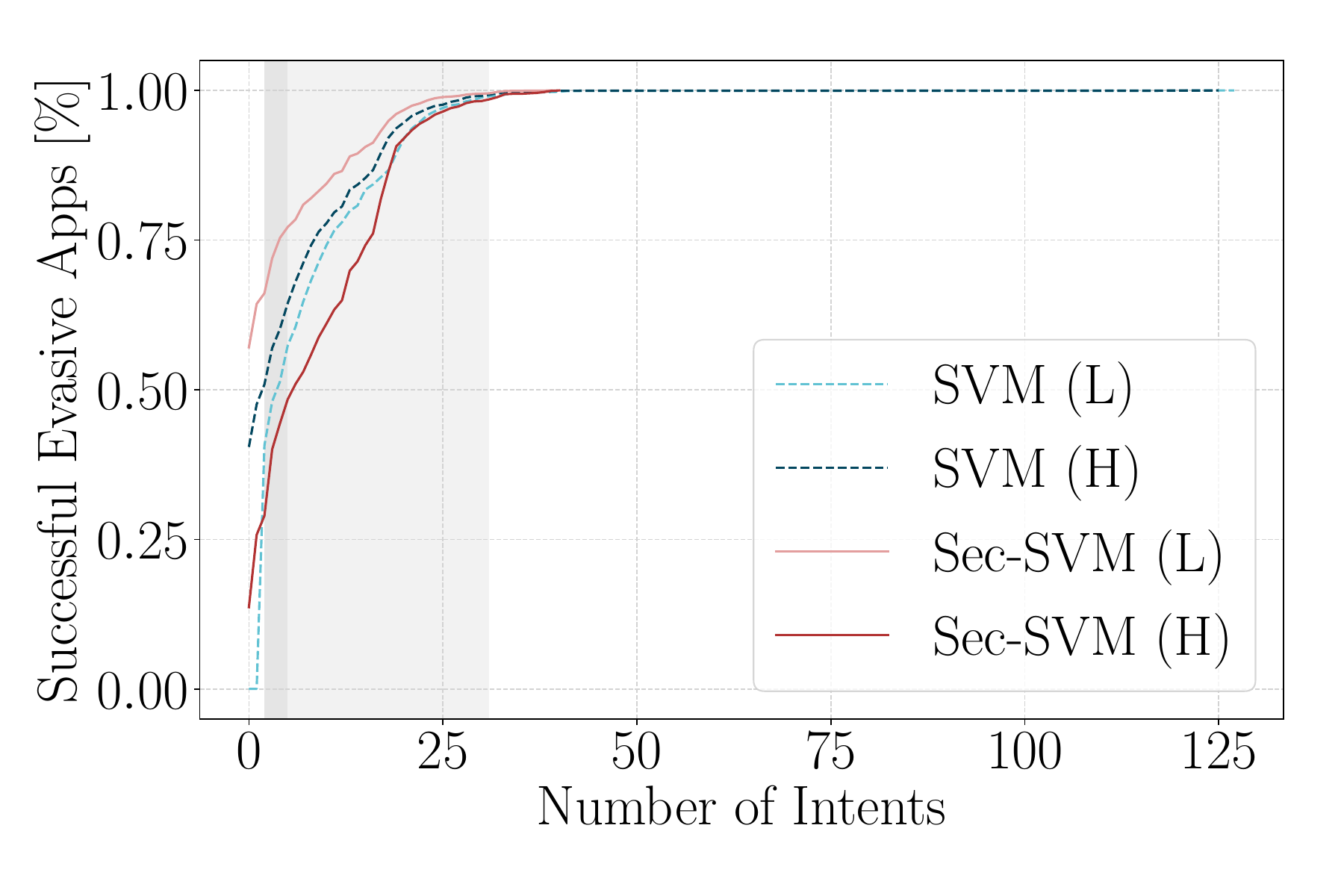}
\end{subfigure}
\begin{subfigure}{.3\columnwidth}
  \centering
  \includegraphics[width=\columnwidth]{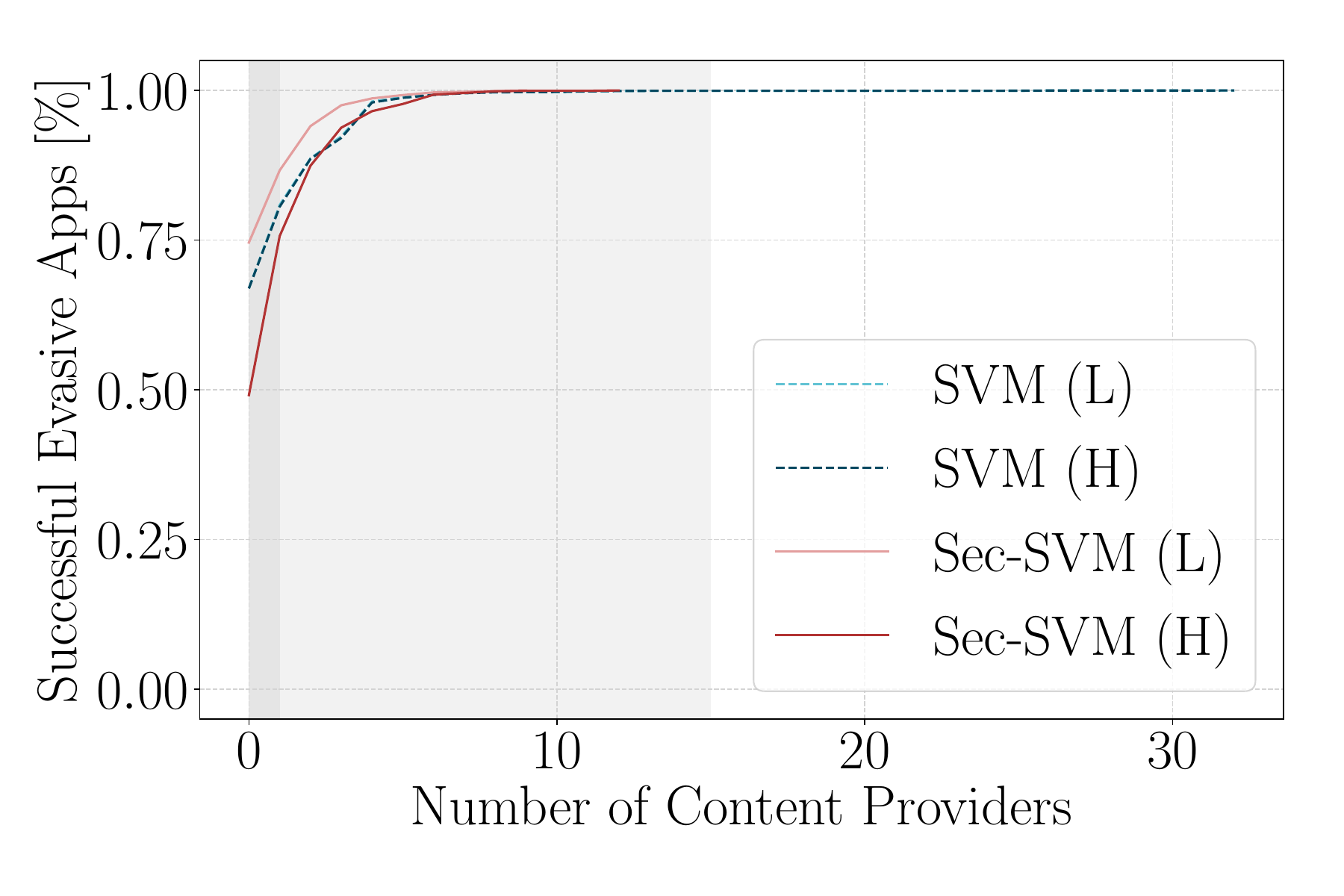}
\end{subfigure}
\begin{subfigure}{.3\columnwidth}
  \centering
  \includegraphics[width=\columnwidth]{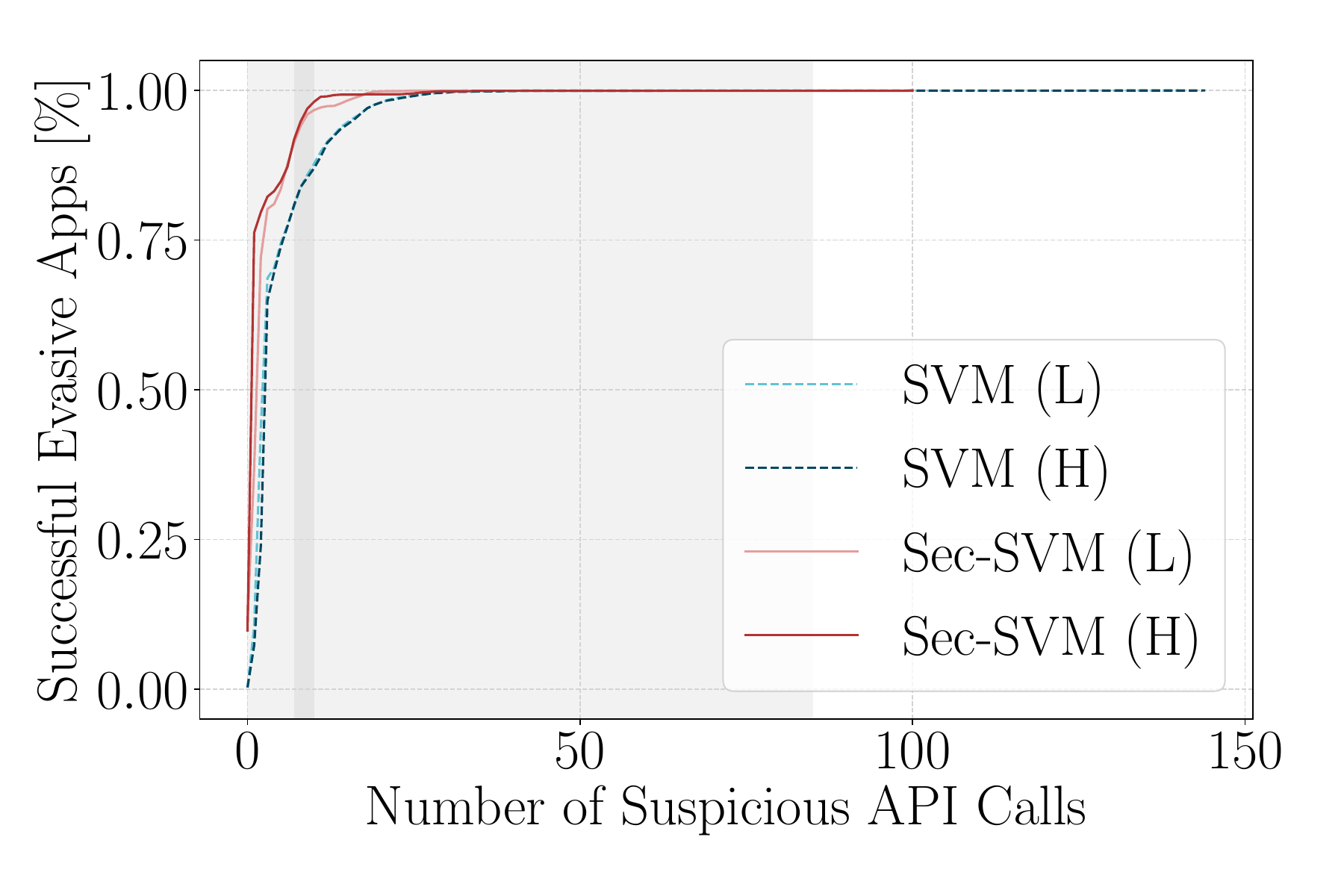}
\end{subfigure}

\caption{Statistics of the evasive malware variants, compared with statistics of benign apps. The dark gray background highlights the area between first and third quartile of benign applications; the light gray background is based on the 3$\sigma$ rule and highlights values benign statistics between $0.15\%$ and $99.85\%$ of the distribution (i.e., spanning $99.7\%$ of the distribution).}
\label{fig:statistics}
\end{figure}

\begin{figure}[t]
	\centering
	\includegraphics[scale=0.2]{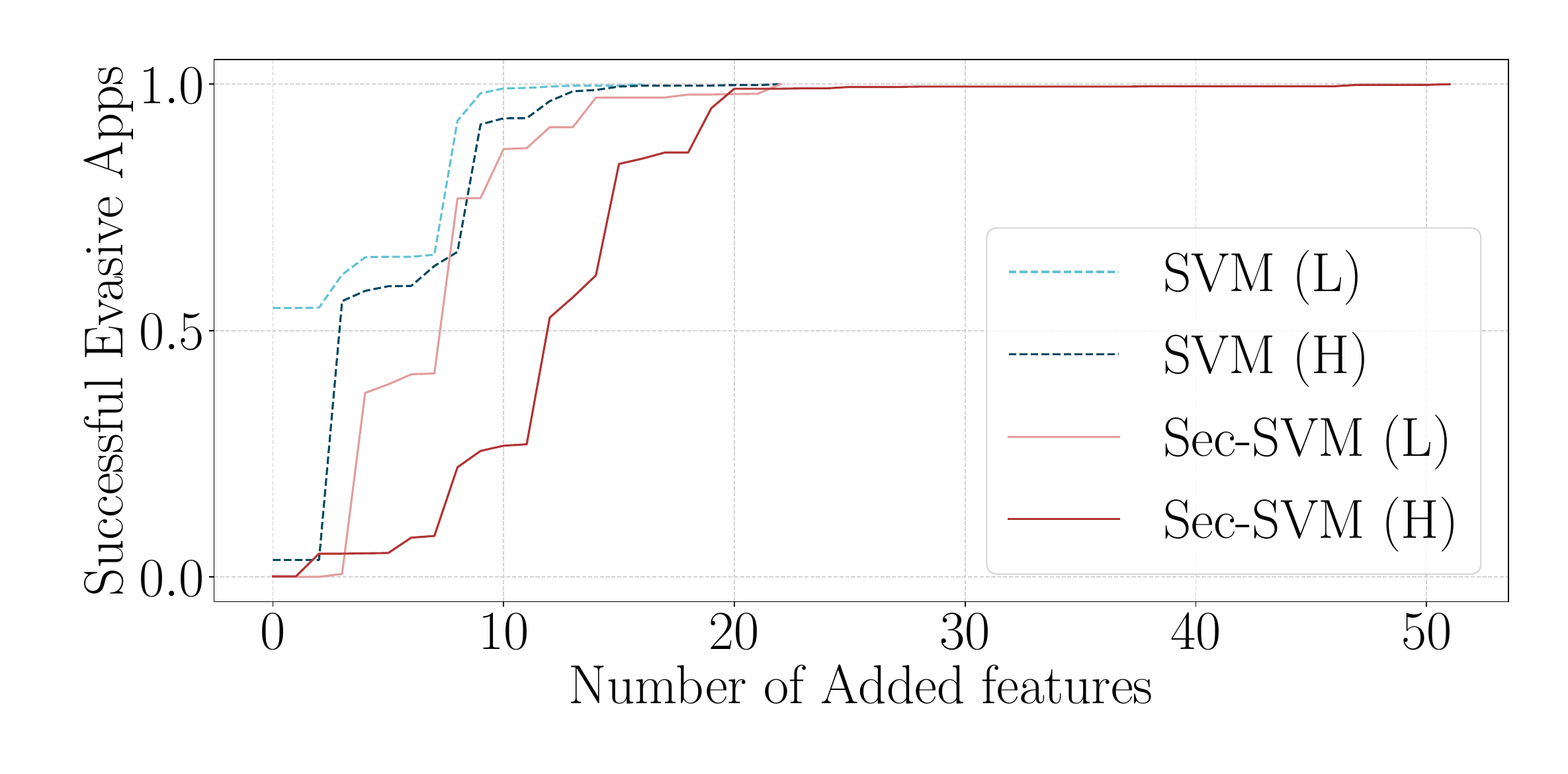}
	\caption{Breakdown of average number of features injected for each considered classifier.}
	\label{fig:injected_features}
\end{figure}

{\bf Attack Success Rate.}
We perform our attack using \emph{true positive} malware from the test set, i.e., all malware objects correctly classified as malware.
We consider four settings depending on the considered model and the attack confidence: SVM~(L), SVM~(H), Sec-SVM~(L), and Sec-SVM~(H).
In absence of FlowDroid exceptions (see the conference version for more details), we are able to create an evasive variant for each malware in all four configurations.
In other words, we achieve a misclassification rate of 100.0\% on the successfully generated apps, where the problem-space constraints are satisfied by construction (as defined in~\autoref{sec:apg-android}).
\autoref{fig:injected_features} reports the cumulative distribution of features added when generating evasive apps for the four different configurations. As expected, Sec-SVM requires the attacker to modify more features, but here we are no longer interested in the feature-space properties, since we are performing a problem-space attack. {This demonstrates that measuring attacker effort with $l_p$ perturbations  as in the original Sec-SVM evaluation~\cite{Battista:SecSVM} \textit{overestimates} the robustness of the defense and is better assessed using our framework~(\autoref{sec:formalization}).} Interesting to note anyway, that compared to our previous version of the attack, the number of needed features has highly decrease due to the improved extraction capabilities of the new implementation.

While the plausibility problem-space constraint is satisfied by design by transplanting only realistic existing code, it is informative to analyze how the statistics of the evasive malware relate to the corresponding distributions in benign apps.
\autoref{fig:statistics} reports the cumulative distribution of app statistics across the four settings: the $X$-axis reports the statistics values, whereas the $Y$-axis reports the cumulative percentage of evasive malware apps.
We also shade two gray areas: a \emph{dark gray area} between the first quartile $q_1$ and third quartile $q_3$ of the statistics for the benign applications; the \emph{light gray area} refers to the $3\sigma$ rule and reports the area within the 0.15\% and 99.85\% of the benign apps distribution.

\autoref{fig:statistics} shows that while evading Sec-SVM tends to cause a shift towards the higher percentiles of each statistic, the vast majority of apps  falls within the gray regions in all configurations. However, thanks to the enhanced extraction capabilities of the framework this is less evident compared to our initial work, making the current attack even more efficient and powerful, especially against Sec-SVM. 
We note that this is just a qualitative analysis to verify that the statistics of the evasive apps roughly align with those of benign apps; it is not sufficient to have an anomaly in one of these statistics to determine that an app is malicious (otherwise, very trivial rules could be used for malware detection itself, and this is not the case). We also observe that there is little difference between the statistics generated by Sec-SVM and by traditional SVM; this means that greater feature-space perturbations do not necessarily correspond to greater perturbations in the problem-space, reinforcing the feasibility and practicality of evading Sec-SVM.

\begin{figure}[H]
\centering
\begin{subfigure}{.4\columnwidth}
  \centering
  \includegraphics[width=\columnwidth]{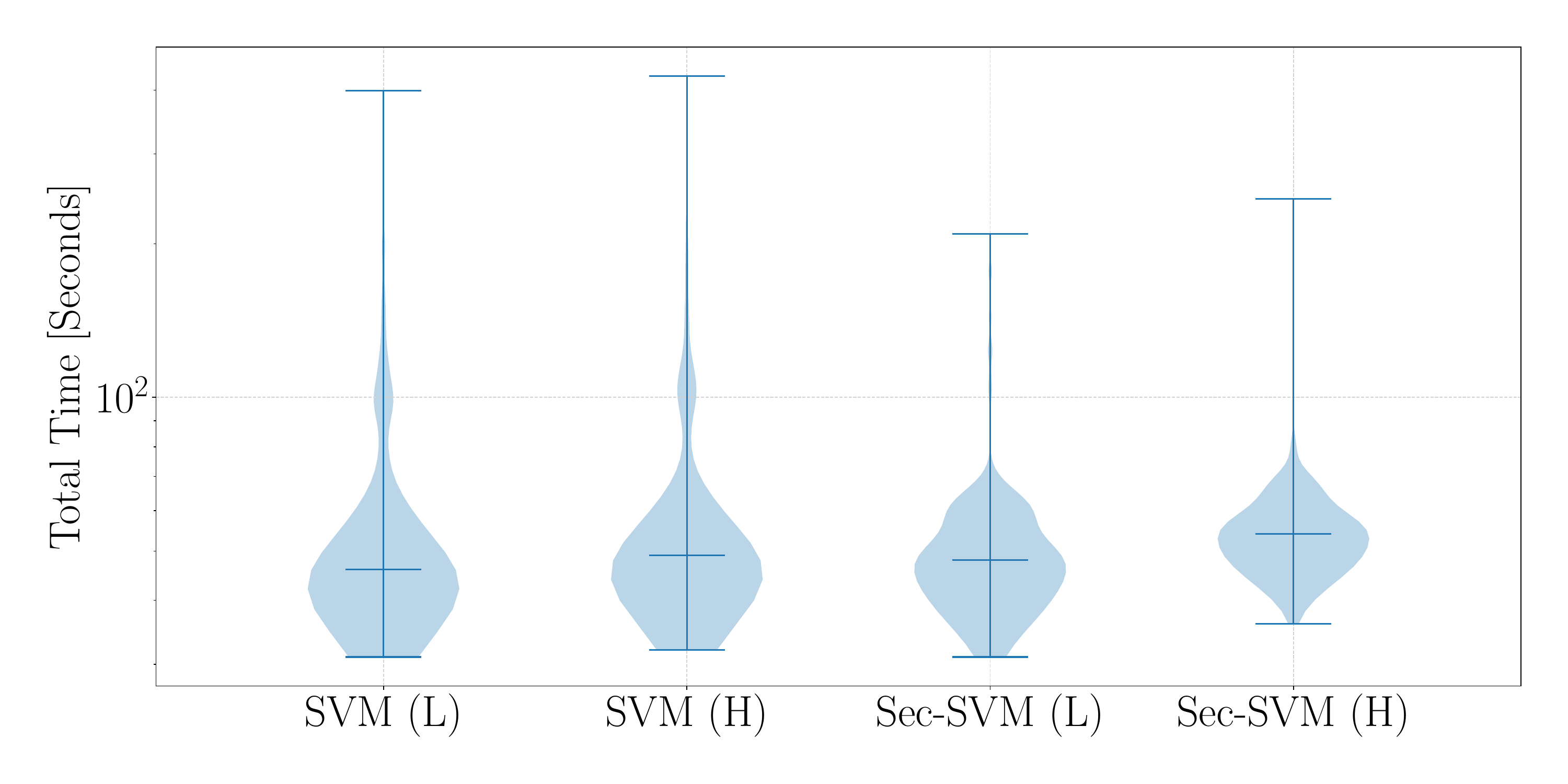}
  \caption{Total time taken to generate the adversarial apk}
\end{subfigure}
\hspace{1.0em}
\begin{subfigure}{.4\columnwidth}
  \centering
  \includegraphics[width=\columnwidth]{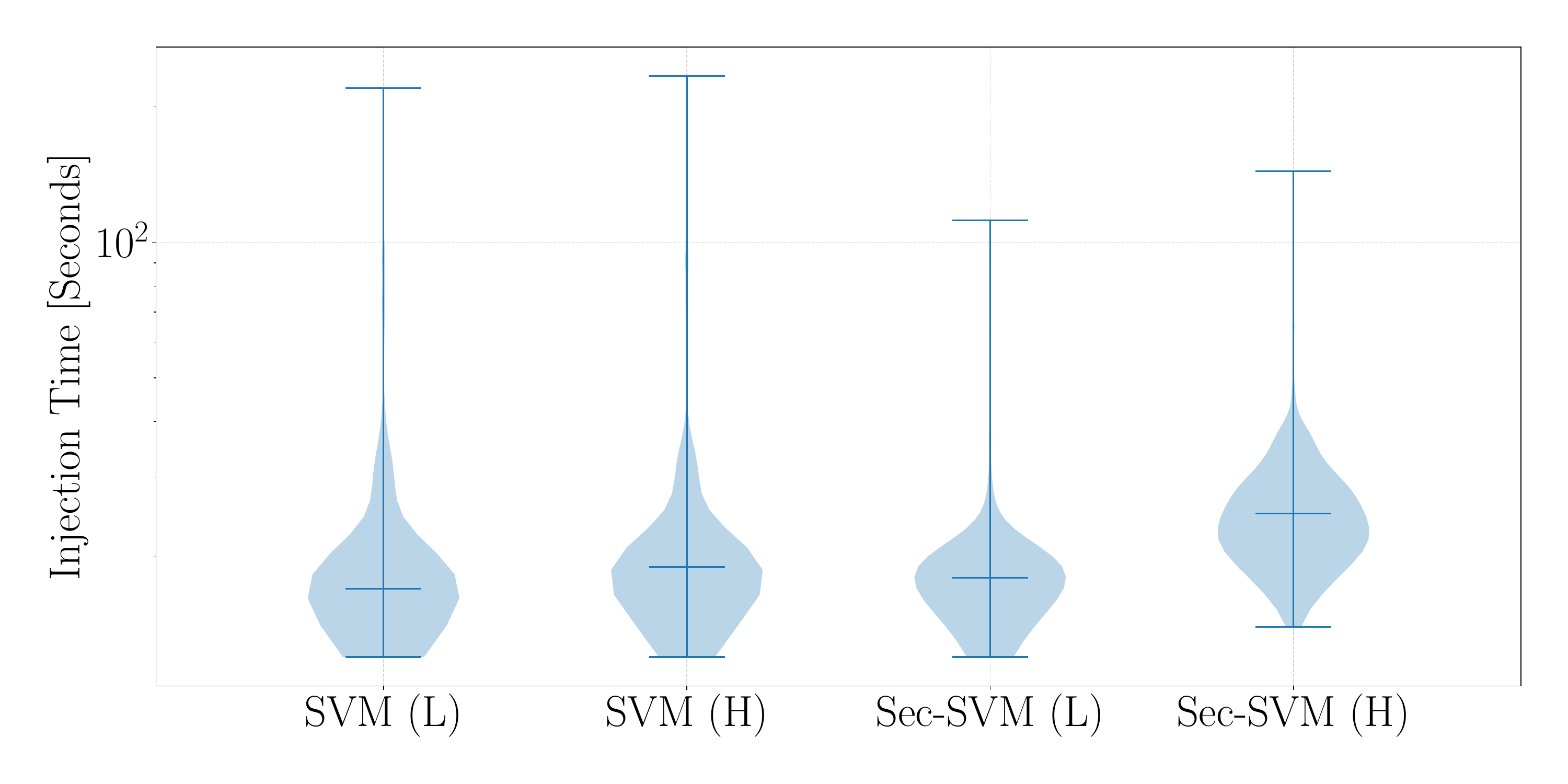}
    \caption{Total time taken to inject gadgets in the original apk}
\end{subfigure}
\caption{Violin plots of times per adversarial app.}
\label{fig:violinplots}
\end{figure}

{\bf Runtime Overhead.}
The time to perform the search strategy occurring in the feature space is almost negligible; the most demanding operation is in the actual code modification.
\autoref{fig:violinplots} depicts the distribution of injection times for our test set malware which is the most expensive operation in our approach while the rest is mostly pipeline overhead.
The time spent per app is low: in most cases, less than 100 seconds, and always less than 2,000 seconds (\texttildelow33 mins).

The low runtime cost suggests that it is feasible to perform this attack at scale and reinforces the need for new defenses in this domain.

\subsection{Discussion}
Through our evaluation we have shown that it is possible to create realistic adversarial samples and depending on the amount of available gadgets to implant the final outcome and difficulty varies. Indeed, with the new set of gadgets, which has far more comprehensive, we show that the average amount of injected features has been significantly reduced.

\section{Model hardening}
A further step we have done in this paper is to explore how feasible is adversarial training in order to harden the target classifiers  against problem space generated adversarial samples. Our objective is to assess the efficacy of various hardening methods in response to the attack strategy we have outlined, as referenced in \autoref{alg:attack}. In this scenario, the attacker faces no restrictions on the number of gadgets they can employ, with their only limitation being the total quantity of gadgets available for harvest. We show that different sets of hardening methodologies may lead to different robustness levels against the attack. We include experiments both focusing on linear SVM and on Sec-SVM\cite{Battista:SecSVM}.

\subsection{Hardening Strategies}
\label{defender_strategies}
As defenders, our goal is to explore and use different approaches in order to harden our considered classifier. In order to do so, we are considering two different approaches: adversarial training and adversarial retraining. 
Adversarial training \cite{szegedy2013intriguing} involves the creation of adversarial examples using an efficient approach. These adversarial examples are then integrated into the model's training phase, where they are blended with the standard dataset throughout the training periods. The most widely accepted technique \cite{madry2017towards} involves consistently generating adversarial examples that result in the maximum possible loss. Subsequently, the model's adjustable parameters are fine-tuned to reduce the error rate when classifying these challenging adversarial examples. Then follows the unchanged equation and its explanation:

\begin{equation}
\min_{\theta} \mathbb{E}{(x, y) \sim D} \left[ \max{\delta \in S} L(\theta, x + \delta, y) \right]
\end{equation}
where \((x, y) \sim D\) represents training data sampled from the distribution \(D\), \(\delta \in S\) is the maximum allowed perturbation, \(L\) is the model loss and \(\theta\) is the model hyperparameters.

On the other hand, adversarial retraining as hardening strategy \cite{chen2020explore}, consists in using attack A over the training dataset D to produce an adversarial training set \(D_a = \{( \hat{x}_i, \hat{y}_i) \}_{i=1}^N\). We join the original and the adversarial training sets and retrain the classifier \(H(x)\) over this augmented set \(D^* = D \cup D_a\). It is to be noted that this is different from classical adversarial training where new adversarial examples are generated and mixed with the original training set continuously. We have chosen to consider this strategy because it is suited for both traditional machine learning algorithms and neural networks, so can be equally applied to both our considered models.

We have hypothesised two different types of attacks that defender could use in order to create adversarial samples for hardening: a feature space attack and our proposed problem space attack \autoref{alg:attack}.  The feature space attack is a variant of the latter in which we inject the most important features of the target classifier without considering ${\eta}$.

We consider and compare the following aspects:
\begin{itemize}
    \item The amount of malware samples used during hardening
    \item The hardening technique adopted
    \item The transformation type for creating adversarial samples (PS vs. FS)
    \item The considered level of confidence both for evasion and for the hardening phase
\end{itemize}
More details about the considered settings can be found at \autoref{app:def_settings}.

\begin{figure}[h!]
\centering
\begin{subfigure}{.48\columnwidth}
  \centering
  \includegraphics[width=\columnwidth]{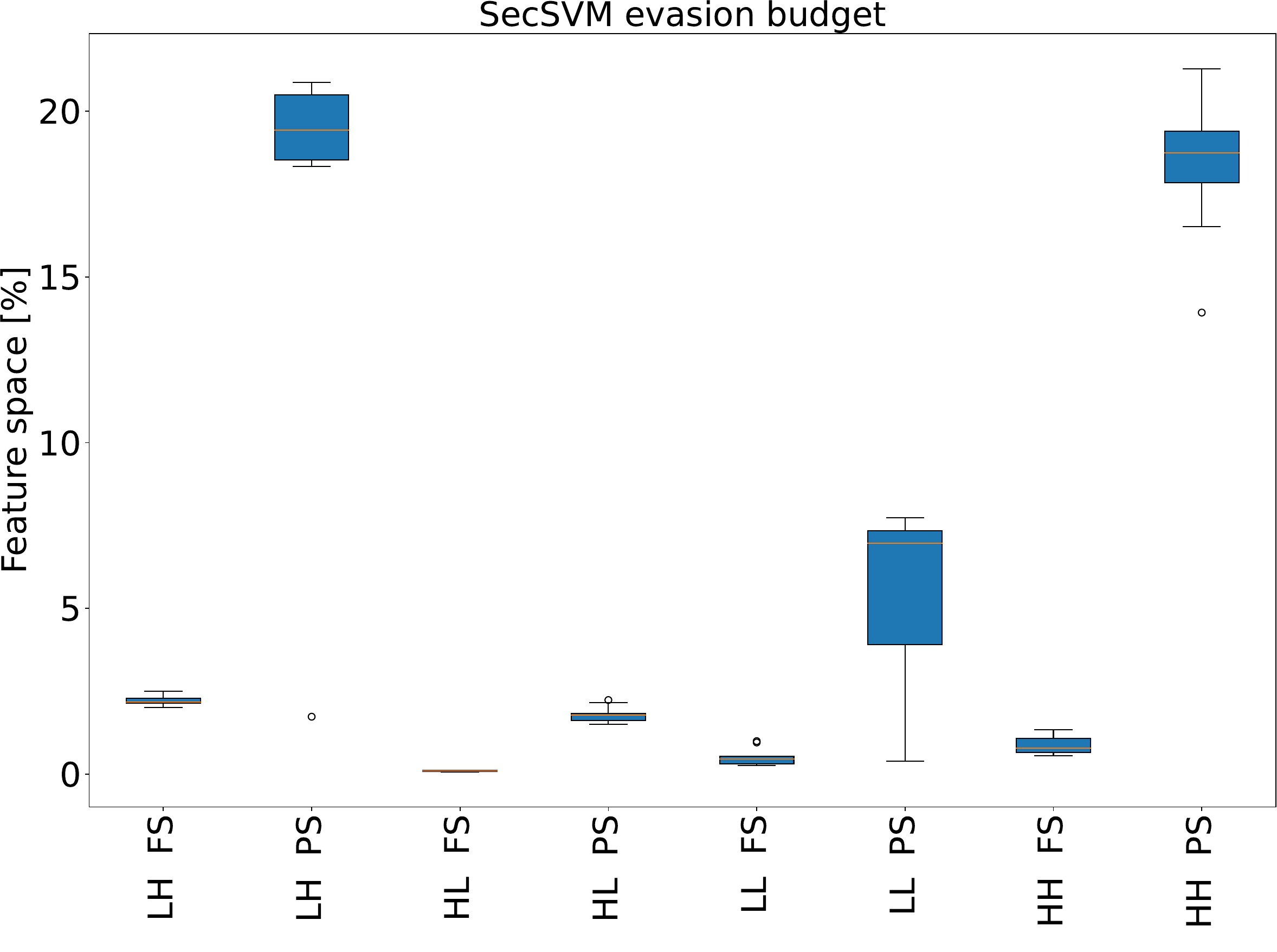}

\end{subfigure}
\begin{subfigure}{.48\columnwidth}
  \centering
  \includegraphics[width=\columnwidth]{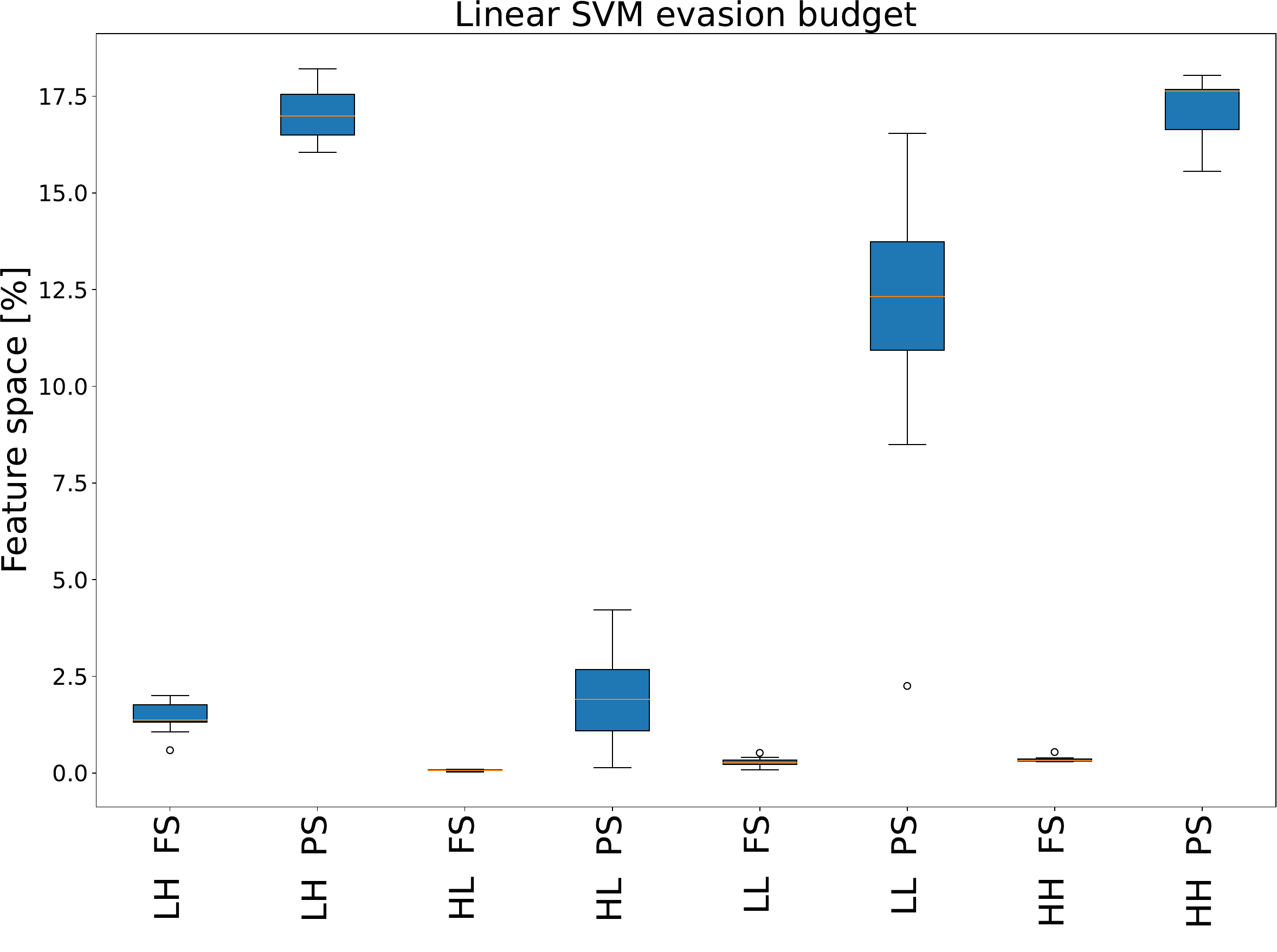}
\end{subfigure}
\caption{Amount of gadgets needed in order to successfully bypass the target classifier in the considered configurations.}
\label{fig:budgets}
\end{figure}

\subsection{Evaluation}

For this evaluation we assume that the defender is using an increasing percentage of the available malware in the training set (cf. \autoref{defender_strategies}), and we focus on the two classification models previously described, a LinearSVM and SecSVM. For implementation details, refer to \autoref{app:secsvm}. Our goal is to  evaluate the robustness against our proposed white-box problem-space attack through the adversarial training~\cite{madry2017towards} vs. adversarial retraining~\cite{chen2020explore}, comparing the impact of using realistic  problem-space samples against unrealistic feature-space samples. We consider two different levels of confidence of the adversarial samples, as explained in \ref{attack_confidence}, which can be set as a target both during hardening and evaluation. For example, a Low-Low configuration means that both during the hardening and during evaluation the target for the generated adversarial samples is low confidence.

In this segment of our discussion, we omit visual representations of adversarial retraining outcomes, as our investigations consistently demonstrate that this technique fails to enhance the resilience of models when faced with the specific threat actors under consideration. Throughout our experiments, attackers invariably succeeded in generating high-confidence adversarial samples targeting models subjected to this hardening strategy. The ASR consistently exceeded 95\%, unequivocally indicating the ineffectiveness of the hardening approach.

A notable observation across all our experimental evaluations is the uniformity of outcomes, regardless of the variations in experimental setups or models tested. This striking consistency highlights a fundamental limitation of adversarial retraining in its current form when it comes to countering the generation of both low and high confidence adversarial samples by the considered threat actors.

\begin{figure}[t]
\centering
\begin{subfigure}{.45\textwidth}
  \centering
  \includegraphics[width=\columnwidth]{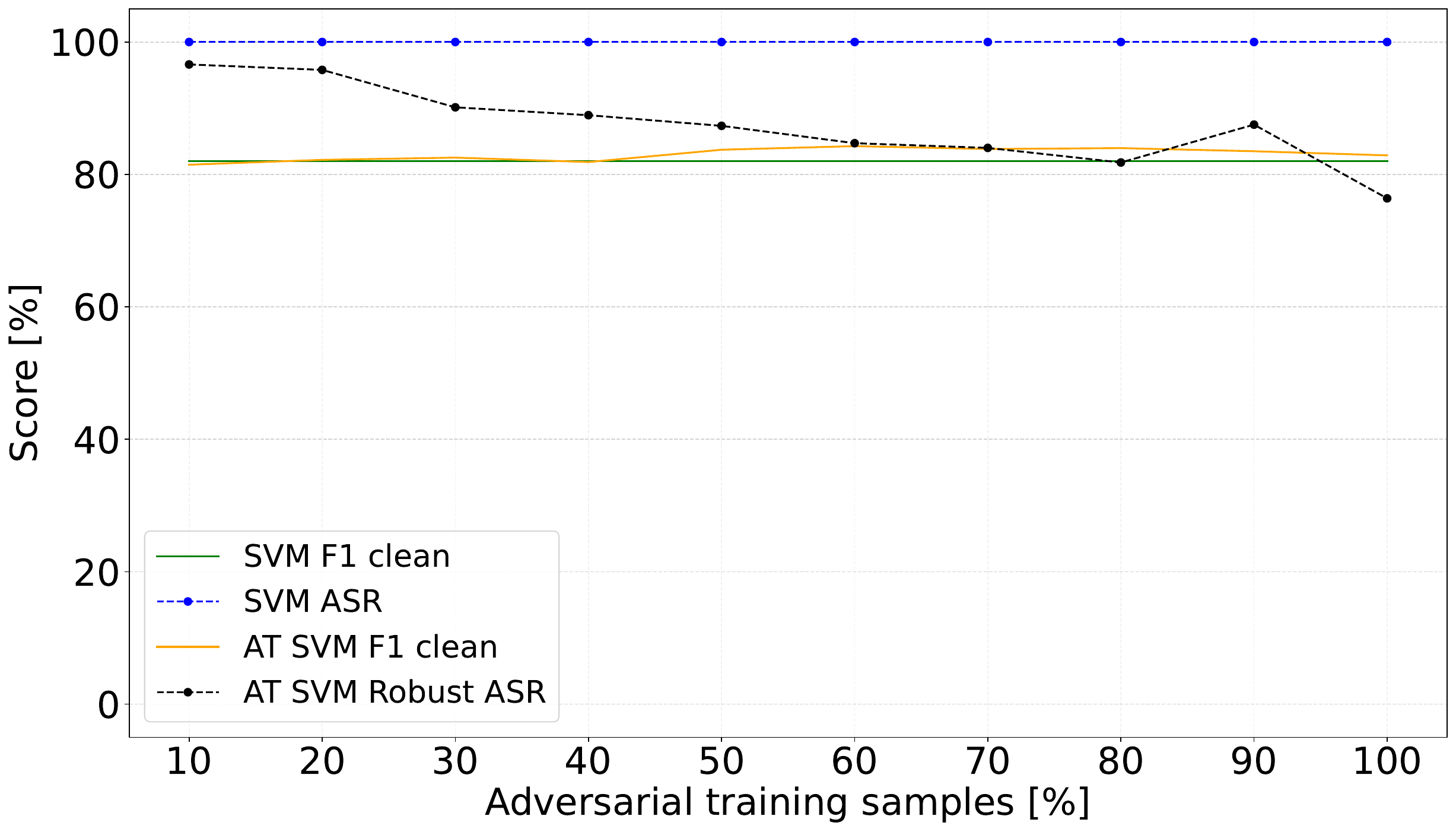}
  \caption{FS Low-Low}
  \label{fig:svm_at_ll_fs}
\end{subfigure}
\begin{subfigure}{.45\textwidth}
  \centering
  \includegraphics[width=\columnwidth]{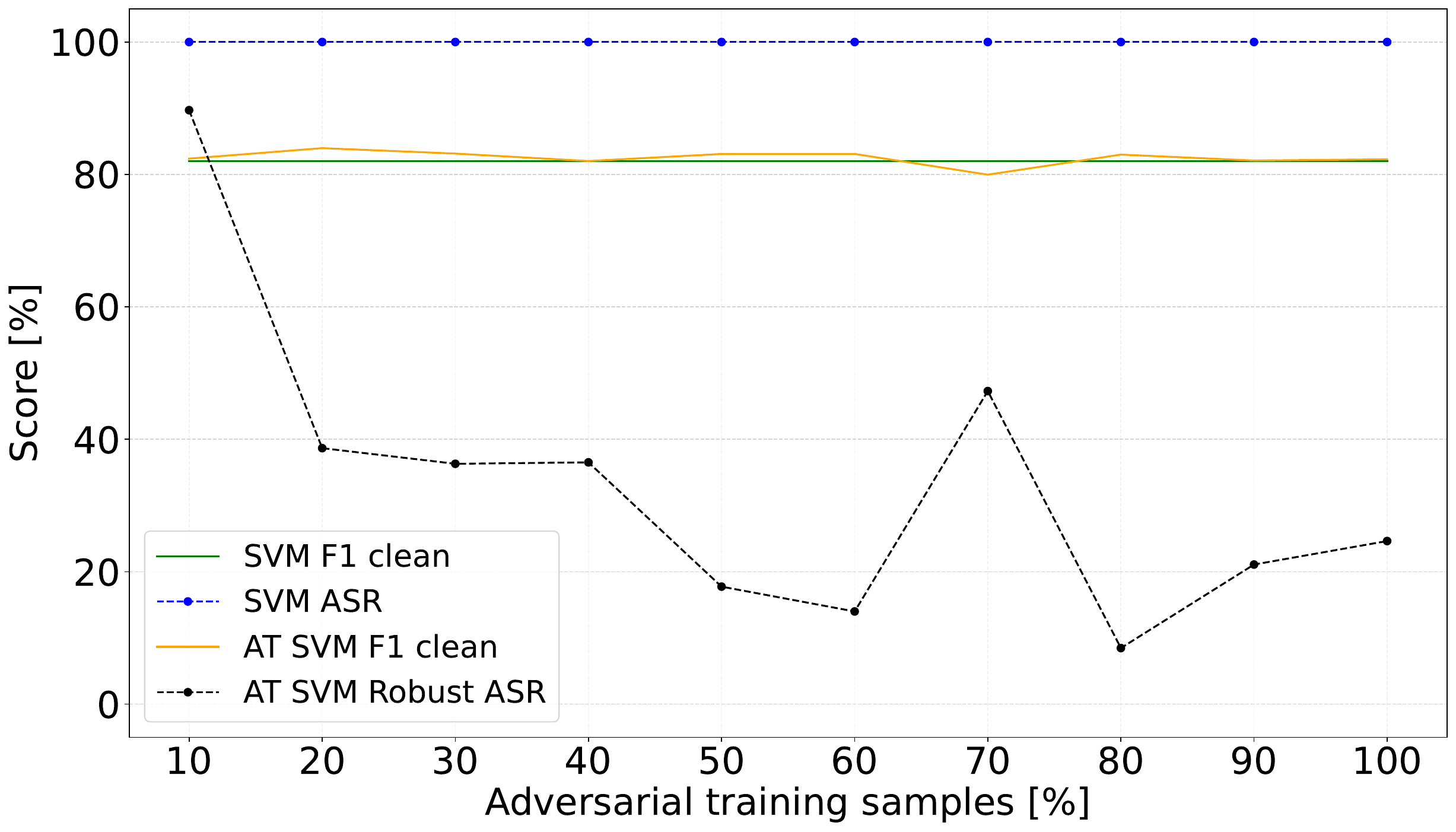}
  \caption{PS Low-Low}
  \label{fig:svm_at_ll_ps}
\end{subfigure}
\begin{subfigure}{.45\textwidth}
  \centering
  \includegraphics[width=\columnwidth]{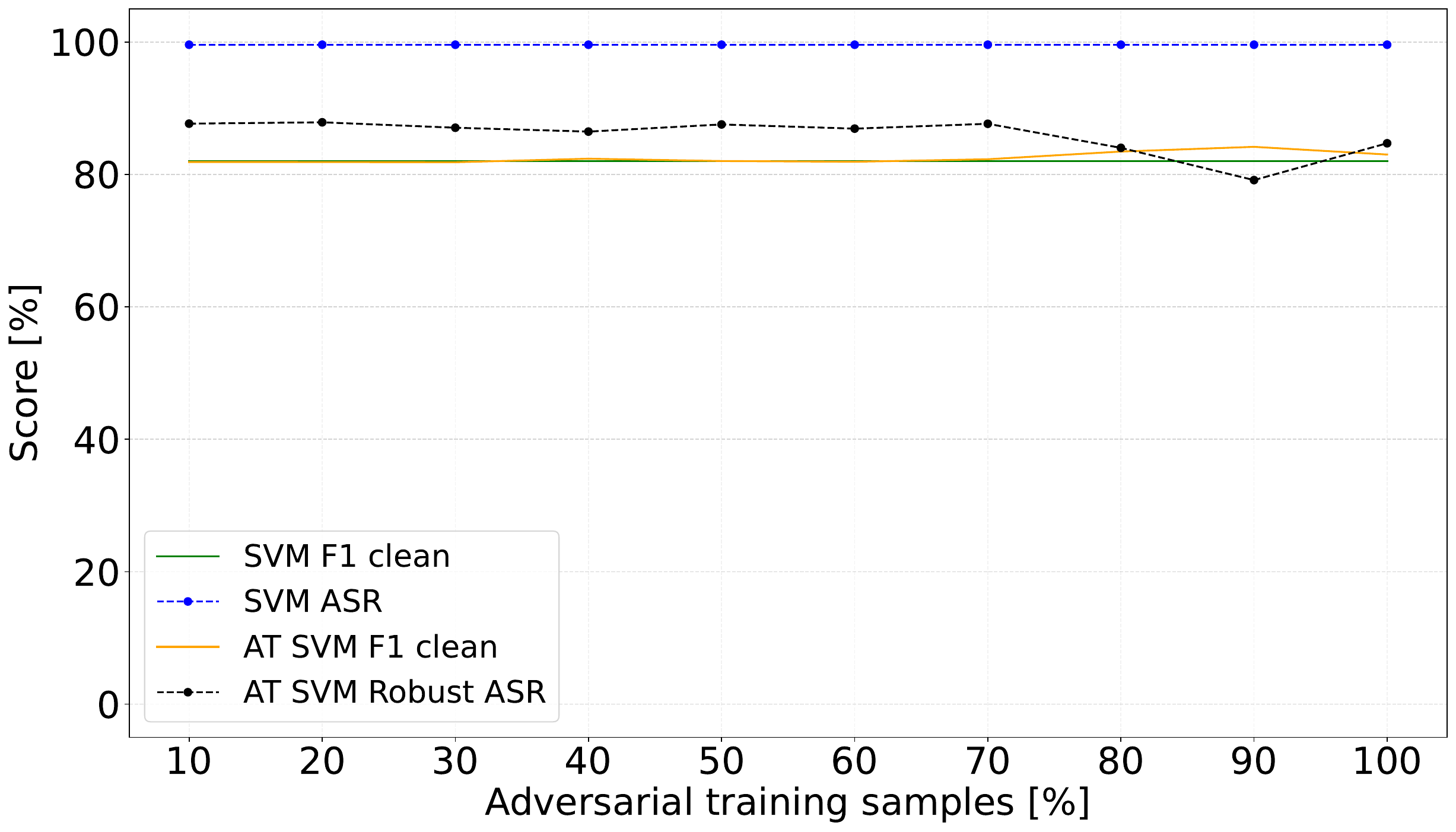}
  \caption{FS High-High}
  \label{fig:svm_at_hh_fs}
\end{subfigure}
\begin{subfigure}{.45\textwidth}
  \centering
  \includegraphics[width=\columnwidth]{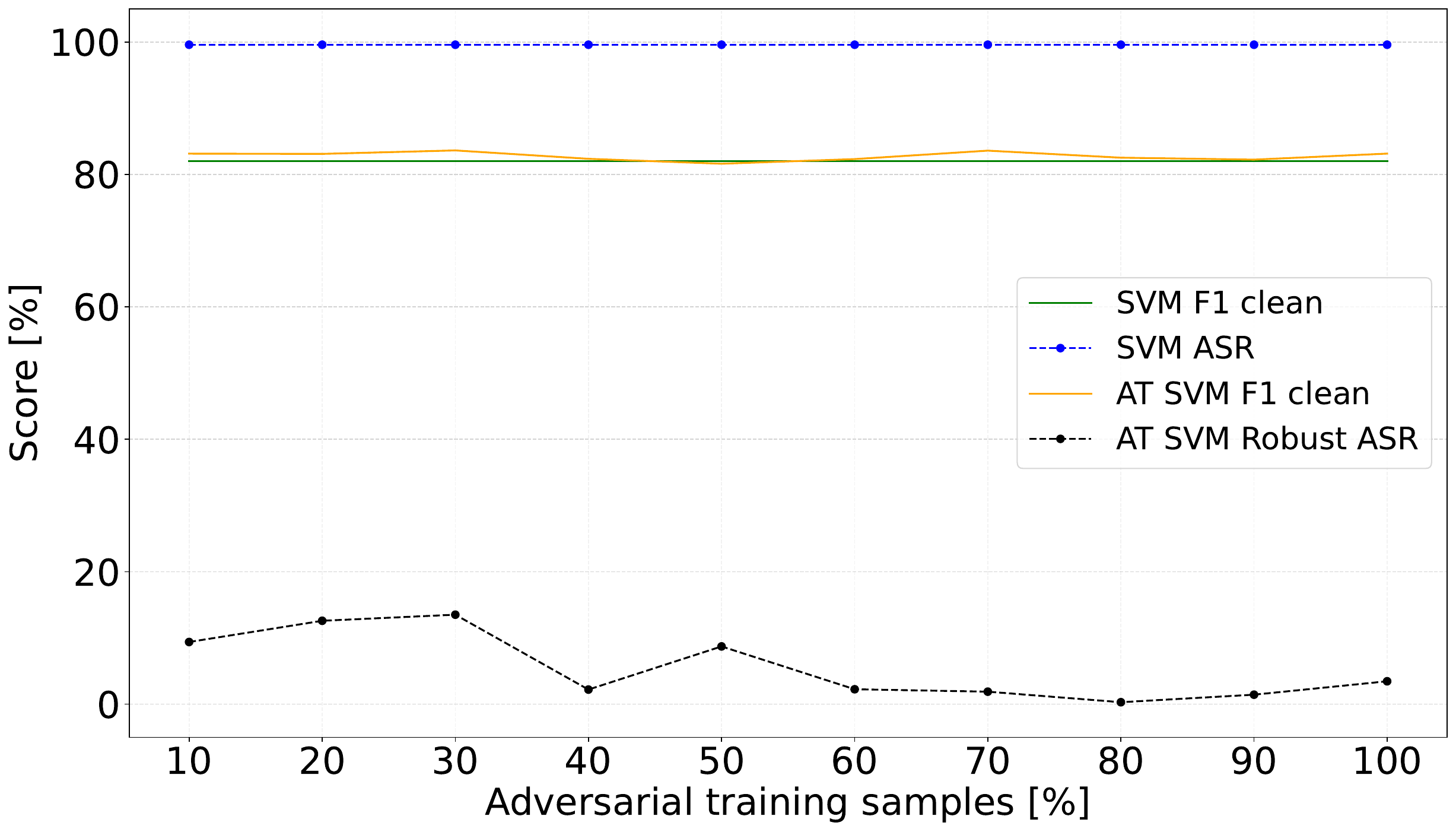}
  \caption{PS High-High}
  \label{fig:svm_at_hh_ps}
\end{subfigure}
\caption{Evaluating the effectiveness of both the original and the adversarially trained LinearSVM model against adversarial examples in the problem space, as well as assessing performance on clean data. The charts illustrate the varied performance outcomes resulting from different hardening approaches. "PS" indicates that the model underwent hardening through the use of adversarial samples in the Problem Space, whereas "FS" signifies hardening via adversarial samples in the Feature Space.}
\label{fig:svm_at}
\end{figure}

\begin{figure}[t]
\centering
\begin{subfigure}{.45\columnwidth}
  \centering
  \includegraphics[width=\columnwidth]{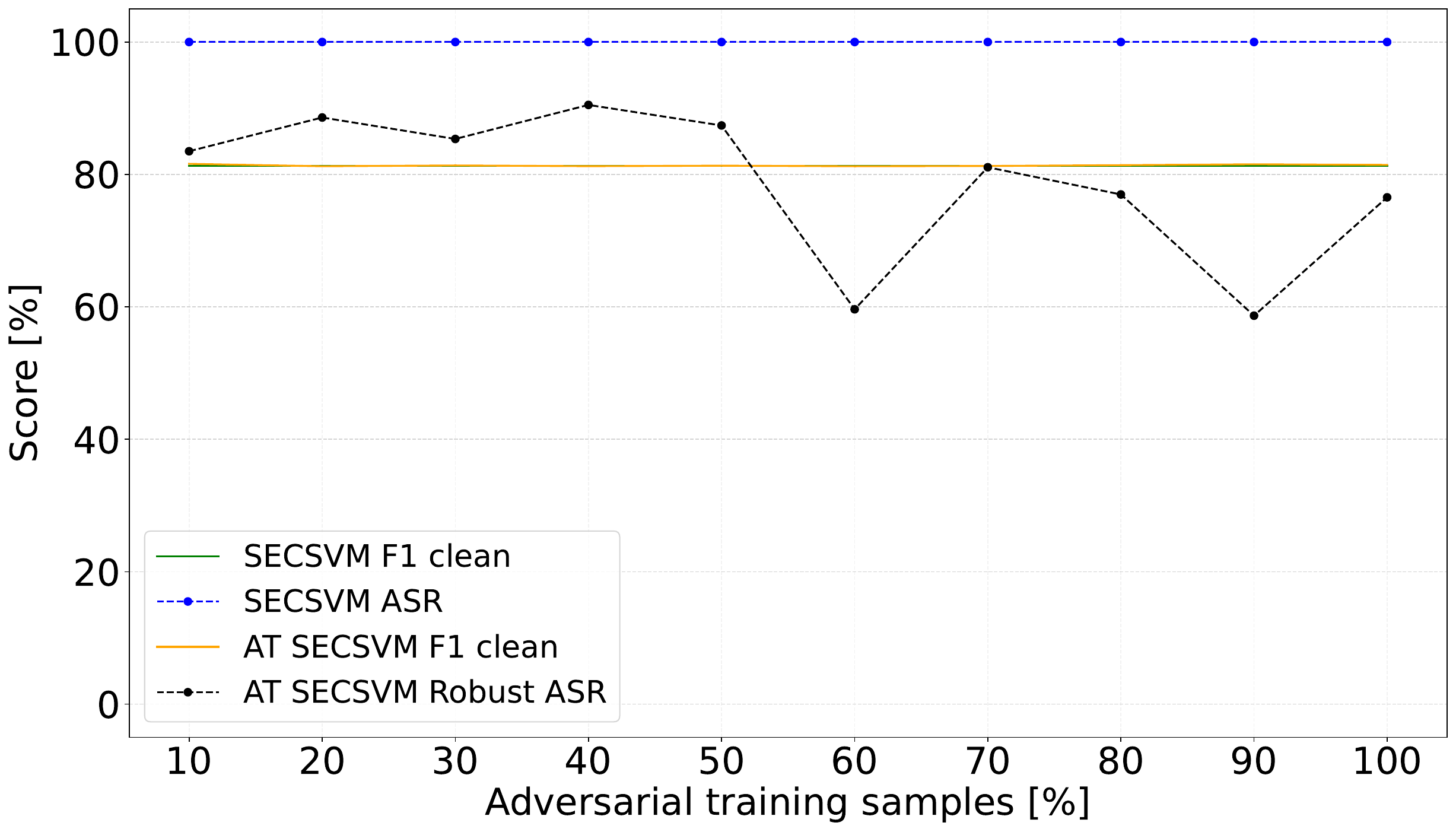}
  \caption{FS Low-Low}
  \label{fig:secsvm_at_ll_fs}
\end{subfigure}
\begin{subfigure}{.45\columnwidth}
  \centering
  \includegraphics[width=\columnwidth]{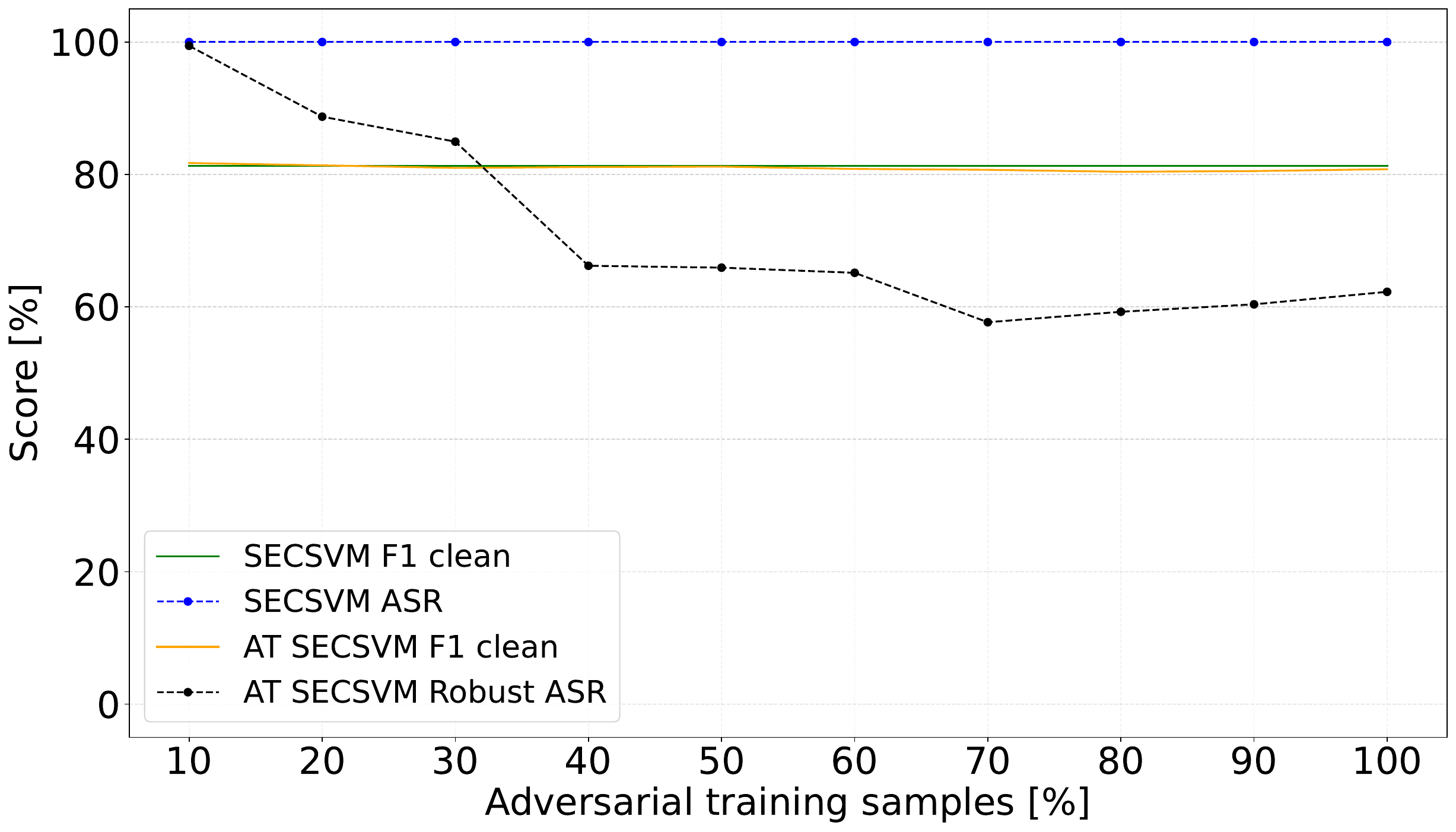}
  \caption{PS Low-Low}
  \label{fig:secsvm_at_ll_ps}
\end{subfigure}
\begin{subfigure}{.45\columnwidth}
  \centering
  \includegraphics[width=\columnwidth]{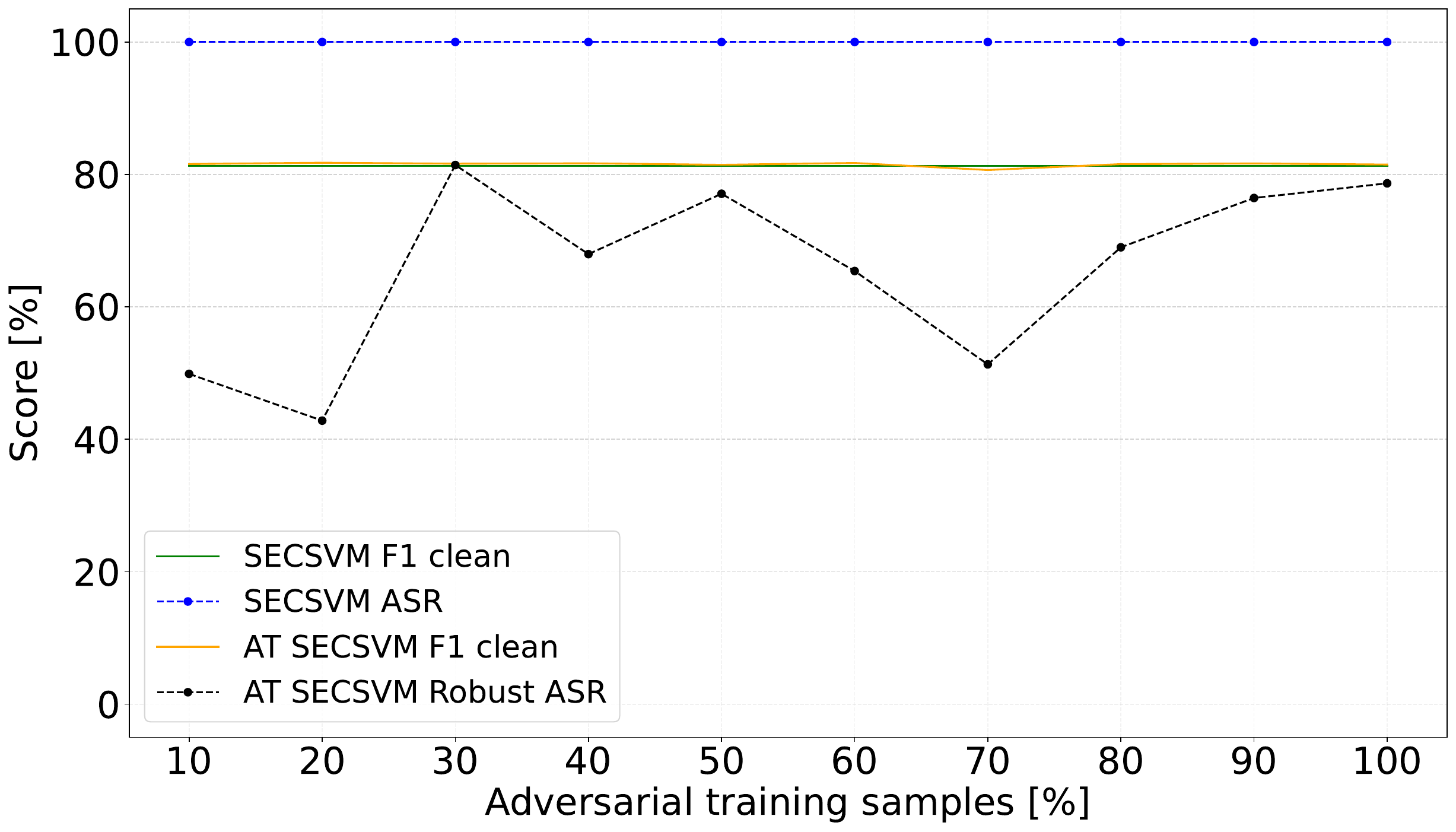}
  \caption{FS High-High}
  \label{fig:secsvm_at_hh_fs}
\end{subfigure}
\begin{subfigure}{.45\columnwidth}
  \centering
  \includegraphics[width=\columnwidth]{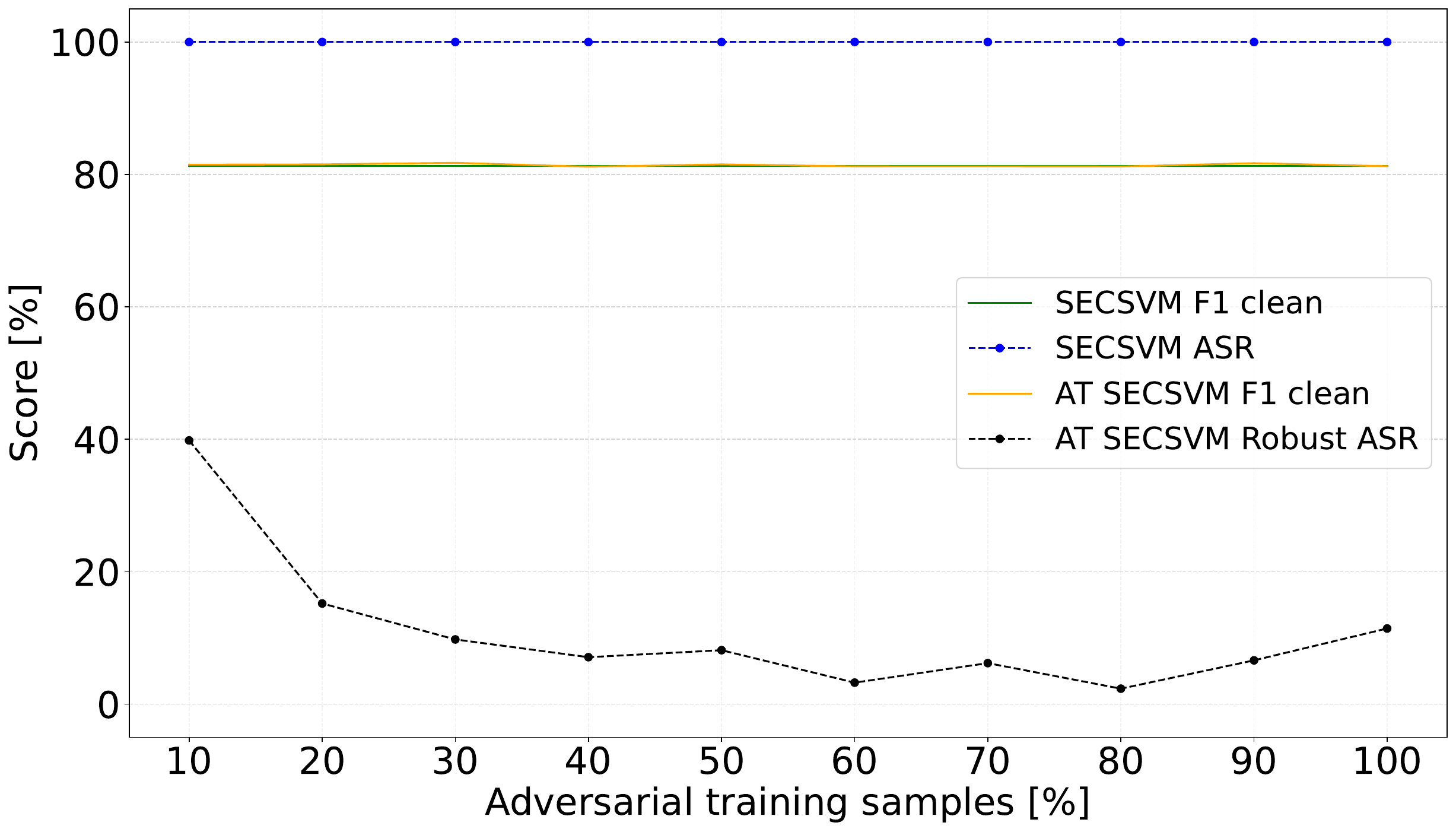}
  \caption{PS High-High}
  \label{fig:secsvm_at_hh_ps}
\end{subfigure}
\caption{Evaluating the effectiveness of both the original and the adversarially trained Sec-SVM model against adversarial examples in the problem space, as well as assessing performance on clean data.The charts illustrate the varied performance outcomes resulting from different hardening approaches. "PS" indicates that the model underwent hardening through the use of adversarial samples in the Problem Space, whereas "FS" indicates hardening via adversarial samples in the Feature Space.}
\label{fig:secsvm_at}
\end{figure}

On the other hand, hardening through adversarial training methodology has been proven more effective, especially on LinearSVM. Indeed, \autoref{fig:svm_at} and \autoref{fig:secsvm_at} show that adversarial training is able to significantly reduce the ASR against the generation of high confidence adversarial samples, while against low confidence adversarial samples we can enforce some level of robustness but always higher than 20\%. We can notice  a huge difference between models trained using only feature space transformations and problem space: models hardened with feature space transformations even in the High-High scenario cannot reach a reasonable level of robustness while the ones hardened with problem space transformations can.

All the considered evasions are unbound, meaning that the attacker can use all the gadgets he was able to harvest, without any need of minimising the perturbation. Specifically for this work, we have been able to extract around $\approx$800 suitable gadgets from goodware. Looking at the number of gadgets used in order to evade \autoref{fig:budgets}, we can notice that in the majority of cases the model hardened using problem space samples significantly raise the bar in the amount of gadgets needed for evasion, especially for SVM.

\subsection{Discussion}

Our investigations underscore that adversarial retraining fails to adequately fortify the model in question within the experimental frameworks we explored. This deficiency in strengthening is evident through the consistent and successful breaches by adversaries against models purportedly made more robust through adversarial retraining. Such findings signal the need for a comprehensive review and possible overhaul of this strategy for similar settings.

On the flip side, our research shows the beneficial impact of adversarial training on enhancing model robustness. Even if using feature space transformations for hardening brings some minor level of robustness, we discovered that adversarial samples created through problem-space transformations substantially increase the resilience of the target classifier. These samples prove to be more effective than feature spaced transformations, offering a comparative advantage for both classifiers subjected to our tests.

A noteworthy observation from our study is the superior potential for improvement exhibited by the SVM classifier over the Sec-SVM when subjected to adversarial training in though problem-space hardening. This difference in potential improvement is likely rooted in the inherent differences between their learning algorithms. The SVM classifier, tends to concentrate on a select set of essential features \cite{melis2018explaining}, while in contrast, the Sec-SVM employs a strategy of weight clamping that aims to distribute significance more uniformly across a broader spectrum of features by imposing upper and lower limits on their influence. While intended to foster a more balanced approach to feature importance, this strategy could inadvertently set a cap on the positive impacts of adversarial training by too strictly regulating the weights.

A consistent and significant theme emerging from our research is the distinct advantage offered by adversarial samples that are crafted through realistic transformations in the problem space, especially in terms of hardening models. This advantage manifests not only in the broadening of the total amount of budget needed in order to craft successful adversarial samples but also in the notable enhancement of model robustness. Practically, we observed improvements in robustness ranging between 10 to 60\% in classifiers that underwent hardening with realistic adversarial samples. This translates into an increased evasion budget requirement of 3 to 15\%, indicating a substantial uplift in model defense capabilities.

Additionally, our analysis reveals the absence of uniform patterns in the ASR metrics, as evidenced in  previous plots \autoref{fig:secsvm_at} and \autoref{fig:svm_at}. This phenomenon may be attributed to the random selection process employed for each batch, where points for transformation are randomly chosen from a specified subset at a predetermined rate. This randomness underscores the critical importance of selecting representative samples for transformation, suggesting that the strategic alteration of the model's weights through these samples is more significant than the sheer number of points utilized. This insight serves to highlight the complex and nuanced nature of adversarial training, shedding light on the myriad of factors that play a crucial role in determining its effectiveness and success.

\section{Discussion on Attack and Defence Results}
\label{sec:discussion}

We provide some deeper discussion on the results of our novel problem-space attack and the investigated hardening methods.

{\bf Android Attack Effectiveness.}
We conclude that it is practically \emph{feasible} to evade the state-of-the-art Android malware classifier DREBIN~\cite{Arp:Drebin} and its hardened variant, Sec-SVM~\cite{Battista:SecSVM}, and that we are able to automatically generate realistic and inconspicuous evasive adversarial applications, often in less than 2 minutes.
This shows for the first time that it is possible to create realistic adversarial applications at scale.

{\bf Obfuscation.}
It could be argued that traditional obfuscation methods can be used to simply hide malicious functionality.
The novel problem-space attack in this work evaluates the feasibility of an ``adversarial-malware as a service'' scenario, where the use of mass obfuscation may raise the suspicions of the defender; for example, antivirus companies often classify samples as malicious simply because they utilize obfuscation or packing~\cite[][]{Balzarotti:Packer, Balzarotti:DeepPacker}. Moreover, some other analysis methods combine static and dynamic analysis to prioritize evaluation of code areas that are likely obfuscated~\cite[e.g.,][]{leslous2017gpfinder}.
On the contrary, our transformations aim to be fully inconspicuous by adding only  legitimate benign code and, to the best of our knowledge, we do not leave any relevant artifact in the process. {While the effect on problem-space constraints may differ depending on the setting, attack methodologies such as ours and traditional obfuscation techniques naturally complement each other in aiding evasion and, in the program domain, code transplantation may be seen as a tool for developing new forms of inconspicuous obfuscation~\cite{Fass:Javascript}.}

{\bf {Other Defense Directions}.}
A promising direction by~\cite{Wagner:Monotonic} studies the use of \emph{monotonic classifiers}, where adding features can only increase the decision score (i.e., an attacker cannot rely on adding more features to evade detection); however, such classifiers require non-negligible time towards manual feature selection (i.e., on features that are harder for an attacker to change), and---at least in the context of Windows malware~\cite{Wagner:Monotonic}---they suffer from high false positives and an average reduction in detection rate of 13\%.
Moreover, we remark that we decide to add goodware parts to malware for practical reasons: the opposite transplantation would be immediate to do if a dataset with \emph{annotated} malicious bytecode segments were available. As part of future work we aim to investigate whether it would still be possible to evade monotonic classifiers by adding only a minimal number of malicious slices to a benign application.

{\bf {Defenses Against Problem-Space Attacks}.}
Unlike settings where feature and problem space are closely related (e.g., images and audio), limitations on feature-space $l_p$ perturbations are often insufficient to determine the risk and feasibility of an attack in the real world.
Our novel problem-space formalization (\autoref{sec:formalization}) paves the way to the study of \emph{practical} defenses that can be effective in settings which lack an inverse feature mapping.
Simulating and evaluating attacker capabilities in the problem space helps define realistic threat models with more constrained modifications in the feature space---which may lead to more robust classifier design.
Our Android evasion attack (\autoref{sec:apg-android}) demonstrates for the first time that it is \emph{feasible} to evade feature-space defenses such as Sec-SVM in the problem-space---and to do so \emph{en masse}.

{\bf {Model hardening through adversarial training}.}
Our experiments demonstrate that depending on the considered strategies the result may highly vary. Adversarial retraining did not succeed in fortifying the classifier in any capacity. Conversely, the effect of adversarial training led to a reduction in the Attack Success Rate (ASR) by up to 80\% when facing the creation of low-confidence adversarial samples, and by 90\% for high-confidence ones. In our settings, problem space hardening revealed to be more effective in all considered scenario if compared to feature-space transformation adversarial samples, highlighting the exploration of more meaningful areas of the feature space. The difference between the two considered models' architectures suggest that the success of adversarial training is closely tied to the training methodology, the considered feature space and the classification model used, shown also by the opposite conclusions of \cite{dyrmishi2023empirical} and \cite{lucas2023adversarial}. Future research could delve into various classifiers and feature spaces to uncover the underlying relationship between representation and learning logic, aiming to boost the effectiveness of adversarial training. Another key aspect is related to the points selected during the hardening process: evidence shows that using all the available points does not always achieve the best robustness.

\section{Related Work}
\label{sec:related}

{\bf Adversarial Machine Learning.}
Adversarial ML attacks have been studied for more than a decade~\cite{Biggio:Wild}. These attacks aim to modify objects either at training time (\emph{poisoning}~\cite{FAIL}) or at test time (\emph{evasion}~\cite{Biggio:Evasion}) to compromise the confidentiality, integrity, or availability of a machine learning model.
Many formalizations have been proposed in the literature to describe feature-space attacks, either as optimization problems~\cite{Biggio:Evasion, Carlini:Robustness} (see also~\autoref{sec:fs-attack} for details)  or game theoretic frameworks~\cite{dalvi2004adversarial}.

{\bf Problem-Space Attacks.}
Recently, research on adversarial ML has moved towards domains in which the feature mapping is not invertible or not differentiable.
Here, the adversary needs to modify the objects in the problem space (i.e., input space) without knowing exactly how this will affect the feature space.
This is known as the \emph{inverse feature-mapping} problem~\cite{Tygar:Adversarial, Biggio:Evasion, Konrad:Attribution}.

Many works on problem-space attacks have been explored on different domains: text~\cite{TextBugger, alzantot2018generating}, PDFs~\cite{Maiorca:Bag, Maiorca:PDF, Laskov:PDF, dang2017morphing,  Evans:EvadeML}, Windows binaries~\cite{Battista:EXE,MalConv,Rosenberg:Generic}, Android apps~\cite{Battista:SecSVM, Papernot:ESORICS, Yang:Malware}, NIDS~\cite{fogla2006polyblending, apruzzese2018evading, apruzzese2019evading, corona2013adversarial}, ICS~\cite{zizzo2019adversarial}, and Javascript source code~\cite{Konrad:Attribution}.

However, each of these studies has been conducted empirically and followed some inferred best practices: while they share many commonalities, it has been unclear how to compare them and what are the most relevant characteristics that should be taken into account while designing such attacks.
Our formalization (\autoref{sec:formalization}) aims to close this gap, {and we show how it can be used} to describe representative feature-space and problem-space attacks from the literature (\autoref{sec:instances}).

{\bf Adversarial Android Malware.}
This paper also proposes a novel adversarial problem-space attack in the Android domain (\autoref{sec:apg-android}); our attack overcomes limitations of existing proposals, which are evidenced through our formalization.
The most related approaches to our novel attack are on attribution~\cite{Konrad:Attribution}, and on adversarial malware generation~\cite{Yang:Malware,Rosenberg:Generic, Papernot:ESORICS}.
\cite{Konrad:Attribution} do \emph{not} consider malware detection, but design a set of simple mutations to change the programming style of an application to match the style of a target developer (e.g., replacing \emph{for} loops with \emph{while} loops).
This strategy is effective for attribution, but is insufficient for malware detection as altering stylometric properties alone would not evade a malware classifier which captures program semantics.
Moreover, it is not feasible to define a hardcoded set of transformations for all possible semantics---which may also leave artifacts in the mutated code.
Conversely, our attack relies on automated software transplantation to ensure plausibility of the generated code and avoids hardcoded code mutation artifacts.

{\cite{Papernot:ESORICS} perform minimal modifications that preserve semantics, and only modify single lines of code in the Manifest; {but these may be easily detected and removed due to unused permissions or undeclared classes}. Moreover, they limit their perturbation to 20 features, whereas our problem-space constraints represent a more realistic threat model.

\cite{Yang:Malware} propose a method for adversarial Android malware generation. Similarly to us, they rely on \emph{automated software transplantation}~\cite{barr2015automated} and evaluate their adversarial attack against the DREBIN classifier~\cite{Arp:Drebin}.
However, they do not formally define which semantics are preserved by their transformation, and their approach is extremely unstable, breaking the majority of  apps they mutate (e.g., they report failures after 10+ modifications on average---which means they would likely not be able to evade Sec-SVM~\cite{Battista:SecSVM} which on average requires modifications of 50+ features).

Moreover, {the code is unavailable, and the paper lacks details required for reevaluating the approach}, including any clear descriptions of {preprocessing robustness}.
Conversely, our attack is resilient to the insertion of a large number of features (\autoref{sec:experiments}), preserves dynamic app semantics through opaque predicates (\autoref{sec:semantics}), {and is resilient against static program  analysis} (\autoref{sec:opaque}).

\cite{Rosenberg:Generic} propose a black-box adversarial attack against  {Windows malware classifiers that rely on API sequence call analysis---an evasion strategy that is also applicable to similar Android classifiers}. In addition to the limited focus on API-based sequence features, their problem-space transformation leaves two major artifacts which could be detected through program analysis: the addition of no-operation instructions (\textit{no-ops}), and patching of the import address table (IAT).

Firstly, the inserted API calls need to be executed at runtime and so contain individual no-ops hardcoded by the authors following a practice of ``security by obscurity'', which is known to be ineffective~\cite{KerckhoffsLaCM, Carlini:Evaluating}; intuitively, they could be detected and removed by identifying the tricks used by attackers to perform no-op API calls (e.g., reading 0 bytes), or by filtering the ``dead'' API calls (i.e., which did not perform any real task) from the dynamic execution sequence before feeding it to the classifier.
Secondly, to avoid requiring access to the source code, the new API calls are inserted and called using IAT patching. However, all of the new APIs must be included in a separate segment of the binary and, as IAT patching is a known malicious strategy used by malware authors~\cite{eresheim2017evolution}, IAT calls to non-standard dynamic linkers or multiple jumps  from the IAT to an internal segment of the binary would immediately be identified as suspicious.
Conversely, our attack does not require hardcoding and by design {is resilient against traditional non-ML program analysis techniques}.

{\bf Defences against adversarial samples}
A different approach from SecSVM which aims to create more robust models can be identified by the strategy described in \cite{tong2019improving}, in which authors leverage on "conserved features" (those that cannot be unilaterally modified without compromising
malicious functionality), in order to alter the used feature space.  However the considered domain is different, they do evaluation on malicious PDF files, which implies that also the considered feature representation is really different from the one we are considering. While PDF feature space includes also structural features, Drebin feature space lack of such and so we could think as all features to be "conserved features". For this reason, such reasoning is hardly applicable to the evaluated scenario of this paper.

{\bf Impact of realistic adversarial samples on adversarial training}
We are not the first researchers to investigate such research field, e.g. \cite{dyrmishi2023empirical} \cite{lucas2023adversarial}. Both work explore the impact of feature space generated adversarial samples compared to problem space ones in the malware domain and end to a different conclusion. 

Focusing on Adversarial Training, \cite{dyrmishi2023empirical} make a wide scope research evaluating the different impact on adversarial training of realistic and unrealistic adversarial training samples on random forest classifiers in multiple domains: text classification, malicious traffic and windows malware domain. They evaluate different attack strategies and make an evaluation of the hardened model and verify if the adversarial training granted robustness against realistic attacks. They claim that unrealistic samples do not suffice to harden the classifier in the malware and text classification domain, while they are enough in the malicious traffic domain.
\cite{lucas2023adversarial} research the same problem focusing entirely in the windows malware domain training a DNN on raw-binary  feature representation. They consider multiple state-of-the-art problem-space attacks - In-Place Replacement (IPR), Displacement (Disp) and Kreuk - and they conclude that it is not necessary to train on the highest-effort attacks to still gain substantial robustness against problem-space attacks, implying that feature space adversarial samples are enough in order to provide robustness. They also claim that early adversarial training harmed natural TPR, and training with more batches is important to recover accuracy on the original data.
This contrasting results can be explained analysing the differences between the two approaches. Indeed, they consider a different classifier and a different feature representation, which underlines the importance of the considered model and the representation used in order to evaluate the impact of adversarial training.

\section{Availability}
\label{sec:availability}

We release the code and data of our approach to other
researchers by responsibly sharing a private repository. The
project website with instructions to request access is at:
\url{https://s2lab.kcl.ac.uk/projects/intriguing}.

\section{Conclusions}

Since the seminal work that evidenced intriguing properties of neural networks~\cite{szegedy2013intriguing}, the community has become more widely aware of the brittleness of machine learning in adversarial settings~\cite{Biggio:Wild}.

To better understand real-world implications across different application domains, {we propose a novel formalization of problem-space attacks as we know them today, that enables comparison between different proposals and lays the foundation for more principled designs in subsequent work}. We uncover new relationships between feature space and problem space, and provide necessary and sufficient conditions for the existence of problem-space attacks. Our novel problem-space attack shows that automated generation of adversarial malware at scale is a realistic threat---taking on average less than 2 minutes to mutate a given malware example into a variant that can evade a hardened state-of-the-art classifier.

We also try to understand the effectiveness of adversarial learning and the impact of leveraging on realistic adversarial samples. The gathered evidence underline the benefits of the usage of problem-space adversarial samples for adversarial training for the considered models, highlighting that the impact may vary depending on the considered representation, model and adopted methodology.

Future work will explore this reasoning operating in different feature spaces, training different models and considering different feature space attacks, pursuing the underline relation between training and the representation taken into account.

\section*{Acknowledgments}

This research has been partially supported by UK EPSRC Grant EP/X015971/1, by the German Federal Ministry of Education and Research under the grant BIFOLD23B, by the Deutsche Forschungsgemeinschaft (DFG, German Research Foundation) under Germany's Excellence Strategy -- EXC 2092 CASA -- 390781972, and by the IFI program of the German Academic Exchange Service (DAAD).

\clearpage

\AtNextBibliography{\small}

\printbibliography

\begin{appendices}

\section{Symbol Table}
\label{app:symbol}

\autoref{tab:symbols} provides a reference for notation and major symbols used throughout the paper.

\begin{table}
\centering
\scriptsize
\caption{Table of symbols.}
\label{tab:symbols}
\begin{tabular}{|l||p{6cm}|}
  \hline
  {\sc Symbol} & {\sc Description} \\
  \hline \hline
  $\mathcal{Z}$ & Problem space (i.e., input space). \\ \hline
  $\mathcal{X}$ & Feature space $\mathcal{X} \subseteq \mathbb{R}^n$. \\ \hline
  $\mathcal{Y}$ & Label space. \\ \hline
  $\varphi$ & Feature mapping function $\varphi: \mathcal{Z} \longrightarrow \mathcal{X}$.\\ \hline
  $h_i$ & Discriminant function $h_i:\mathcal{X}\longrightarrow \mathbb{R}$ that assigns object $\bm{x} \in \mathcal{X}$ a score in $\mathbb{R}$ (e.g., distance from hyperplane) that represents fitness to class $i \in \mathcal{Y}$.  \\ \hline
  $g$ & Classifier $g:\mathcal{X}\longrightarrow \mathcal{Y}$ that assigns object $\bm{x} \in \mathcal{X}$ to class $y\in \mathcal{Y}$. Also known as \emph{decision function}. It is defined based on the output of the discriminant functions $h_i, \forall i \in \mathcal{Y}$ .  \\ \hline
  $\mathcal{L}_y$ & Loss function $\mathcal{L}_y:\mathcal{X} \times \mathcal{Y} \longrightarrow \mathbb{R}$ of object $\bm{x} \in \mathcal{X}$ with respect to class $y \in \mathcal{Y}$. \\\hline
  $f_{y,\kappa}$ & Attack objective function $f_{y,\kappa}: \mathcal{X} \times \mathcal{Y} \times \mathbb{R} \longrightarrow \mathbb{R}$ of object $\bm{x}\in \mathcal{X}$ with respect to class  $y \in \mathcal{Y}$ with maximum confidence $\kappa \in \mathbb{R}$. \\ \hline
  $f_{y}$ & Compact notation for $f_{y,0}$. \\ \hline
  $\Omega$ & Feature-space constraints. \\ \hline
  $\bm{\delta}$ & $\bm{\delta} \in \mathbb{R}^n$ is a symbol used to denote a feature-space perturbation vector. \\ \hline
  $\bm{\eta}$ & Side-effect feature vector. \\ \hline \hline
  $T$ & Transformation $T:\mathcal{Z} \longrightarrow \mathcal{Z}$. \\ \hline
  $\seqT$ & Transformation sequence $\seqT = T_n \circ T_{n-1} \circ \dots \circ T_1$. \\ \hline
  $\mathcal{T}$ & Space of available transformations. \\ \hline
  $\Upsilon$ & Suite of automated tests $\tau \in \Upsilon$ to verify preserved semantics. \\ \hline
  $\Pi$ & Suite of manual tests $\pi \in \Pi$ to verify plausibility. In particular, $\pi(z)=1$ if $z \in \mathcal{Z}$ is plausible, else $\pi(z)=0$. \\ \hline
  $\Lambda$ & Set of {preprocessing operators $\artifactremoval \in \Lambda$ for which $z \in \mathcal{Z}$ should be resistant (i.e., $\artifactremoval(\seqT(z))=\seqT(z)$)}. \\ \hline

  $\Gamma$ & Problem-space constraints $\Gamma$, consisting of $\{ \Pi, \Upsilon, \mathcal{T}, \Lambda \}$. \\ \hline \hline
  $\mathcal{D}$ & Training dataset. \\ \hline
  $\bm{w}$ & Model hyper-parameters. \\ \hline
  $\Theta$ & Knowledge space. \\ \hline
  $\theta$ & Threat model assumptions $\theta \in \Theta$; more specifically, $\theta = (\mathcal{D}, \mathcal{X}, g, \bm{w})$. A \emph{hat} symbol is used if only estimates of parameters are known. See \autoref{app:threatmodel} for details.\\
  \hline \hline
\end{tabular}
\end{table}

\section{Threat Model}
\label{app:threatmodel}

The threat model must be defined in terms of attacker \emph{knowledge} and \emph{capability}, as in related literature~\cite{Biggio:Wild, FAIL, Carlini:Evaluating}. While the attacker knowledge is represented in the same way as in the traditional feature-space attacks, their capability also includes the problem-space constraints $\Gamma$. For completeness, we report the threat model formalization proposed in Biggio and Roli~\cite{Biggio:Wild}.

{\bf Attacker Knowledge.}
	We represent the knowledge as a set $\theta \in \Theta$	which may contain (i) training data $\mathcal{D}$, (ii) the feature set $\mathcal{X}$, (iii) the learning algorithm $g$, along with the loss function $\mathcal{L}$ minimized during training, (iv) the model parameters/hyperparameters $\bm{w}$. A parameter is marked with a \emph{hat} symbol if the attacker knowledge of it is limited or only an estimate (i.e., $\hat{\mathcal{D}}$, $\hat{\mathcal{X}}$, $\hat{g}$, $\hat{\bm{w}}$).
	There are three major scenarios~\cite{Biggio:Wild}:
	\begin{itemize}
	\item \emph{Perfect Knowledge (PK) white-box attacks}, in which the attacker knows all parameters and $\theta_{PK} = (\mathcal{D},\mathcal{X},g,\bm{w})$.
	\item \emph{Limited Knowledge (LK) gray-box attacks}, in which the attacker has some knowledge on the target system. Two common settings are LK with Surrogate Data (LK-SD), where $\theta_{LK-SD}=(\hat{\mathcal{D}},\mathcal{X},g,\hat{\bm{w}})$, and LK with Surrogate Learners, where $\theta_{LK-SL}=(\hat{\mathcal{D}},\mathcal{X},\hat{g},\hat{\bm{w}})$. {Knowledge of the feature space and the ability to collect surrogate data, $\theta \supseteq (\hat{\mathcal{D}}, \mathcal{X})$, enables the attacker to perform \emph{mimicry attacks} in which the attacker manipulates examples to resemble the high density region of the target class~\cite{Biggio:Evasion,fogla2006polyblending}.}
	\item \emph{Zero Knowledge (ZK) black-box attacks}, where the attacker has no information on the target system, but has some information on which kind of feature extraction is performed (e.g., only static analysis in programs, or structural features in PDFs). In this case, $\theta_{LK}=(\hat{\mathcal{D}},\hat{\mathcal{X}},\hat{g},\hat{\bm{w}})$.
	\end{itemize}
Note that $\theta_{PK}$ and $\theta_{LK}$ imply knowledge of any defenses used to secure the target system against adversarial examples, depending on the degree to which each element is known~\cite{carlini2017bypassingten}.

{\bf Attacker Capability.} The capability of an attacker is expressed in terms of his ability to modify feature space and problem space, i.e., the attacker capability is described through feature-space constraints~$\Omega$ and problem-space constraints~$\Gamma$.

We observe that the attacker's knowledge and capability can also be expressed according to the {\sf FAIL}~\cite{FAIL} model as follows: knowledge of \emph{Features} $\mathcal{X}$ ({\sf F}), the learning \emph{Algorithm} $g$ ({\sf A}), \emph{Instances} in training $\mathcal{D}$ ({\sf I}), \emph{Leverage} on feature space and problem space with $\Omega$ and $\Gamma$ ({\sf L}).

More details on the threat models can be found in~\cite{Biggio:Wild,FAIL}.

\section{Theorem Proofs}
\label{app:proofs}

{\bf Proof of \autoref{eq:nc}.}
We proceed with a proof by contradiction.
Let us consider a problem-space object $z \in \mathcal{Z}$ with features $\bm{x} \in \mathcal{X}$, which we want to misclassify as a target class $t \in \mathcal{Y}$.
Without loss of generality, we consider a low-confidence attack, with desired attack confidence $\kappa=0$ (see Equation~\autoref{eq:objfun}).
We assume by contradiction that there is no solution to the feature-space attack; more formally, that there is no solution $\bm{\delta}^*=\arg\min_{\bm{\delta} \in \mathbb{R}^n : \bm{\delta} \models \Omega} f_{t}(\bm{x} + \bm{\delta})$ that satisfies $f_{t}(\bm{x}+\bm{\delta}^*) < 0$.
We now try to find a transformation sequence $\seqT$ such that $f_{t}(\varphi(\seqT(z))) < 0$.
Let us assume that $\seqT^*$ is a transformation sequence that corresponds to a successful problem-space attack. By definition, $\seqT^*$ is composed by individual transformations: a first transformation $T_1$, such that $\varphi(T_1(z)) = \bm{x} + \bm{\delta}_1$; a second transformation $T_2$ such that $\varphi(T_2(T_1(z)) = \bm{x} + \bm{\delta}_1 + \bm{\delta}_2$; a $k$-th transformation $\varphi(T_k( \cdots T_2(T_1(z)))) = \bm{x} + \sum_k \bm{\delta}_k$.
We recall that the feature-space constraints are determined by the problem-space constraints, i.e., $\Gamma \vdash \Omega$, and that, with slight abuse of notation, we can write that $\Omega \subseteq \Gamma$; this means that the search space allowed by $\Gamma$ is smaller or equal than that allowed by $\Omega$.
Let us now replace $\sum_k \bm{\delta}_k$ with $\bm{\delta}^\dagger$, which is a feature-space perturbation corresponding to the problem-space transformation sequence $\seqT$, such that $f_t(\bm{x}+\bm{\delta}^\dagger) < 0$ (i.e., the sample is misclassified). However, since the constraints imposed by $\Gamma$ are stricter or equal than those imposed by $\Omega$, this means that $\bm{\delta}^\dagger$ must be a solution to  $\arg\min_{\bm{\delta} \in \Omega} f_{t}(\bm{x} + \bm{\delta})$ such that $f_{t}(\bm{x}+\bm{\delta}^\dagger) < 0$. However, this is impossible, because we hypothesized that there was no solution for the feature-space attack under the constraints $\Omega$. Hence, having a solution in the feature-space attack is a \emph{necessary condition} for finding a solution for the problem-space attack.

{\bf Proof of \autoref{eq:nsc}.}
The existence of a feature-space attack (\autoref{eq:necessary-condition}) is the necessary condition, which has been already proved for~\autoref{eq:nc}.
Here, we need to prove that, with \autoref{eq:sufficient-condition}, the condition is  sufficient for the attacker to find a problem-space transformation that misclassifies the object.
Another way to write \autoref{eq:sufficient-condition} is to consider that the attacker knows transformations that affect individual features only (modifying more than one feature will result as a composition of such transformations). Formally, for any object $z \in \mathcal{Z}$ with features $\varphi(z)=\bm{x} \in \mathcal{X}$, for any feature-space dimension $X_i$ of $\mathcal{X}$, and for any value $v \in domain(X_i)$, let us assume the attacker knows a valid problem-space transformation sequence $\seqT:\seqT(z) \models \Gamma, \varphi(\seqT(z))=\bm{x}'$, such that:
\begin{align}
x_i' = x_i+v, \qquad  &x_i \in \bm{x}, x_i' \in \bm{x}'\\
x_j' = x_j, \qquad  &\forall j \neq i, x_j \in \bm{x}, x_j' \in \bm{x}'
\end{align}
Intuitively, these two equations refer to the existence of a problem-space transformation $\seqT$ that affects only one feature $X_i$ in $\mathcal{X}$ by any amount $v \in domain(X_i)$. In this way, given \emph{any} adversarial feature-space perturbation $\bm{\delta}^*$, the attacker is sure to find a transformation sequence that modifies each individual feature step-by-step. In particular, let us consider $idx_0, \dots, idx_{q-1}$ corresponding to the $q > 0$ values in $\bm{\delta}^*$ that are different from 0 (i.e., values corresponding to an actual feature-space perturbation). Then, a transformation sequence $\seqT:\seqT(z) \models \Gamma, \seqT = \seqT^{idx_q-1} \circ \seqT^{idx_{q-2}} \circ \dots \circ \seqT^{idx_0}$ can always be constructed by the attacker to  satisfy $\varphi(\seqT(z))=\bm{x}+\bm{\delta}^*$. We highlight that we do not consider the existence of a specific transformation in $\mathcal{Z}$ that maps to $\bm{x}+\bm{\delta}^*$ because that may not be known by the attacker; hence, the attacker may never learn such a specific transformation. Thus, \autoref{eq:sufficient-condition} must be valid for all possible perturbations within the considered feature space.

\section{Opaque Predicates Generation}
\label{app:opaque}

We use opaque predicates~\cite{Moser:Limits_of_static_analysis} as inconspicuous conditional statements always resolving to {\tt False} to preserve dynamic semantics of the Android applications.

To ensure the intractability of such an analysis, we follow the work of~\cite{Moser:opaque} and build opaque predicates using a formulation of the 3-SAT problem such that resolving the truth value of the predicate is equivalent to solving the NP-complete 3-SAT problem.

The $k$-satisfiability ($k$-SAT) problem asks whether the variables of a Boolean logic formula can be consistently replaced with \texttt{True} or \texttt{False} in such a way that the entire formula evaluates to \texttt{True}; if so the formula is \emph{satisfiable}. Such a formula is easily expressed in its conjunctive normal form:
$$\textstyle\bigwedge\nolimits_{i=1}^m (V_{i1} \vee V_{i2} \vee ... \vee V_{ik})\,,$$
where $V_{ij} \in \{v_1, v_2, ..., v_n\}$ are Boolean variables and $k$ is the number of variables per clause.

To consistently generate NP-Hard k-SAT problems we use \emph{Random k-SAT}~\cite{selman:ksat} in which there are 3 parameters: the number of variables $n$, the number of clauses $m$, and the number of literals per clause $k$.

To construct a 3-SAT formula, $m$ clauses of length 3 are generated by randomly choosing a set of 3 variables from the $n$ available, and negating each with probability $50\%$.
An empirical study by \cite{selman:ksat} showed that $n$ should be at least 40 to ensure the formulas are hard to resolve. Additionally, they show that formulas with too few clauses are \emph{under-constrained} while formulas with too many clauses are \emph{over-constrained}, both of which reduce the search time. These experiments led to the following conjecture. For further details please refer to the conference version.

\section{DREBIN and models Implementation Details}
\label{app:secsvm}

We have access to a working Python implementation of DREBIN based on \emph{sklearn}, \emph{androguard}, and \emph{aapt}, and we rely on \emph{LinearSVC} classifier with C=1.

We now describe the details of our implementation of the Sec-SVM approach~\cite{Battista:SecSVM}.
To have have full control of the training procedure, we approximate the linear SVM as a \emph{single-layer} neural network (NN) using PyTorch~\cite{paszke2017automatic}.
We recall that the main intuition behind Sec-SVM is that classifier weights are distributed more evenly in order to force an attacker to modify more features to evade detection.
Hence, we modify the training procedure so that the Sec-SVM weights are bounded by a \emph{maximum weight value} $k$ at each training optimization step.
Similarly to~\cite{Battista:SecSVM}, we perform feature selection for computational efficiency, since PyTorch does not support sparse vectors. We use an $l_2$ (Ridge) regularizer to select the top 10,000 with negligible reduction in AUROC.
This performance retention follows from recent results that shows SVM tends to overemphasize a subset of features~\cite{melis2018explaining}.
To train the Sec-SVM, we perform an extensive hyperparameter grid-search: with Adam and Stochastic Gradient Descent (SGD) optimizers; training epochs of 5 to 100; batch sizes from $2^{0}$ to $2^{12}$; learning rate from $10^{0}$ to $10^{-5}$.
We identify the best single-layer NN configuration for our training data to have the following parameters: Stochastic Gradient Descent (SGD), batch size 1024, learning rate $10^{-4}$, and 75 training epochs.
We then perform a grid-search of the Sec-SVM hyperparameter $k$ (i.e., the maximum weight absolute value~\cite{Battista:SecSVM}) by clipping weights during training iterations.
We start from $k=w_{max}$, where $w_{max}=\max_i(w_i)$ for all features $i$; we then continue reducing $k$ until we reach a weight distribution similar to that reported in~\cite{Battista:SecSVM}, while allowing a maximum performance loss of $2\%$ in AUROC. In this way, we identify the best value for our setting as $k=0.2$.

In \autoref{sec:experiments}, \autoref{fig:roc} reported the AUROC for the DREBIN classifier~\cite{Arp:Drebin} in SVM and Sec-SVM modes.
The SVM mode has been evaluated using the \texttt{LinearSVC} class of scikit-learn~\cite{scikit-learn} that utilizes the LIBLINEAR library~\cite{fan2008liblinear}; as in the DREBIN paper~\cite{Arp:Drebin}, we use hyperparameter C=1.
The performance degradation of the Sec-SVM compared to the baseline SVM shown in \autoref{fig:roc} is in part related to the defense itself (as detailed in~\cite{Battista:SecSVM}), and in part due to minor convergence issues (since our single-layer NN converges less effectively than the LIBLINEAR implementation of scikit-learn).
We have verified with~\cite{Battista:SecSVM} the correctness of our Sec-SVM implementation and its performance, for the analysis performed in this work.
Due to limitations of the developing libraries, in order to apply correctly adversarial training on LinearSVM we have re-implemented it using PyTorch as we could not rely on the sklearn implementation which would not grant access to the training epoch easily. As an implementation detail, we decided to not generate adversarial samples during the first epochs in order to facilitate the convergence of the model. In the context of hyperparameters have applied the same reasoning as SecSVM.

\section{Defender evaluation settings}
\label{app:def_settings}

In \autoref{defender_strategies}, we examine different strategies for enhancing classifier robustness, which include varying approaches to generating adversarial examples, levels of confidence, and hardening methods. We adjust the proportion of adversarial examples from the training set, in 10\% increments from 10 to 100\%, to see how it affects robustness. The study also considers other variables such as the model used, hardening approach, and transformation type.

Here in details:
\begin{itemize}
    \item Confidence level: We consider the two confidence levels defined in \autoref{attack_confidence} and employ them both for the hardening and test phase:
    \begin{itemize}
        \item High-Low: the model is hardened using high confidence adversarial points  and tested against a low confidence evasion 
        \item Low-Low : the model is hardened using low confidence adversarial points and tested against a low confidence evasion
        \item High-High : the model is hardened using high confidence adversarial points  and tested against a high confidence evasion
        \item Low-High : the model is hardened using low confidence adversarial points and tested against a high confidence evasion    
    \end{itemize}
    \item Model: we are considering both LinearSVM and Sec-SVM
    \item Hardening strategy: we are considering both adversarial training and adversarial retraining 
    \item Transformations set: we are evaluating the hardening through feature space transformation and problem space transformations while always evaluating against adversarial points generated through realistic transformations.

\end{itemize}

\section{Attack Algorithms}
\label{app:algos}

As previously described in the paper, we adopt a greedy approach in order to create adversarial samples, both in the feature space and in the problem space. Algorithm~\ref{alg:initialization} and Algorithm~\ref{alg:attack} describe in detail the two main phases of our search strategy: organ harvesting and adversarial program generation.
For the sake of simplicity, we describe a low-confidence attack, i.e., the attack is considered successful as soon as the classification score is below zero.
It is immediate to consider high-confidence variations (as we evaluate in~\autoref{sec:experiments}).

\vspace{1em}
\begin{algorithm}[H]
\caption{Initialization (Ice-Box Creation)}
\tiny
\label{alg:initialization}
\DontPrintSemicolon
\SetAlgoNoEnd
\KwIn{Discriminant function $h(\bm{x})=\bm{w}^T\bm{x}+b$, which classifies $\bm{x}$ as malware if $h(\bm{x}) > 0$, otherwise as goodware. Minimal app $z_{min} \in \mathcal{Z}$ with features $\varphi(z_{min})=\bm{x}_{min}$.}
\SetKwInOut{Parameter}{Parameters}
\Parameter{Number of features to consider $n_f$; number of donors per-feature $n_d$.}
\KwOut{Ice-box of harvested organs with feature vectors.}
\vspace{2pt}
\hrule
\vspace{2pt}
ice-box $\gets$ \{\} {Empty key-value dictionary.}\\
$L \gets$ List of pairs ($w_i$, $i$), sorted by increasing value of $w_i$.\\
$L' \gets$ First $n_f$ elements of $L$, then remove any entry with $w_i \geq 0$.  \\
\For{$(w_i, i)$ in $L'$}{
	ice-box[i] $\gets$ [] {Empty list for gadgets with feature $i$.} \\
	\While{\emph{length(ice-box[i])}$<n_d$}{
		$z_j \gets$ Randomly sample a benign app with feature $x_i=1$.\\
		Extract gadget $\rho_j \in \mathcal{Z}$ with feature $x_i=1$ from $z_j$. \\
		$s \gets $ Software stats of $\rho_j$\\
		$z' \gets$ Inject gadget $\rho_j$ in app $z_{min}$.\\
		$(\bm{x}_{min} \vee \bm{e}_i \vee \bm{\eta}_j) \gets \varphi(z')$ {$\bm{e}_i$ is a one-hot vector.}\\
		$\bm{r}_j \gets (\bm{e}_i \vee \bm{\eta}_j) \gets \varphi(z') \wedge \neg \bm{x}_{min}${Gadget features obtained through set difference.}\\
		\If{$h(\bm{r}_j) > 0$}{
			Discard the gadget;
		}\Else{
		Append $(\rho_j, \bm{r}_j, s)$ to ice-box[i]. {Store gadget}
		}
	}
}

\Return ice-box;
\end{algorithm}

\vspace{0.5em}
\begin{algorithm}[H]
\caption{Attack (Adv. Program Generation)}
\tiny
\label{alg:attack}
\DontPrintSemicolon
\SetAlgoNoEnd
\KwIn{Discriminant function $h(\bm{x})=\bm{w}^T\bm{x}+b$, which classifies $\bm{x}$ as malware if $h(\bm{x}) > 0$, otherwise as goodware. Malware app $z \in \mathcal{Z}$. Ice-box $G$.}
\SetKwInOut{Parameter}{Parameters}
\Parameter{Problem-space constraints.}
\KwOut{Adversarial app $z' \in \mathcal{Z}$ such that $h(\varphi(z'))<0$.}
\vspace{2pt}
\hrule
\vspace{2pt}
$\mathcal{T} \gets$ Transplantation through gadget addition. \\
$\Upsilon \gets$ Smoke test through app installation and execution in emulator.\\
$\Pi \gets$ Plausibility by-design through code consolidation.\\
$\Lambda \gets$ Artifacts from last column of \autoref{tab:instantiations}. \\
$\Gamma \gets \{ \mathcal{T}, \Upsilon, \Pi, \Lambda \}$\\

$s_z \gets$ Software stats of $z$\\
$\bm{x} \gets \varphi(z)$\\
$L \gets $ [] {Empty list.}\\
$\seqT(z) \gets $ Empty sequence of problem-space transformations.\\
\For{$(\rho_j, \bm{r}_j, s)$ in $G$}{
	$\bm{d}_j \gets \bm{r}_j \wedge \neg \bm{x}$ {Feature-space contribution of gadget $j$.}\\
	$score_j \gets h(\bm{d}_j)$ {Impact on decision score.}\\
	Append the pair $(score_j,i,j)$ to $L$ {Feature $i$, Gadget $j$.}
}
\For{$(score_j, i, j)$ in $L'$}{
	\If{z has $x_i=1$}{
		Do nothing; {Feature $i$ already present.}
	}\ElseIf{z has $x_i=0$}{
		$(\rho_j, \bm{r}_j, s) \gets$ element $j$ in ice-box $G$\\
		\If{\emph{check\_feasibility(}$s_z$, $s$\emph{)} is True}{
				$\bm{x} \gets (\bm{x} \vee \bm{e}_i \vee \bm{\eta}_j)$ {Update features of $z$.} \\
				Append transplantation $T \in \mathcal{T}$ of gadget $\rho_j$ in $\seqT(z)$.\\
				\If{$h(\bm{x}) < 0 $}{
					Exit from cycle; {Attack gadgets found.}
				}
		}

	}
}
$z' \gets$ Apply transformation sequence $\seqT(z)$ {Inject chosen gadgets.}\\
\If{$h(\varphi(z')) < 0$ {\bf and} $\seqT(z) \models \Gamma$}{
	\Return z'; {Attack successful.}
}\Else{
	\Return Failure;
}

\end{algorithm}

\end{appendices}

\end{document}